\documentclass[]{aastex631}
\usepackage{CJK}

\usepackage{graphicx}	
\usepackage{amsmath}	
\usepackage{amssymb}	
\usepackage{float}
\usepackage{subfigure}
\usepackage{multirow}
\usepackage{enumerate}
\usepackage{color}
\usepackage{amsmath}
\usepackage{tabularx}
\usepackage{booktabs}
\usepackage{threeparttable}

\submitjournal{ApJS}

\shorttitle{46 new Open Clusters Identified by Hybrid pyUPMASK and RF}
\shortauthors{Chi et al.}

\graphicspath{{./figures}{/}}

\begin{document}
\begin{CJK*}{UTF8}{gbsn}

\begin{sloppypar}
\title{Identify 46 New Open Clusters Candidates In Gaia EDR3 Using pyUPMASK and Random Forest Hybrid Method}

\correspondingauthor{Shoulin Wei, Feng Wang and Zhongmu Li}
\email{weishoulin@astrolab.cn, fengwang@gzhu.edu.cn}

\author[0000-0001-7343-7332]{Huanbin Chi ({\CJKfamily{gbsn}迟焕斌})}
\affiliation{School of Management and Economics, Faculty of Information Engineering And Automation, \\ Kunming University of Science and Technology, Kunming 650500, China}

\author[0000-0002-3547-4025]{Shoulin Wei ({\CJKfamily{gbsn}卫守林})}
\affiliation{School of Management and Economics, Faculty of Information Engineering And Automation, \\ Kunming University of Science and Technology, Kunming 650500, China}

\author[0000-0002-9847-7805]{Feng Wang ({\CJKfamily{gbsn}王锋})}
\affiliation{Center for Astrophysics and Great Bay Center of National Astronomical Data Center, \\ Guangzhou University, Guangzhou 510006 , China}
\affiliation{Peng Cheng Laboratory, Shenzhen, 518000, China}

\author[0000-0002-0240-6130]{Zhongmu Li ({\CJKfamily{gbsn}李忠木})} 
\affiliation{Institute of Astronomy, Dali University,  Dali, 671003,  China}



\begin{abstract}

Open clusters (OCs) are regarded as tracers to understand stellar evolution theory and validate stellar models. In this study, we presented a robust approach to identifying OCs. A hybrid method of pyUPMASK and RF is first used to remove field stars and determine more reliable members. An identification model based on the RF algorithm built based on 3714 OC samples from Gaia DR2 and EDR3 is then applied to identify OC candidates. The OC candidates are obtained after isochrone fitting, the advanced stellar population synthesis (ASPS) model fitting, and visual inspection.
Using the proposed approach, we revisited 868 candidates and preliminarily clustered them by the friends-of-friends algorithm in Gaia EDR3. Excluding the open clusters that have already been reported, we focused on the remaining 300 unknown candidates. From high to low fitting quality, these unrevealed candidates were further classified into Class A (59), Class B (21), and Class C (220), respectively.
 As a result, 46 new reliable open cluster candidates among classes A and B are identified after visual inspection.

\end{abstract}

\keywords{open cluster --- GaiaEDR3 --- FoF --- pyUPMASK --- Random Forest}


\section{Introduction}
\label{sec:intro}

Open clusters (OCs) were simultaneously formed from the same molecular cloud, and gravitationally bound stellar systems were born in the same starburst. Therefore, OCs are a kind of natural laboratory and valuable tracers for studying galaxies' structure, chemical composition, and dynamical evolution, as well as providing validation and constraints on the model of evolutionary astrophysical~\citep{Spina2022}.
For example, young OCs are often assigned to analyze the structure of galaxies. They are also used as testbeds to study stellar evolution, allowing us to investigate the boundary conditions necessary for new star formation~\citep{Cantat-Gaudin2018A&A}. 
Besides providing information about the height and radial extension of the galactic disk, old OCs also provide information about the chemical history of the galaxy, e.g., the relationship between age and metallicity, mixing processes, and cluster destruction processes caused by interactions with other clusters.

However, due to the limitations of Galactic dust extinction and contamination of field stars (foreground and background stars), identifying OCs is still a challenging issue~\citep{Deb2022}. Many efforts have been made to hunt OC candidates from the Gaia Second Data Release (Gaia DR2,~\cite{gaiadr2}) and Gaia (Early) Data Release 3 (Gaia (E)DR3, ~\cite{gaiaedr3}). 

Various methods based on an unsupervised machine learning clustering algorithm have been used to search for OCs.
One of the most successful searching methods is the Density-Based Spatial clustering of Applications with Noise algorithm (DBSCAN)~\citep{Ester1996A}. A series of DBSCAN variants based on DBSCAN is capable of effectively identifying OCs~\citep{Castro-Ginard2018A, Castro-Ginard2019, Castro-Ginard2020, He2021, Castro-Ginard2022}.
In addition,~\cite{Kounkel, Kounkel2020, Hunt2021} used the improved method Hierarchical Density-Based Spatial Clustering of Applications with Noise (HDBSCAN) based on DBSCAN to detect many new clusters.
~\cite{Cantat-Gaudin2018} applied an unsupervised membership assignment code (UPMASK) to the Gaia DR2 data contained within the fields of those clusters. 
~\citet{Gao2018AJ, Gao2020} identified cluster members using a Gaussian mixture model (GMM) clustering method. 

Besides DBSCAN and its variants, the friends-of-friends (FoF) algorithm ~\citep{fof} is also applied to identify OCs. 
~\cite{Liu&Pang2019} found 76 candidate star clusters from Gaia DR2 using the FoF algorithm.
~\cite{paper1} also used the FoF algorithm to perform a blind search for OCs in Gaia EDR3 within 25 degrees of the Galactic plane. As a result, 61 new OCs were reported among the 868 candidates.
The advantage of the FoF algorithm to group stars is that clustering considers a five-dimensional weighted parameter space of parallax, position, and velocity.
However, its disadvantage is that it is not sensitive to the size of the cluster radius and the uneven distribution of star density which changes at different distances
$b_{FoF}$ is a unique hyperparameter of the FoF algorithm, significantly impacting the clustering results.

\citet{Liu&Pang2019} proposed a high-performance approach (i.e., SHiP) to calculate $b_{FoF}$ in each data region, which has been successful in finding many open clusters \citep{paper1}. $b_{FoF}$ is defined as 
\begin{center}
    $b_{FoF}=0.2 \times\left(\frac{1}{N_{star}}\right)^{0.2}$
\end{center}
where ${N_{star}}$ is the number of stars in each region. However, the approach has some minor deficiencies. For example, it is relatively ineffective in searching for member stars in some sparse spaces of star clusters. 
On the other hand, the minimum number of clusters is set to a predetermined fixed value, e.g. 50. This may lead to some member stars being incorrectly included in a cluster during the merging process of the method.

In this study, we referred to the study of \cite{paper1} to obtain a data set of open cluster candidates using the FoF algorithm. We then presented an improved hybrid algorithm to identify OCs from open cluster candidates found by FoF more robustly.
The rest of the paper is organized as follows. 
We described the method in Section ~\ref{sec:2}, including the membership determination method and identification model for OCs.
The results are presented in Section ~\ref{sec:3}, which includes sample handling, isochrone-fitting, cross-matching, and visual inspection.
We discussed the results in Section ~\ref{sec:discussion}.
Finally, a conclusion is covered in Section ~\ref{sec:conclusions}.

\section{An Improved Identification Method For Open Cluster}
\label{sec:2}

\begin{figure}[htbp]
\centering
\includegraphics[width=120mm]{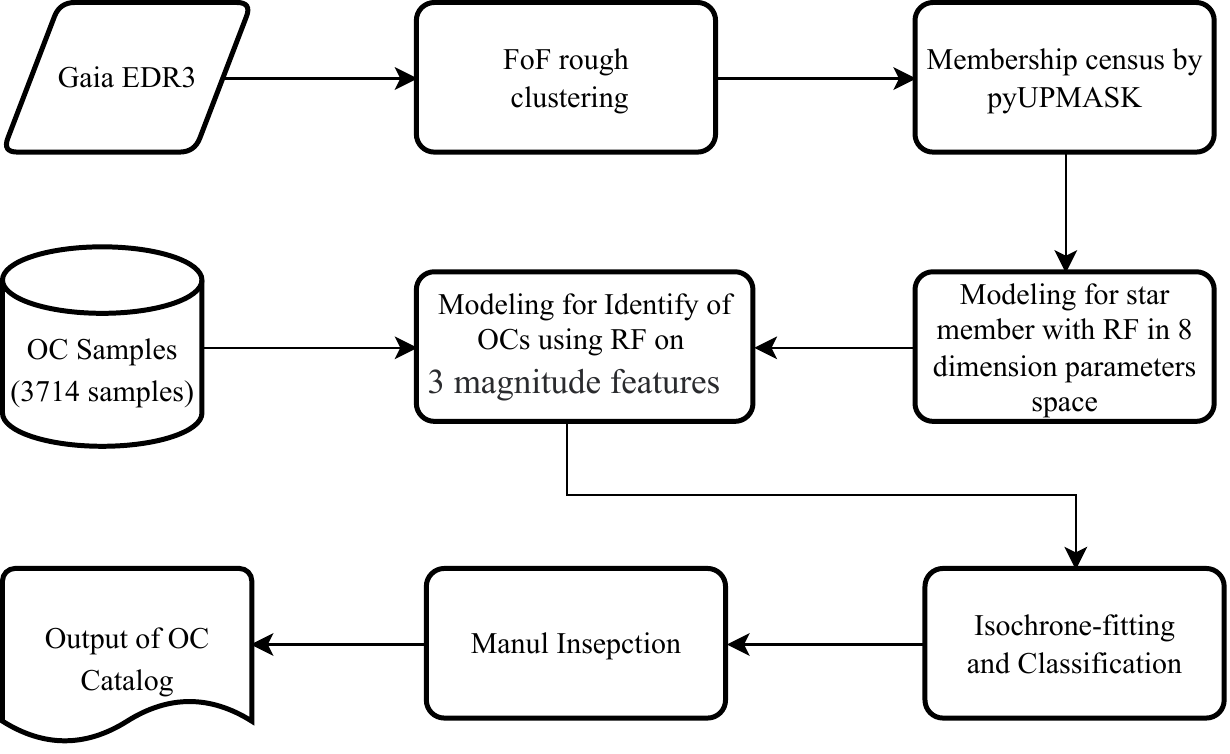}
\caption{Flowchart of the hybrid identification approach. The process of this approach consists of 3 stages. Stage 1 deals with rough FoF clustering within Gaia EDR3. Then, each OC's membership is identified in stage 2. After refining the star member, we implement recognition of OCs, including a submodule of modeling for identifying OCs, isochrone fitting, and visual inspection in stage 3.}
\label{flowchart}
\end{figure}

In light of the previous literature mentioned in the first section, we realized that improving OC recognition accuracy with machine learning methods faces two problems. One is improving the quality of samples, and the other is optimizing the final OC identification model. Therefore, we presented a robust identification approach for OCs and showed the flow chart in Figure \ref{flowchart}.

\subsection{Member Star Determination Method}
Determining which of the stars in a cluster candidate are member stars and which are field stars is a challenging task.
We presented a hybrid algorithm based on pyUPMASK~\citep{Pera2021} and random forest (RF) to eliminate false member stars (field stars) among those star clusters. We created a dataset for each cluster candidate, respectively. Each star in the dataset was labeled as a member star or field star using PyUPMASK. RF algorithm modeled star data to predict more credible member stars.

1.pyUPMASK is a Python package for Unsupervised Photometric Membership Assignment in Stellar
Clusters (UPMASK)~\cite{Krone-Martins2014} used to estimate the membership probability of each input star. 
pyUPMASK has been widely used in the determination of member stars of OC based on astrometric parameters \citep{Cantat-Gaudin2019A&A, He2022ApJS, Bai2022RAA, Dias2022}.

The essence of the UPMASK algorithm is to calculate the kernel density estimation likelihood (KDE) of the member stars of the OC candidates.
The membership probability of stars is iteratively calculated by
\begin{center}
    $P_{star}=\dfrac{K D E_{m}}{\left(K D E_{m}+K D E_{n m}\right)} $,
\end{center}
 where $P_{star}$, $KDE_{m}$, and $KDE_{n m}$ are the membership probability of star, KDE of the members, and field star, respectively. 
In general, besides coordinates, pyUPMASK requires at least two dimensions of data of any type to estimate membership probabilities. We chose coordinates and proper motion as the input of pyUPMASK for some distant open cluster candidates (excluding parallax and photometric data), which has been confirmed effective in \cite{Perren2022}.
To speed up the computing process, we performed parallel computing using Mpi4py \citep{2008MPI}. 
It allows us to process a volume of data that does not fit in the memory of a single machine as well.

After the pyUPMASK calculation, we refer to the \cite{Gao2018AJ} and set the value of membership probability as 0.8. The stars with a membership probability greater than 0.8 are labeled member stars, and others are labeled field stars, which is different from the practices of \cite{He2021}.


2. Based on the label and eight parameters extracted from the catalog, i.e., $l, b, \varpi, \mu_{a^{*}}, \mu_{\delta}, G, G_{RP}$, and $G_{BP}$, we built an RF model for each OC. We regarded the identification of cluster members as a supervised binary classification problem. The RF model predicted the final member stars for each OC.

\subsection{Identification model for Open Cluster}\label{Section_identification_model}
We applied an RF classifier to detect the OCs among the potential candidates. An RF classifier model was built with 1229 positive OC samples collected from Gaia DR2~\citep{Cantat-Gaudin2020} and 628 positive OC samples collected from Gaia EDR3 \citep{Castro-Ginard2018A, Castro-Ginard2019, Castro-Ginard2020, Castro-Ginard2021, Castro-Ginard2022}.
The negative OC samples, which have the same number of positive OC samples, are synthesized with stars, assuming their spatial distribution is a random uniform distribution. 
We finally obtained a trainset of 3714 OC samples for modeling.

\begin{figure}
     \centering
     \includegraphics[scale=0.6]{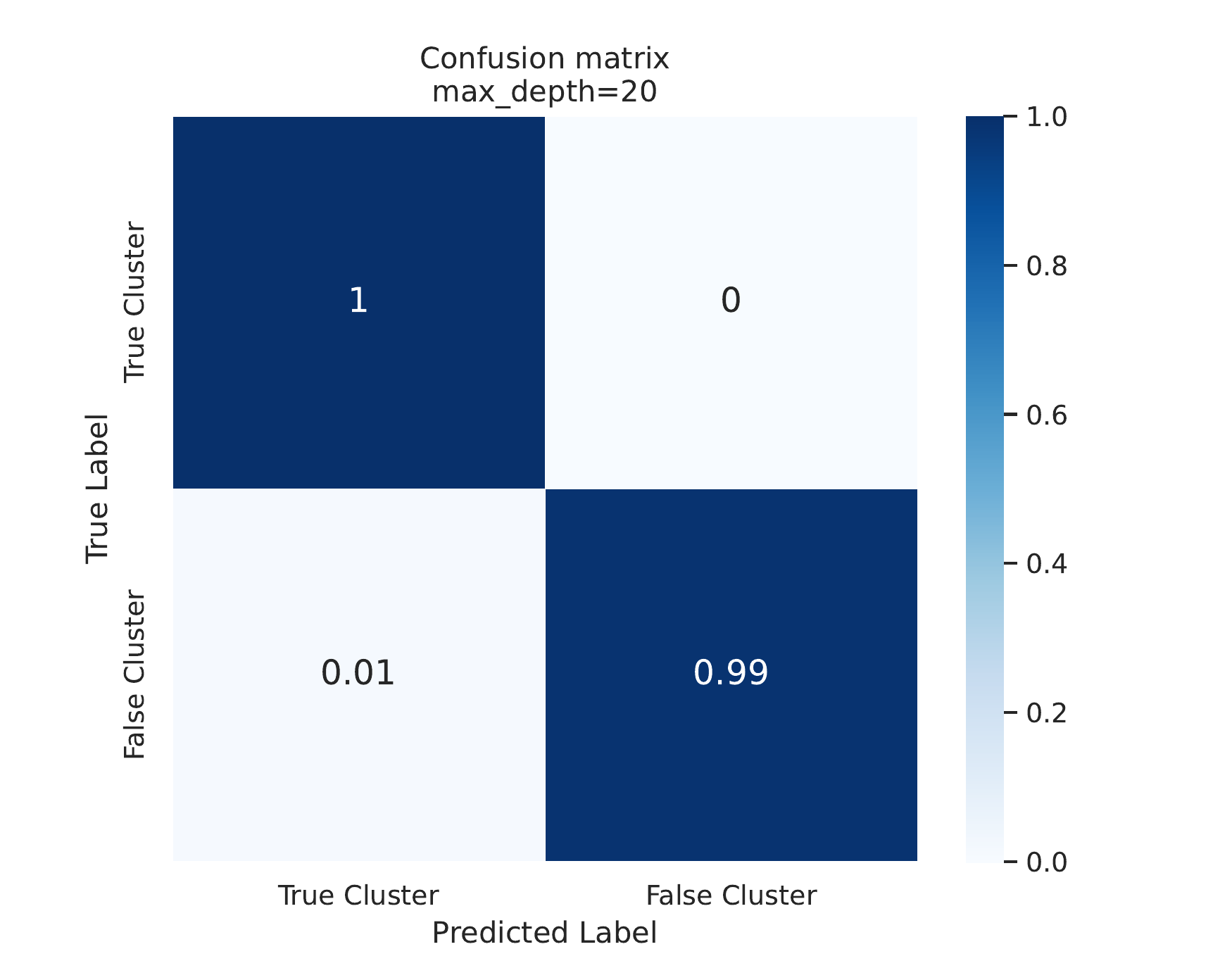}
     \caption{The confusion matrix.}
     \label{fig:cm}
 \end{figure}

We calculated a confusion matrix (see Figure~\ref{fig:cm}) as evaluation metrics to evaluate the model. The precision of the model is 99.35\%. 
 {}

\section{Identification of Open Cluster And Results}
\label{sec:3}
\subsection{Data Preparation For Open Cluster Candidates}

We strictly followed the data preparation method of \cite{paper1} to generate the OC candidate dataset.
First, we filter suitable samples to exclude observational artifacts due to faintness in Gaia EDR3 based on the stellar position, self, and parallax parameters ($\varpi$ between 0.2 mas and 0.7 mas, $G$ $<$ 18 mags and $\mu_\alpha c o s\delta$ $<$ 30 mas yr$^{-1}$, $\mu_\delta$ $<$ 30 mas yr$^{-1}$). Meanwhile, considering most OCs are centered near the galactic disc, we set $\left| b \right| <$ 25 degrees. 

Second, based on 180 million sources extracted, 
to facilitate this procedure, we divide the entire search volume into multiple data regions,
we roughly divided the data into many data regions according to galactic longitude ($l$), galactic latitude ($b$), and parallax ($\varpi$).
The number of divisions for $\varpi$, $b$, and $l$ are 8, 8, and 64, respectively. To avoid splitting the clusters into different regions as much as possible, each of the data regions must not be smaller than two times the typical cluster size  (20 pc) ~\citep{portegies2010young}. To deal with potential clusters located at the boundaries of the region, we set an overlapping region for the two adjacent regions with size ($\varpi$ size 0.2 mas, $l$, and $b$ size 10 pc). 
After carrying out the above scheme, the whole search volume is divided into 4091 data regions.

We used FoF clustering for each data region to find local clusters and aggregate them to obtain 3597 candidate clusters. After cross-matching, we got 807 new candidates for the cluster. 
Using the membership determination RF model, we removed the field stars from each of these 807 candidates. We then classified these 807 candidates using an identification RF model. 801 candidates were classified as open clusters, and the model rejected the other 6 candidates. 

\subsection{Validation Of Open Cluster Candidates}
After obtaining 801 open cluster candidates, we cross-matched our candidates with open cluster catalogs published. 
We consider an OC to be positionally matched to a cataloged one if their centers lie within a circle of radius r = 0.5 degrees and the rest of the astrometric mean parameters are compatible within 5$\sigma$, which is consistent with~\cite{Hunt2021, Castro-Ginard2021}. Here, $\sigma$ is the uncertainties quoted in both catalogs for each quantity. 

We gathered most of the known star cluster catalogues and labelled them as MWSC, CG2017, Hao3794, UBC series, CWNU, Dias1743, and Hao704, respectively. 
The pre-Gaia cluster catalogs (MWSC) contain 3006 star clusters gathered by ~\cite{Dias2012, Kharchenko2013}. CG20 and CG series include 2017 OCs~\citep{Cantat-Gaudin2020}. ~\cite{Hao2021} reports a much more extensive catalog containing 3794 OCs (Hao3794).  
Moreover, the UBC series, hunted by~\cite{Castro-Ginard2018A, Castro-Ginard2019, Castro-Ginard2020, Castro-Ginard2021}, totally reported 1274 OCs so far.
The CWNU~\citep{He2022} has reported 541 newly discovered open cluster candidates found on Galactic Disk Using Gaia DR2/EDR3 Data. 
~\cite{Dias2021} had reported a catalog that updated the parameters of 1743 open clusters based on Gaia DR2 in 2021 (Dias1743).
~\cite{Hao2022} reported 704 newly detected open clusters in the Galactic disk using Gaia EDR3 (Hao704). 

Similarly, we have the same cross-match method to some previous catalogs, i.e., ~\cite{Liu&Pang2019}, ~\cite{Ferreira2020}, ~\cite{Hao2020PASP},~\cite{Hunt2021}, ~\cite{Li2021}.
In particular, using the same method, we performed an integral crossover of 46 recently reported star clusters by ~\cite{He2022ApJS}. 
The 46 newly discovered clusters are located in high-Galactic latitude regions with $\left| b \right| >$20 degrees. In contrast, ours are located at $\left| b \right| < $25 degrees. Therefore, there cannot be any match between our cluster and any of the 46 newly discovered clusters.
MWSC catalogs contain 2976 cataloged objects local to galactic latitude 25 degrees gathered from different data sources.
Because they do not allow for sufficiently accurate comparisons in proper motion space, we only performed a 0.5 degree positional cross-match based on sky coordinates.

After cross-matching, 501 of the 801 candidates were already identified. We obtained 300 candidate clusters that have not been identified and reported, which is the data set for subsequent OC identification. 

\subsection{Color-Magnitude Diagrams fitting}
\label{cmd-fitting}

The color-magnitude diagrams (CMDs) were fitted by two independent approaches, i.e., the fitting based on isochrone models~\citep{Bressan2012} and the fitting based on the advanced stellar population synthesis (ASPS) model~\citep{Li2016A,2017RAA}. We first fitted CMDs by isochrone fitting method and then validated the results with ASPS. 

The isochrone fitting method is a mature and classic fitting method. In theory, member stars in the OCs are born from the same gas cloud in a single episode of star formation. Most of them are expected to follow a single isochrone in color-magnitude diagrams (CMDs) and have the same metallicity and age. 
Therefore, we used the isochrone-fitting method with the PARSEC theoretical isochrone models ~\citep{Bressan2012} updated by the Gaia EDR3 passbands using the photometric calibrations from ESA/Gaia to derive their physical parameters (age and metallicity). 
We applied a log-normal initial mass function~\cite{Chabrier2003} to generate an isochrone library from $log(\dfrac{t}{yr})=$ 6.0 to 11.13 at steps of $\Delta(log t ) = 0.03 $ while metal fractions from 0.002 to 0.042 with a step of 0.002.
Table~\ref{tab1} presents  parameters ranges and steps for isochrone fitting.
An objective fitting function
\begin{center}
    $\bar{d}^{2}=\dfrac{\sum_{k=1}^{n}\left(\mathbf{x}_{k}-\mathbf{x}_{k, n n}\right)^{2}}{n}$
\end{center}
was applied to 300 OC candidates, where $n$ is the
number of selected members in a cluster candidate, and ${x}_{k}$ and
${x}_{k, n n}$ are the positions of the member stars and the points on the
isochrone that are closest to the member stars, respectively.

ASPS is a model that contains different kinds of stellar populations, including single-star simple stellar populations (ssSSP); binary star stellar populations (bsSSP); single-star composite stellar populations (ssCSP); binary-star composite stellar populations (bsCSP); the simple stellar population of single, binary, and rotating stars (sbrSSP); and composite stellar population of single, binary, and rotating stars (sbrCSP). ASPS can be used to fit the CMDs and determine the cluster properties. Due to the large number of parameters considered, the fit of ASPS for CMDs takes a longer time.


\begin{figure*}

\begin{center}
\subfigure {
\includegraphics[width=7in,height=1.7in]{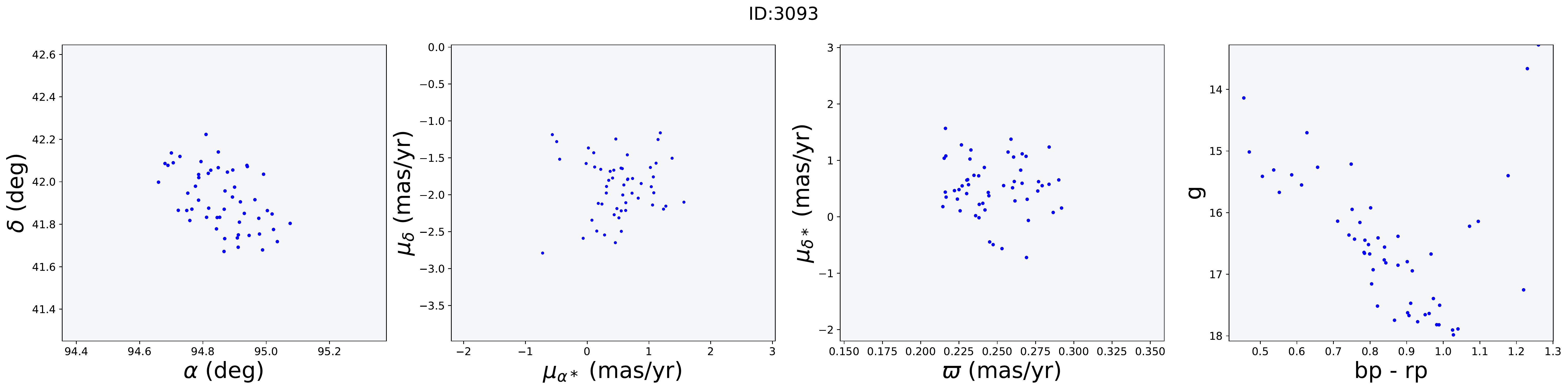}
}
\end{center}
\begin{center}
\subfigure {
\includegraphics[width=7in,height=1.7in]{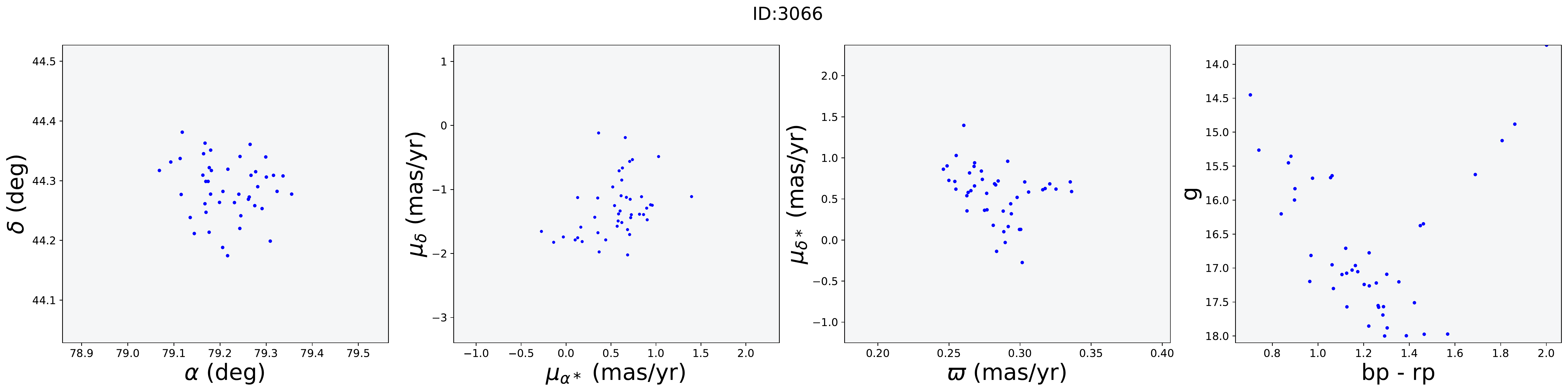}
}
\end{center}
\begin{center}
\subfigure {
\includegraphics[width=7in,height=1.7in]{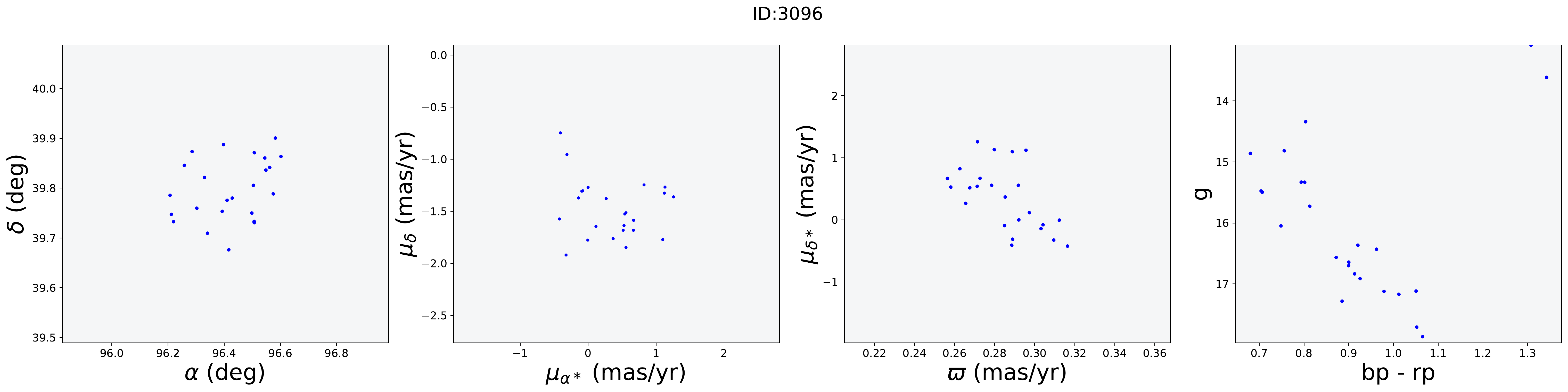}
}
\end{center}
\begin{center}
\subfigure {
\includegraphics[width=7in,height=1.7in]{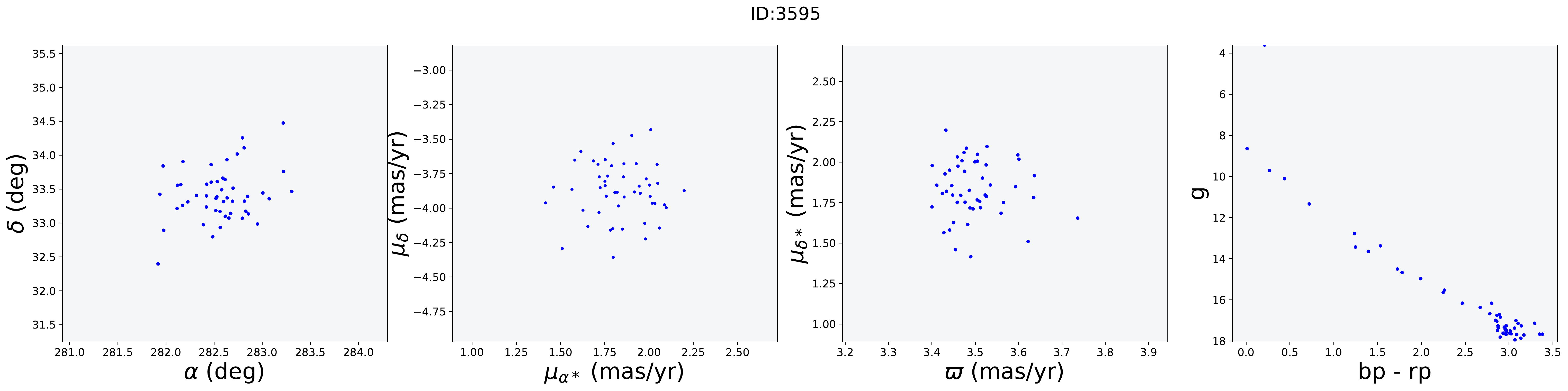}
}
\end{center}

\caption{Preliminary data check for our 801 OCs. ID 3093, 3066, 3096, and 3595 are listed, respectively. Leftmost plots:
position of the OC in ($\alpha$, $\delta$). Inner left plots: ($\varpi$ , $\mu_{a^{*}}$ ) distribution, whilst inner right plots: ($\mu_{a^{*}}$, $\mu_{\delta}$ ) distribution. Rightmost plots: CMD of OC.}

\label{precheck}
\end{figure*}

\subsection{Maunal Verification And Final Results}

After isochrone fitting, we classified OC candidates based on their CMD morphology and isochrone fitting results to facilitate our manual visual inspection. Referring to some previous literature~\citep{Liu&Pang2019, Castro-Ginard2020, He2022ApJS, He2022}, we applied the following parameters to the classification.

1. $n\_{star}$: Number of member stars that brighter than $G<17$ mag.

2. ${d}^2$: The average square of the distance between cluster stars and an isochrone used to measure the isochrone-fitting error.

3. $r_{\mathrm{n}}$: The narrowness of the MS in the CMD calculated as $\left|\dfrac{v_1}{v_2} \right|$.
$v_{1}$
and $v_{2}$ are the two eigenvalues of the covariance matrix of the distribution of stars in the CMD. 

We classified 300 open cluster candidates into three classes:

1. Class $A: n_{star} \geqslant 20, r_{\mathrm{n}}<0.1, \bar{d}^2<0.05$;

2. Class $B: n_{star} \geqslant 20, r_{\mathrm{n}}<0.1$

3. Class $C:$ other cases.

As a result, class A (20\%, 59/300) has a sufficient number of member stars and clear CMDs. The examples of class A are shown in the first row of Figure  See Figure A.1 in Appendix)).
Class B (7\%, 21/300) includes candidates with unclear isochrone fitting and loose CMD distribution (see examples second row in Figure A.1 of Appendix ).
Some candidates of class C are shown in the third row of Figure A.1 of Appendix. Given parameters of age and metallicity, class C obtained by isochrone fitting has a large uncertainty.




We performed a manual visual inspection of each of these candidates using the distribution of position distribution (PD), vector point diagram (VPD), CMD, and parallax with $\mu_{a^{*}}$ as in Figure~\ref{precheck}. Among the 80 candidates (Class A and Class B) mentioned above, 46 candidates were finally considered as possible real OCs (see Table~\ref{tab1}). Parts of these 46 candidates are shown in Figure ( See Figure A.2 in Appendix).

To further validate the results, we fitted the CMDs of these 46 OC candidates using the ASPS model. Comparing the results of ASPS-based fitting and isochrone fitting, 42 OCs are consistent (see Figure 3-6 in Appendix A).


4 cluster candidates with the ID of 3512, 3526, 3567, and 3595 are not well suited to ASPS models. However, these four cluster candidates give very reasonable results in isochrone fitting and also have a high probability of being clusters.
Figure A.7 in Appendix shows the results of isochrone fitting.

\begin{figure}
\centering
\includegraphics[scale=1.0]{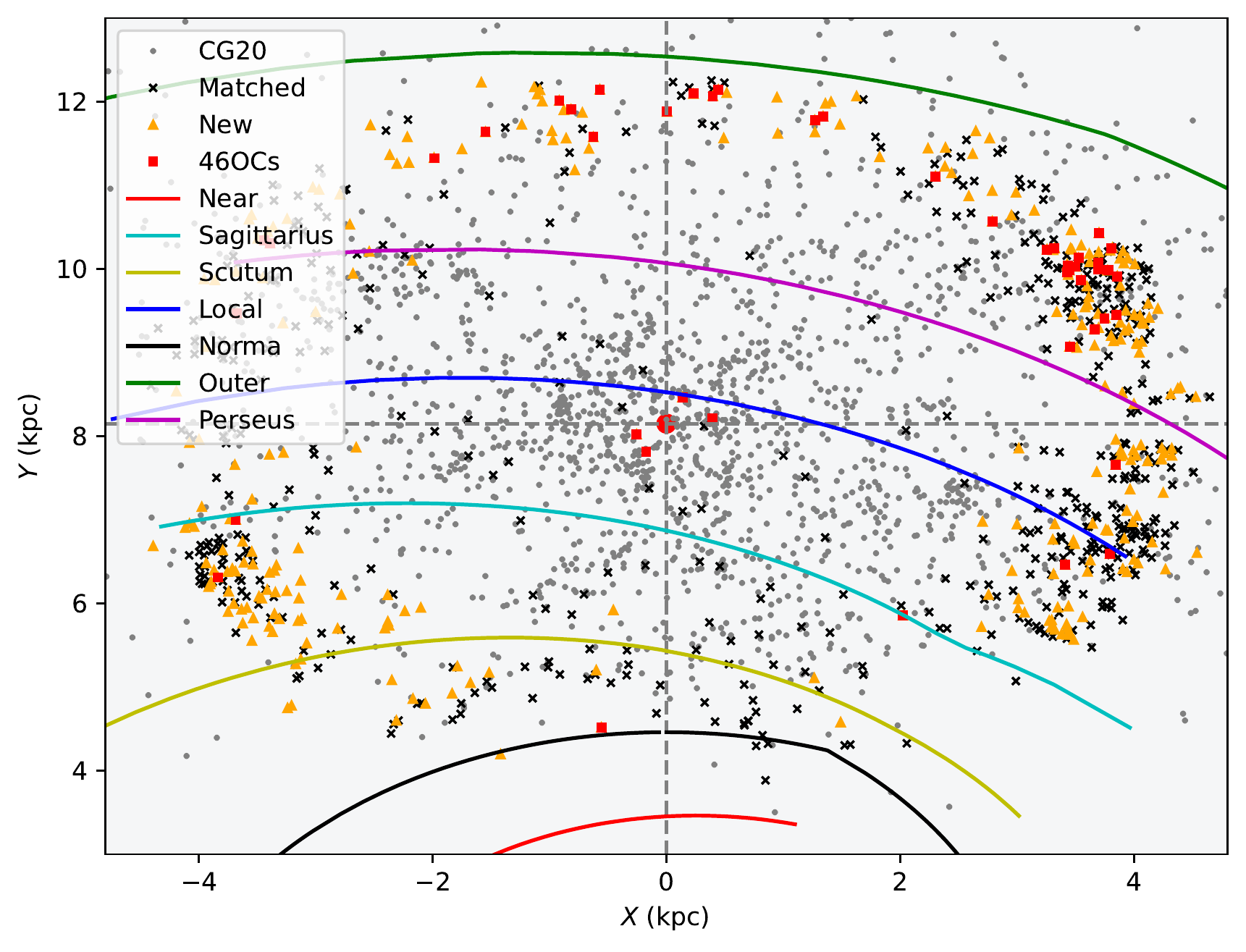}
\caption{Distribution of matched and new OCs in the Galactic $X-Y$ plane with an indication of spiral arms. Images viewed from the north pole of the galaxy, from which it rotates clockwise. We adopted the ~\cite{Reid2019ApJ} recommended values of $R\odot=8.15$ kpc and $Z\odot=5.5$ pc. The red point represents the position of the Sun.}
\label{XY_distribution}
\end{figure}

\begin{figure}
\centering
\includegraphics[scale=0.7]{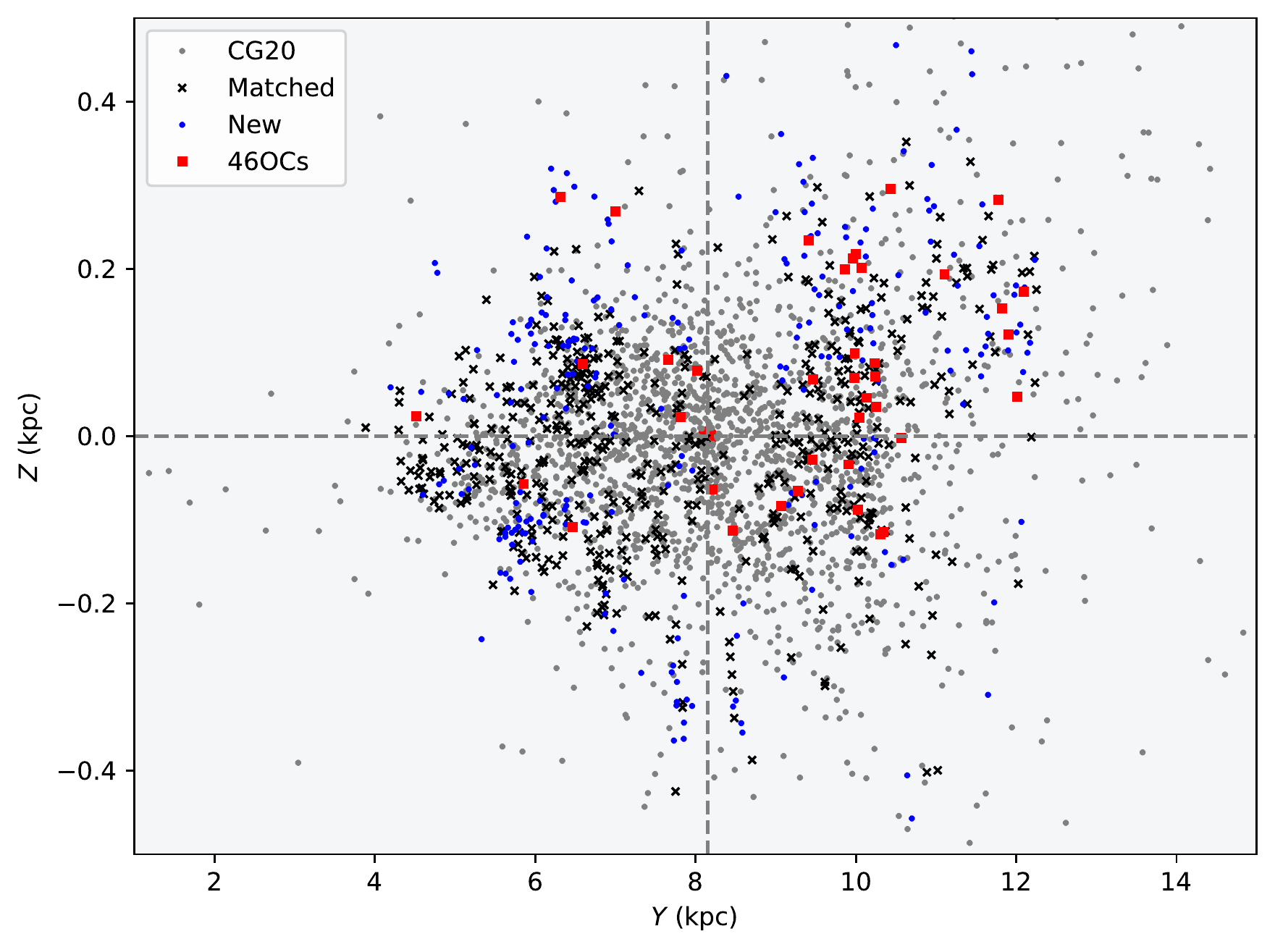}
\caption{Same as Figure ~\ref{XY_distribution}, but with an edge-on view.}
\label{YZ_distribution}
\end{figure}

\begin{figure}[htbp]
\centering  
\subfigure[Metallicity (Z) distribution of 46 newly identified cluster. Age unit: $log(Z/Z\odot)$]{   
\begin{minipage}{7cm}
\centering    
\includegraphics[scale=0.4]{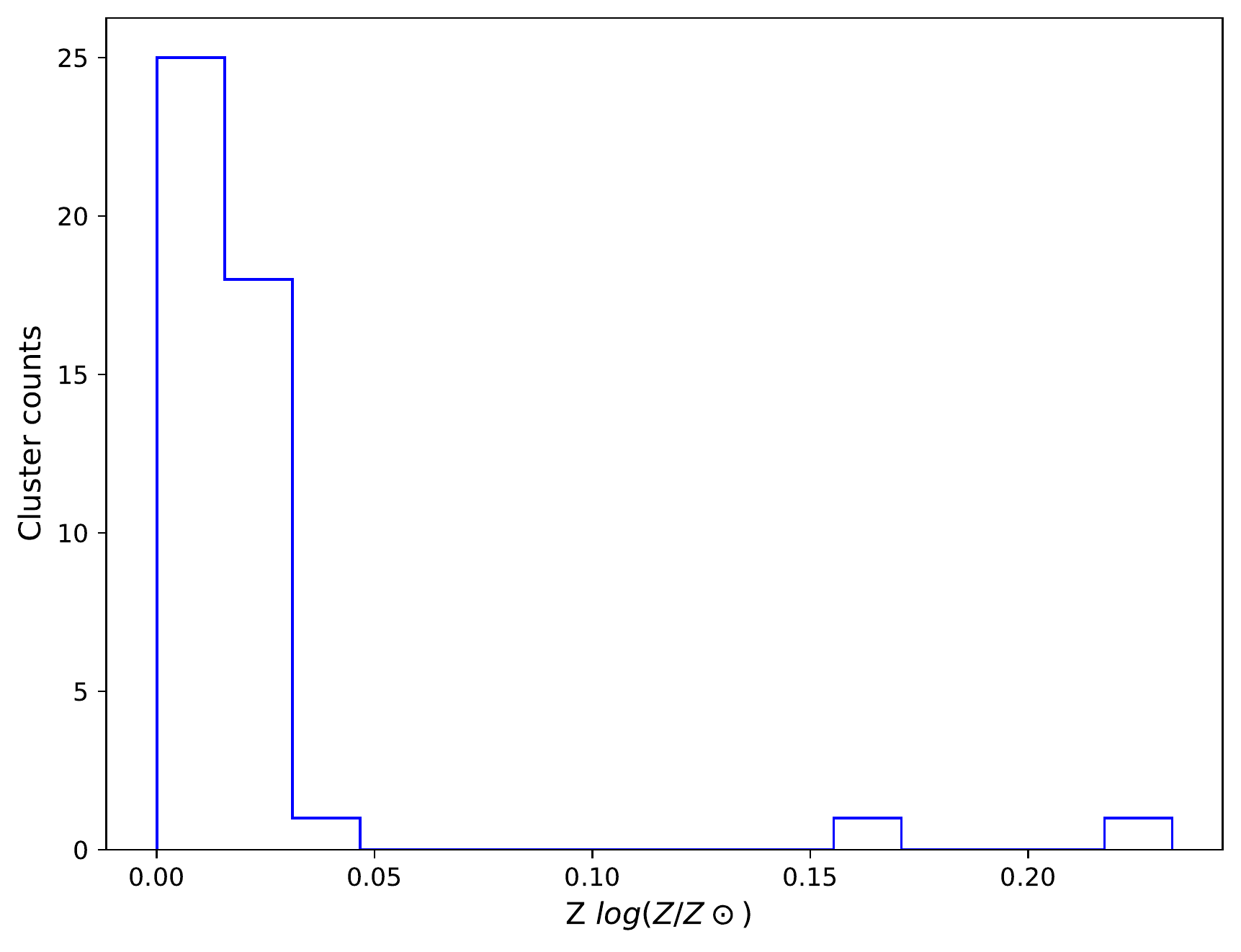}  
\end{minipage}
}
\subfigure[Age distribution of 46 newly identified clusters. Unit: $log(age/yr)$]{ 
\begin{minipage}{7cm}
\centering    
\includegraphics[scale=0.4]{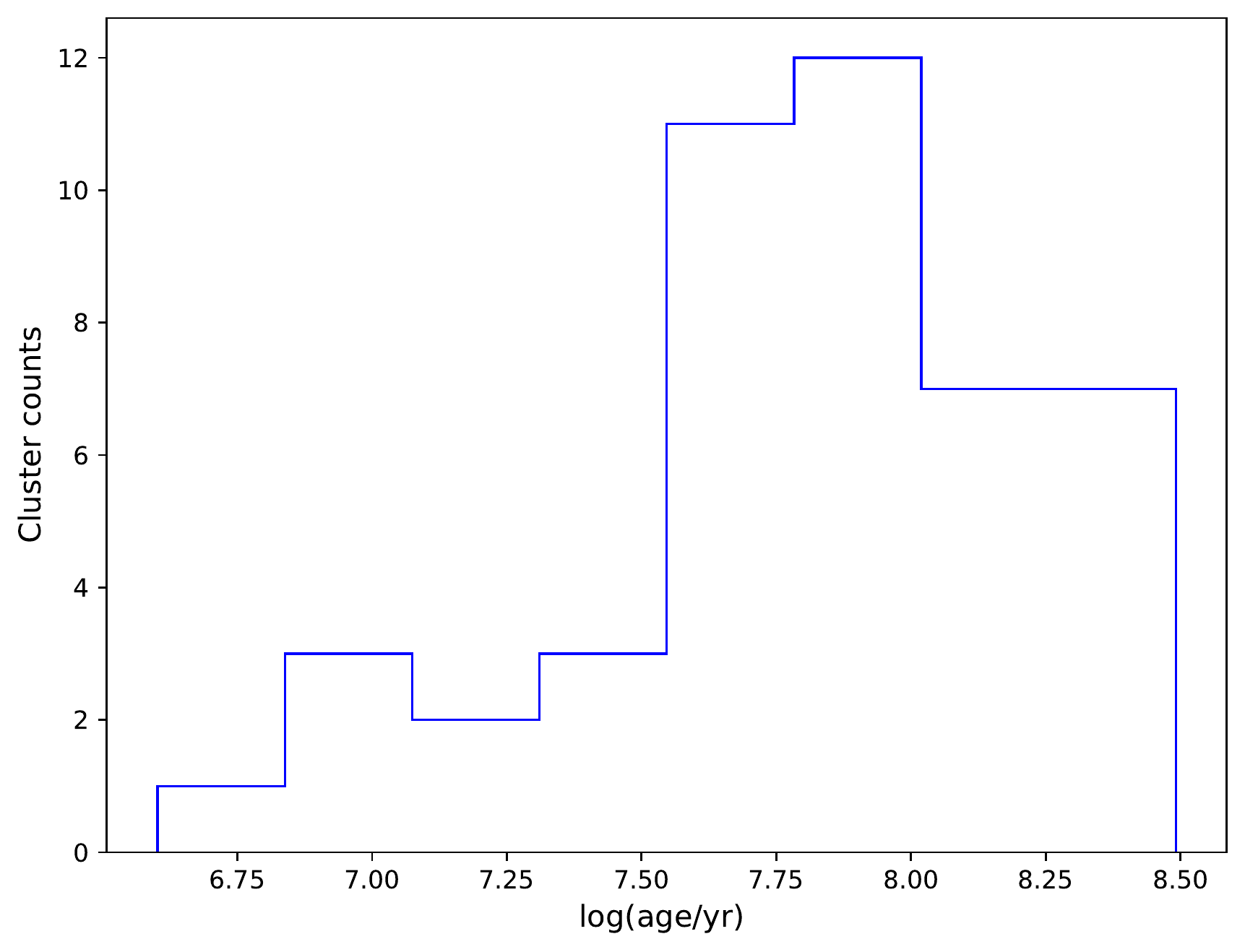}
\end{minipage}
}
\caption{Hisogram of Z and age for 46 OCs.}   
\label{fig:hisogram_age_z}   
\end{figure}

Figure~\ref{XY_distribution} and ~\ref{YZ_distribution} indicate most new candidates distributed between the Norma and Near arms. 
Figure~\ref{fig:hisogram_age_z}  shows the distributions of age and metallicity of the newly
identified OCs. We found that these OCs are younger than 3.0 Gyr (see Figure~\ref{fig:hisogram_age_z} (b)).
Additionally, most of them are metal-poor ( see Figure~\ref{fig:hisogram_age_z} (a)).

\begin{table}[htbp]
    \centering
    \caption{ Parameters of the final 46 New OCs in this work. Some of the best-fit and observed CMDs are compared in Figure A.3 of Appendix}
    \label{tab1}
    \begin{tabular}{l c c c c c c c c c c c c r}
        \toprule
        ~ & ID & ra & std\_ra & dec & std\_dec & ... & pmdec & std\_pmdec & N\_mem & $N_{17}$ & Z & age & class  \\ 
        
        ~ & ~ & {[}deg{]} & {[}deg] & {[}deg] & {[}deg] & ... & {[}mas $yr^{-1}$ ] & {[}mas $yr^{-1}$] & ~ & ~ & & {[}Gyr] & ~   \\ 
        \midrule
        0 & 0976 & 102.651 & 0.105 & 29.270 & 0.072 & ... & -1.971 & 0.349 & 27 & 21 & 0.0300 & 0.7000 & A \\ 
        1 & 1144 & 115.668 & 0.066 & -26.113 & 0.085 & ... & 3.367 & 0.561 & 31 & 22 & 0.0300 & 0.7000 & B \\ 
        2 & 1161 & 123.797 & 0.087 & -37.456 & 0.105 & ... & 3.199 & 0.417 & 71 & 28 & 0.0001 & 0.6000 & A \\ 
        3 & 1275 & 177.764 & 0.082 & -63.818 & 0.021 & ... & 0.860 & 0.335 & 40 & 21 & 0.0200 & 0.4000 & B \\  
        4 & 1382 & 12.025 & 0.223 & 61.186 & 0.023 & ... & -0.576 & 0.371 & 65 & 34 & 0.0001 & 1.2000 & B \\  
        5 & 1384 & 11.851 & 0.149 & 61.115 & 0.028 & ~ & -0.469 & 0.507 & 45 & 27 & 0.0100 & 0.9000 & A \\  
        6 & 1428 & 110.273 & 0.049 & -14.385 & 0.094 & ... & 1.429 & 0.574 & 81 & 49 & 0.0300 & 0.5000 & A \\  
        7 & 1438 & 116.845 & 0.051 & -24.774 & 0.098 & ... & 1.908 & 0.603 & 72 & 37 & 0.0040 & 0.9000 & A \\  
        8 & 1460 & 118.543 & 0.051 & -28.847 & 0.102 & ... & 2.833 & 0.430 & 54 & 26 & 0.0300 & 0.6000 & A \\  
        9 & 1481 & 122.233 & 0.080 & -33.798 & 0.097 & ... & 3.071 & 0.520 & 80 & 43 & 0.0100 & 0.8000 & A \\  
        10 & 1500 & 122.701 & 0.081 & -35.461 & 0.061 & ... & 2.890 & 0.472 & 45 & 26 & 0.0200 & 0.8000 & A \\  
        11 & 1746 & 225.492 & 0.032 & -60.023 & 0.015 & ... & -2.792 & 0.181 & 32 & 21 & 0.0360 & 0.040 & A \\  
        12 & 1991 & 90.700 & 0.067 & 27.281 & 0.086 & ... & -1.286 & 0.537 & 49 & 25 & 0.0040 & 2.3000 & A \\  
        13 & 2009 & 98.861 & 0.061 & 12.457 & 0.140 & ... & -0.503 & 0.437 & 52 & 28 & 0.0080 & 0.7000 & B \\  
        14 & 2060 & 118.380 & 0.041 & -24.904 & 0.082 & ... & 2.612 & 0.312 & 56 & 31 & 0.0040 & 0.5000 & A \\  
        15 & 2073 & 116.970 & 0.096 & -24.682 & 0.061 & ... & 2.117 & 0.500 & 42 & 25 & 0.0200 & 0.6000 & B \\  
        16 & 2075 & 116.885 & 0.048 & -24.181 & 0.104 & ... & 2.268 & 0.507 & 66 & 43 & 0.0300 & 0.8000 & A \\  
        17 & 2083 & 115.650 & 0.022 & -21.430 & 0.044 & ... & 2.733 & 0.406 & 35 & 27 & 0.0300 & 0.4000 & A \\  
        18 & 2084 & 115.716 & 0.078 & -21.029 & 0.057 & ... & 2.656 & 0.700 & 55 & 40 & 0.0040 & 0.1000 & B \\  
        19 & 2091 & 120.071 & 0.042 & -24.196 & 0.103 & ... & 1.799 & 0.638 & 72 & 41 & 0.0080 & 1.4000 & A \\  
        20 & 2100 & 119.923 & 0.107 & -24.668 & 0.066 & ... & 2.694 & 0.575 & 45 & 24 & 0.0100 & 0.4000 & A \\  
        21 & 2110 & 121.050 & 0.052 & -26.050 & 0.067 & ... & 2.961 & 0.572 & 39 & 24 & 0.0080 & 0.6000 & A \\  
        22 & 2130 & 119.134 & 0.069 & -26.984 & 0.057 & ... & 3.061 & 0.391 & 50 & 24 & 0.0040 & 1.2000 & B \\  
        23 & 2239 & 147.722 & 0.073 & -52.375 & 0.060 & ... & 3.398 & 0.391 & 54 & 25 & 0.0010 & 1.8000 & B \\  
        24 & 2324 & 171.305 & 0.080 & -59.957 & 0.086 & ... & 1.912 & 0.429 & 67 & 40 & 0.0080 & 1.2000 & B \\  
        25 & 2457 & 271.006 & 0.029 & -21.275 & 0.047 & ... & -1.940 & 0.366 & 47 & 35 & 0.0300 & 0.1000 & B \\  
        26 & 2761 & 344.905 & 0.131 & 60.889 & 0.078 & ... & -1.710 & 0.255 & 59 & 30 & 0.0300 & 0.3000 & B \\  
        27 & 2942 & 80.027 & 0.083 & 40.160 & 0.091 & ... & -1.703 & 0.633 & 49 & 24 & 0.0003 & 0.9000 & A \\  
        28 & 2944 & 77.990 & 0.116 & 40.402 & 0.130 & ... & -1.509 & 0.513 & 75 & 43 & 0.0010 & 1.0000 & B \\  
        29 & 2971 & 100.312 & 0.032 & 14.030 & 0.109 & ... & -1.002 & 0.358 & 48 & 29 & 0.0080 & 1.9000 & A \\  
        30 & 2972 & 107.696 & 0.054 & -3.101 & 0.106 & ... & 0.243 & 0.466 & 61 & 32 & 0.0300 & 0.4000 & A \\  
        31 & 2986 & 118.648 & 0.108 & -20.539 & 0.073 & ... & 1.891 & 0.530 & 55 & 31 & 0.0040 & 1.1000 & A \\  
        32 & 2987 & 120.702 & 0.060 & -25.329 & 0.090 & ... & 2.731 & 0.538 & 81 & 50 & 0.0300 & 0.3000 & B \\  
        33 & 2999 & 126.065 & 0.076 & -31.786 & 0.089 & ... & 2.781 & 0.539 & 53 & 22 & 0.0300 & 0.4000 & A \\  
        34 & 3012 & 294.733 & 0.072 & 29.585 & 0.051 & ... & -5.352 & 0.794 & 77 & 33 & 0.0300 & 1.1000 & A \\  
        35 & 3019 & 299.549 & 0.071 & 36.770 & 0.029 & ... & -4.633 & 0.734 & 71 & 30 & 0.0300 & 0.8000 & A \\  
        36 & 3078 & 71.701 & 0.087 & 60.871 & 0.041 & ... & -0.083 & 0.164 & 38 & 24 & 0.0080 & 1.4000 & A \\  
        37 & 3079 & 99.408 & 0.209 & 34.733 & 0.082 & ... & -1.757 & 0.468 & 41 & 23 & 0.0100 & 2.8000 & A \\  
        38 & 3082 & 102.310 & 0.231 & 29.745 & 0.114 & ... & -1.657 & 0.535 & 52 & 38 & 0.0100 & 2.9000 & A \\  
        39 & 3086 & 83.970 & 0.187 & 55.647 & 0.075 & ... & -1.304 & 0.486 & 46 & 21 & 0.0040 & 0.9000 & A \\  
        40 & 3093 & 94.860 & 0.101 & 41.919 & 0.141 & ... & -1.893 & 0.397 & 53 & 36 & 0.0080 & 1.9000 & A \\  
        41 & 3095 & 94.462 & 0.284 & 44.190 & 0.082 & ... & -1.812 & 0.444 & 48 & 29 & 0.0040 & 3.1000 & A \\  
        42 & 3512 & 80.643 & 0.378 & -1.214 & 0.746 & ... & -0.443 & 0.369 & 135 & 99 & 0.014 & 0.1700 & A \\  
        43 & 3526 & 118.280 & 1.352 & -47.348 & 1.316 & ... & 8.762 & 0.455 & 520 & 358 & 0.016 & 0.1480 & A \\  
        44 & 3567 & 277.912 & 0.136 & -3.912 & 0.206 & ... & -9.011 & 0.268 & 50 & 38 & 0.022& 0.0980 & A \\  
        45 & 3595 & 282.582 & 0.337 & 33.407 & 0.381 & ... & -3.878 & 0.204 & 51 & 22 & 0.026 & 0.2750 & A \\  
        \bottomrule
    \end{tabular}
    \begin{tablenotes}
        \footnotesize
        \item{Note: For each cluster, age and metallicity (Z) are determined. The $N_{mem}$ means the total number of member star in each cluster.
        The  $ N_{17}$ means the total number of member star with G $<$ 17 mag each cluster. The member stars of each OC can be downloaded online.}
      \end{tablenotes}
\end{table}

\section{Discussions And Future Works}
\label{sec:discussion}
\subsection{Approach Limitations}
We only use pyUPMASK on the 2-dimensional proper motion parameters space on the higher dimensional feature space instead of using the member star probability census for most clusters. This is because as the distance of the star increases, the uncertainty of its parameters, such as parallax, will increase significantly, introducing more uncertainty.
 
During the member stars census, not all the member stars of the cluster are identified by the RF model. Because in some cases, the membership probability after the census is greater than the threshold we set after running pyUPMASK. For such star clusters, we do not use the RF model. Instead, we use pyUPMASK for membership probability filtering in a 5-dimensional ($l, b, \varpi, \mu_{a^{*}}, \mu_{\delta}$) space. In another case, for those star clusters whose number of member stars is less than 10 after the calculation of the member probability satisfies the filtered condition of the probability threshold, we consider these star clusters as false and discard them.
 
A point that needs to be explained is that when training the RF model for the recognition of member stars, we adopt the weighted RF algorithm. This is because the training set of label samples is unbalanced, produced by pyUPMASK.

\begin{table}[htbp]
\label{parameters_set}
\centering
\caption{Parameters Ranges and Steps for Isochrone Fitting}
\begin{tabular}{lccr}
\toprule
    Parameter & Range& Step & Unit \\ 
\midrule
Age    &   6.000 -- 11.13  & 0.03 & log(t/yr)\\
Z      &  0.002 -- 0.042       & 0.002&    $[\mathrm{Fe} / \mathrm{H}]$  \\

\bottomrule
\end{tabular}
\end{table}

\subsection{Performance Analysis Of Member stars Determination}
To validate the proposed hybrid method, we first test it on a well-studied open cluster, i.e., M67 (NGC 2682), which is a well-studied open cluster whose members are publicly available in many studies~\citet{Castro-Ginard2020, Jadhav2021, Agarwal2021, Ghosh2022}.
We downloaded sources from Gaia EDR3 in a cone around the open cluster center within a radius of 50 pc (hereafter, all sources). And then, we applied our method to detect the member of those OCs according to the step in Figure \ref{flowchart}. 
~\citet{Jadhav2021} used combinations of astrometric, photometric, and systematic parameters to train and supervise a machine learning algorithm along with a Gaussian mixture model for the determination of cluster membership of M67 using the Ultra Violet Imaging Telescope (UVIT) aboard ASTROSAT and Gaia EDR3 (hereafter VV21). 
Since this is the latest representative research result, we mainly focus on this result for comparative analysis.

Compared with the other three studies, i.e., 766 member stars found by FoF, 484 member stars in CG20, and 746 member stars in VV21, we obtained 1131 M67 member stars from all sources. Figure ~\ref{fig:vv21} shows that our results agree with VV21 and FoF for the most part.

\begin{figure}[htbp]

\centering
\includegraphics[scale=0.28]{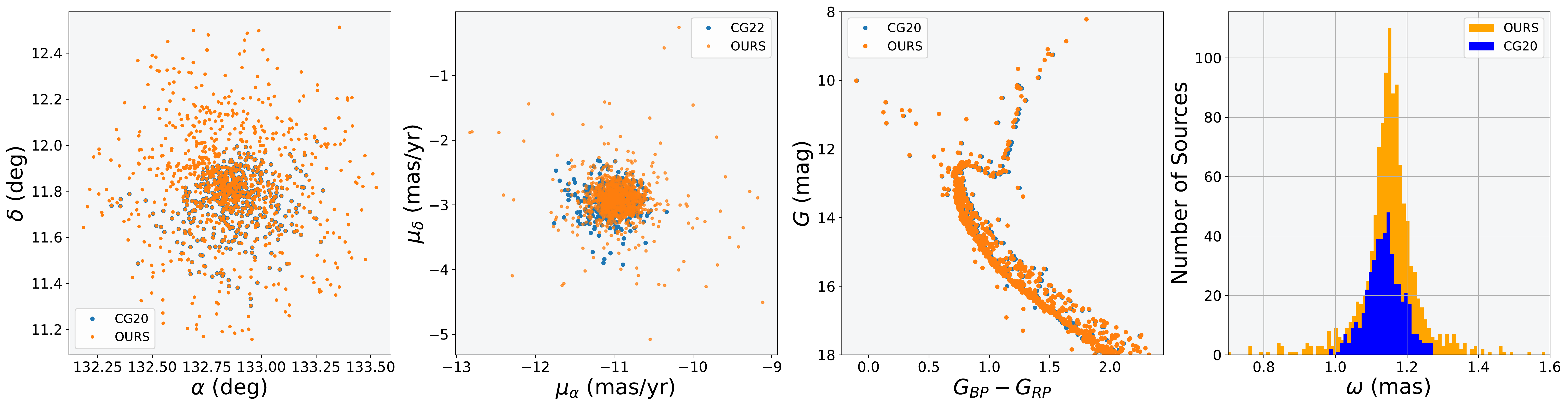}
\caption{The CMDs of members identified for the cluster M67 compared with CG20.}
\label{fig:m67}
\end{figure}

\begin{figure}[htbp]

\centering
\includegraphics[scale=0.28]{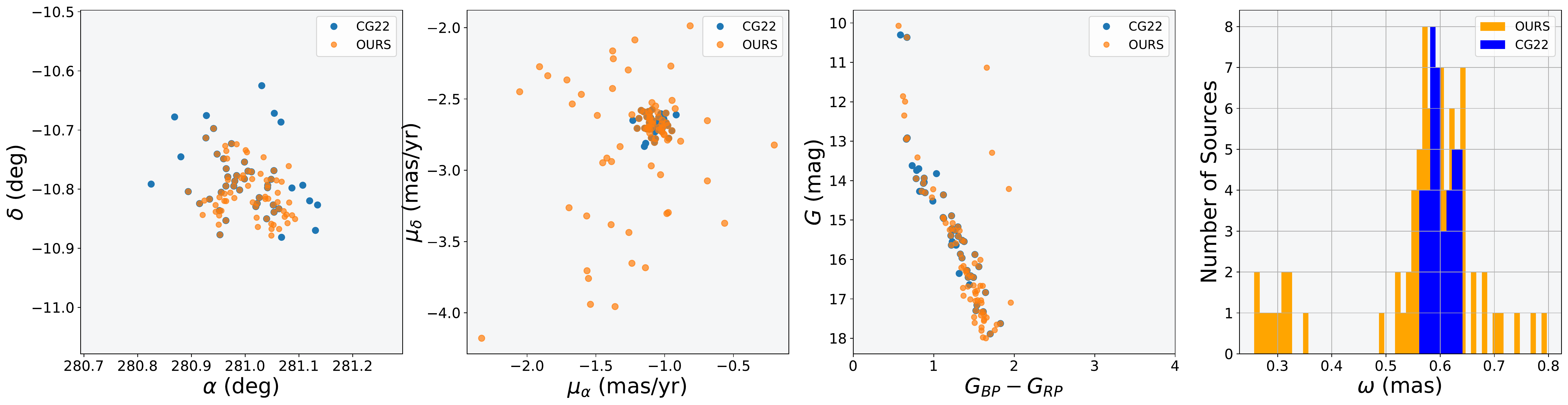}
\caption{The CMDs of members identified for the cluster UBC1029 compared with CG22.}
\label{fig:ubc1029}
\end{figure}

\begin{figure}[htbp]

\centering
\includegraphics[scale=0.28]{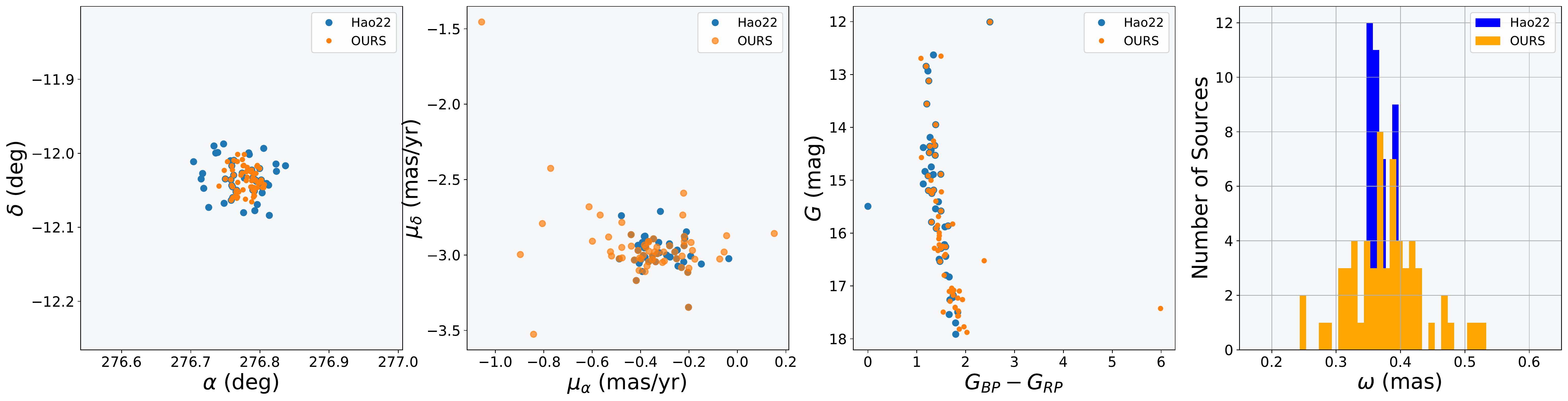}
\caption{The CMDs of members identified for the cluster OC0033 compared with Hao22 and FoF method.}
\label{fig:oc0033}
\end{figure}

\begin{figure}[htbp]

\begin{center}
\subfigure {
\includegraphics[width=3.0in,height=3in]{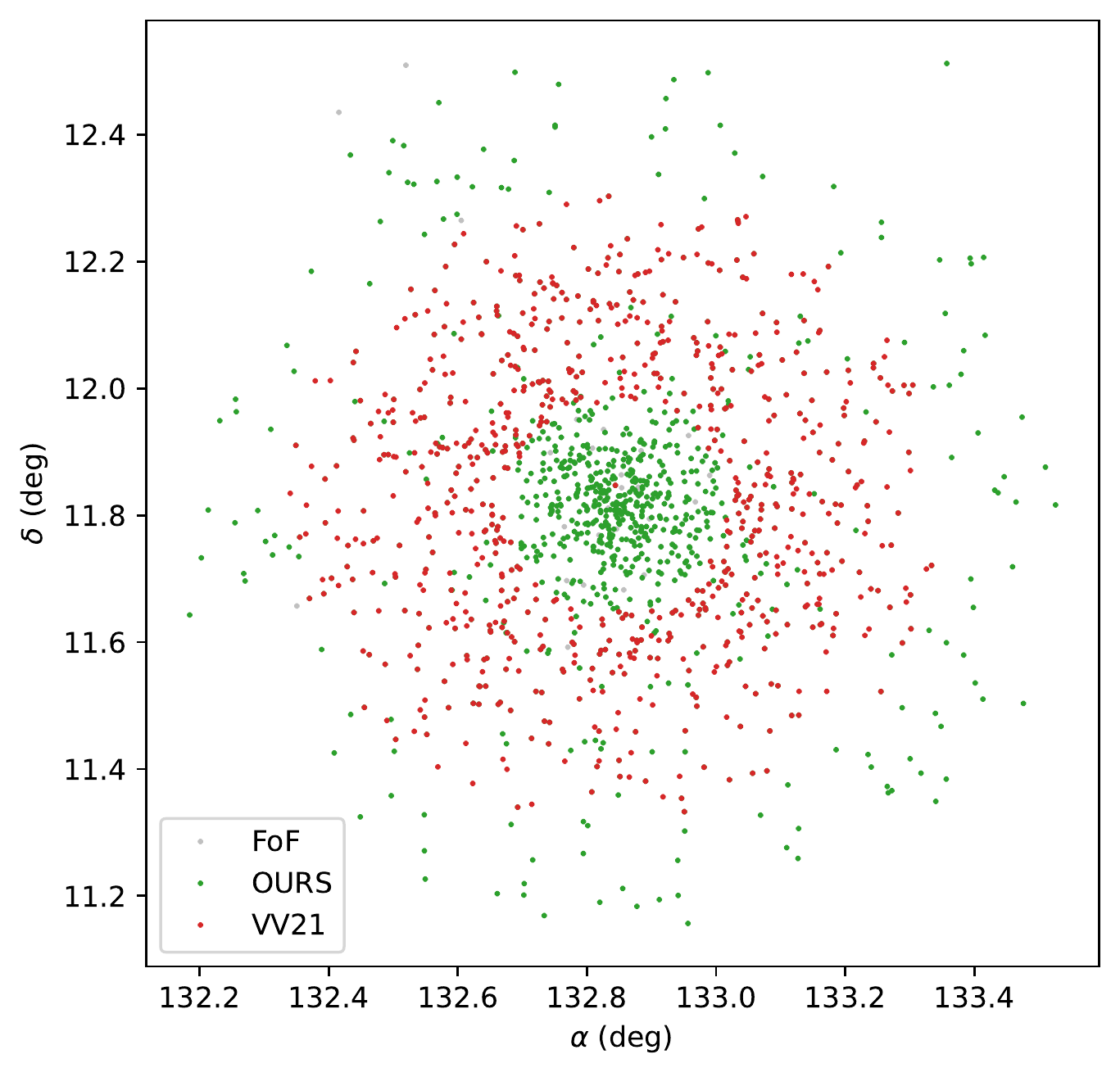}
}
\subfigure {
\includegraphics[width=3.0in,height=3in]{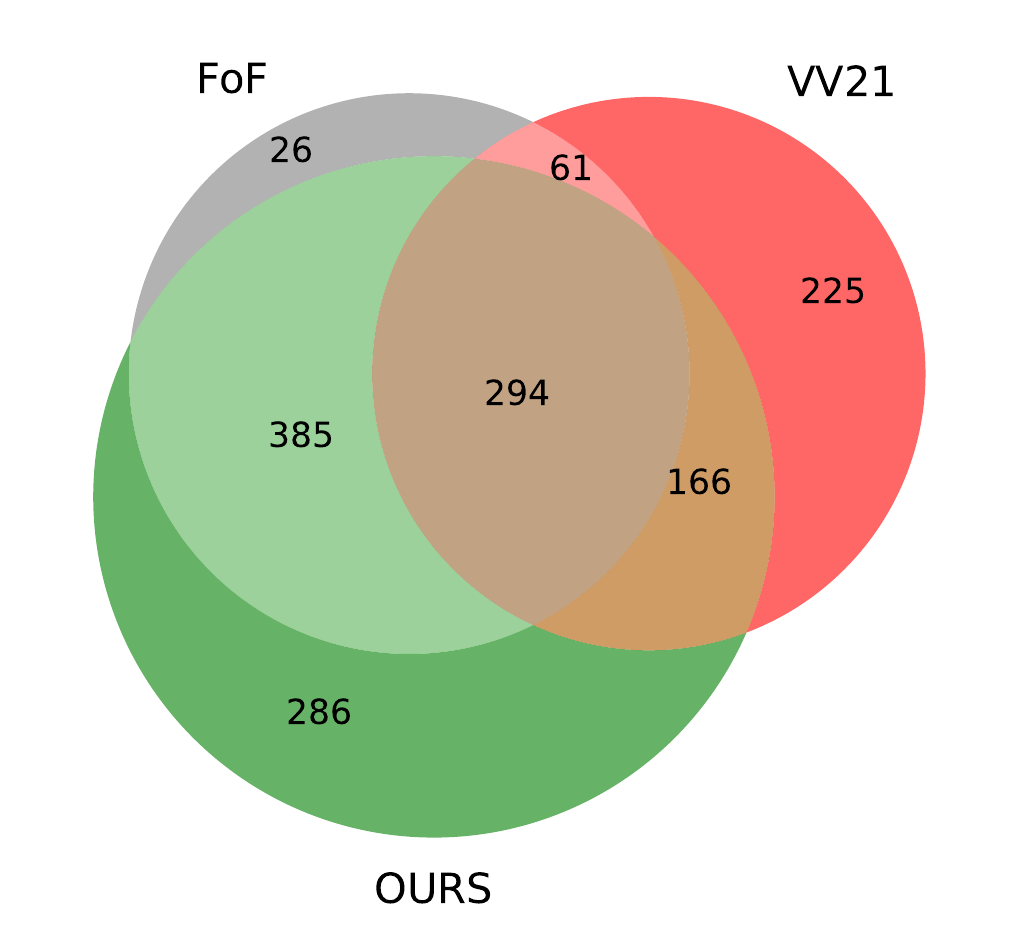}
}
\end{center}

\caption{The spatial distribution of M67 cluster members compared to VV21 and FoF methods. It should be noted that VV21 appears to have a more comprehensive census. This is because VV21 uses the Ultraviolet Imaging Telescope 210 (UVIT) on ASTROSAT and Gaia EDR3 to determine a combination of astrometric, photometric, and systematic parameters of the M67 cluster members \citep{Jadhav2021}. While our work only used the astrometric and photometric data from Gaia EDR3. The Venn diagram of members is shown in the rightmost subplot. VV21 and ours have 460 identical member stars (over 61\%), while FoF and VV21 only have 355 identical members ( 47.5 percent). Our approach resulted in a 14.5 percent increase in joint membership, suggesting that our method is valuable.}
\label{fig:vv21}
\end{figure}

We further tested the robustness of our method in smaller known clusters. We chose two clusters that were smaller in size. One cluster is UBC1029 with 40 members~\citep{Castro-Ginard2022}. Another is OC0033 with 47 members~\citep{Hao2022}. 
From left to right, Figure~\ref{fig:m67},~\ref{fig:ubc1029} and~\ref{fig:oc0033} show the experimental results of equatorial coordinates spatial distribution, proper motion distribution, CMD, and parallax distribution hist, respectively. The mean and variance of each astrometric value (position, parallax, and proper motion) and the number of member stars are presented in Table~\ref{tab:m67}. 
The results show that the proposed method could accurately identify nearby massive and smaller distant clusters. Compared with other OC studies using Gaia EDR3 data, the member stars we identified are significantly more concentrated. This is because they have a more focused spatial distribution, a clear isochrone feature, and more member stars.

Some discrepancies in the results for the member stars are reasonable. This is because the performance is poor for clusters smaller than 50 Myr, which tend to be embedded in their star-forming regions. On the other hand, clusters at distances larger than 1.5 kpc have larger astrometric errors, which makes the member star analysis less reliable ~\citep{Tarricq2022}.

\begin{table}[htbp]

	\centering
	\caption{Comparison Statistic Information of M67, UBC1029 and OC0033.}
	\begin{tabular}{cccccccccccc}
		\toprule   
		ID&ra&std\_ra&dec&dec\_std&plx&std\_plx&pmra&pmra\_std&pmdec&std\_pmdec&Nmember \\ 
		\midrule  
		M67 (CG20)     &132.846&-     &11.814 &-    & 1.135 &0.051 & -10.986& 0.193&-2.964& 0.201&598\\
		M67 (This Work)&132.853& 0.215&11.819 &0.218& 1.151 &0.0866& -10.954& 0.305&-2.913& 0.320&1131\\
		UBC1029 (CG22)&281.0&0.07&-10.78&0.06&0.61&0.03&-1.08&0.07&-2.68&0.06&47 \\
		UBC1029 (This Work)&281.003&0.048&-10.803&0.0418&0.568&0.119&-1.181&0.312&2.791&0.420&84 \\
			OC0033 (Hao22)&276.773&0.03180&-12.034&0.0245&0.372&0.0182&-0.3193&0.0935& -2.981&0.104&47 \\
		OC0033 (This Work)&276.778&0.016&-12.036& 0.0166&0.380& 0.0614& -0.374&0.205& -2.942& 0.245&62 \\
		
		\bottomrule  
	\end{tabular}
	\label{tab:m67}
\end{table}

\subsection{PDs, Parallax and proper motion dispersions}  
\begin{figure*}
\centering
\includegraphics[scale=0.5]{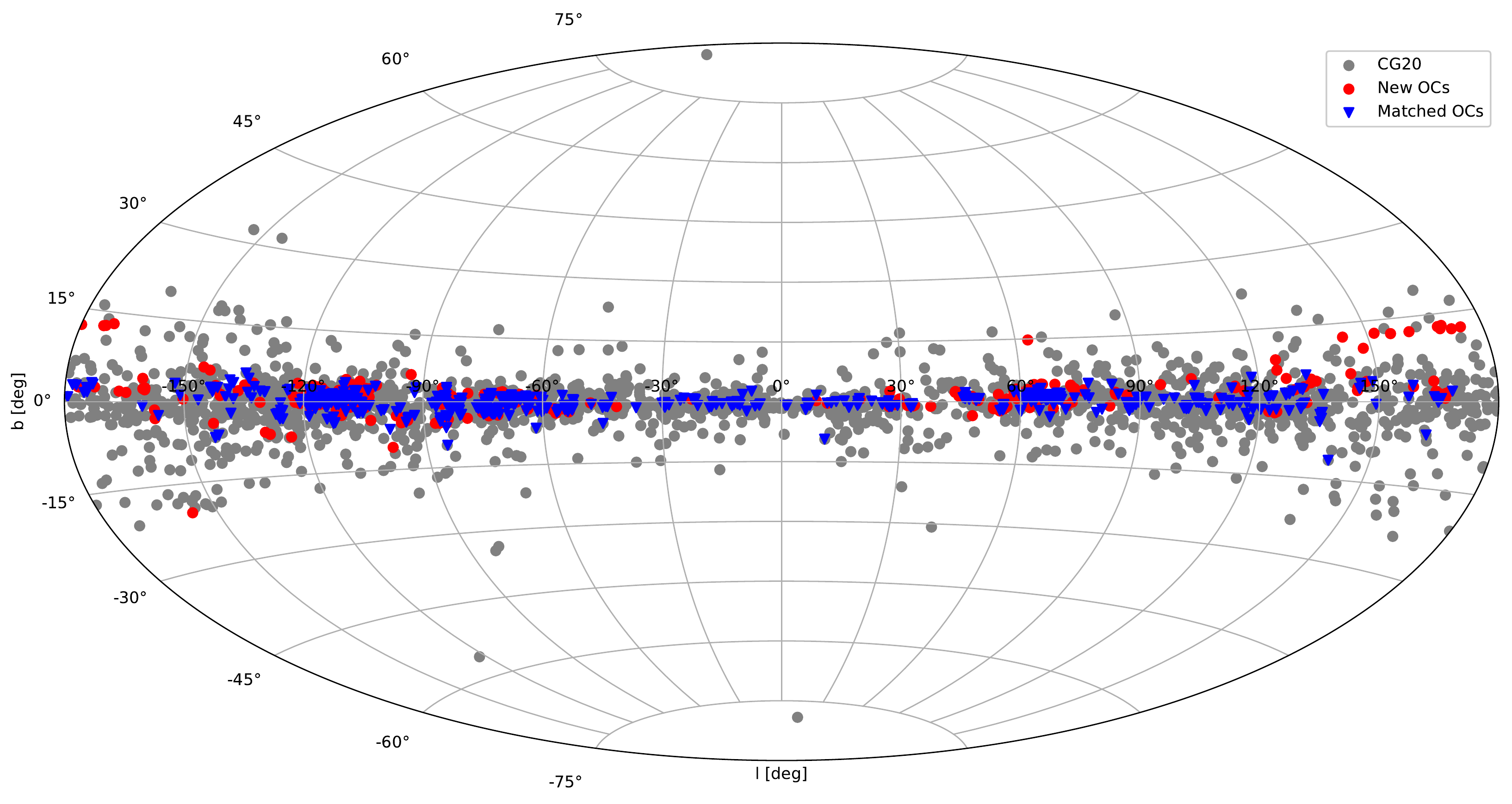}
\caption{Comparison of distribution of the OC population in $l$, $b$ Galactic coordinates. The black circles present the OCs known prior to this study, reported in
~\cite{Castro-Ginard2020}. Blue triangles represent the  OCs found in this work matched with known. Red points represent the new OCs found in this work using Gaia EDR3.}
\label{distribution}
\end{figure*}

We compared the newly discovered cluster candidates with known ones based on CG20 \citep{Cantat-Gaudin2020}. Figure ~\ref{distribution} shows the distribution of OC candidates proposed in the study, including 46 newly identified OCs, and over 500 matched OCs. The location of the proposed candidates matches that of the previous OCs. The vast majority of the new OC candidates (except one) are located at $\left| b \right|<15$ degree, and ~ 95 \% of them within $\left| b \right|<10$ degree.

\begin{figure}[htbp]
\centering
\includegraphics[scale=0.7]{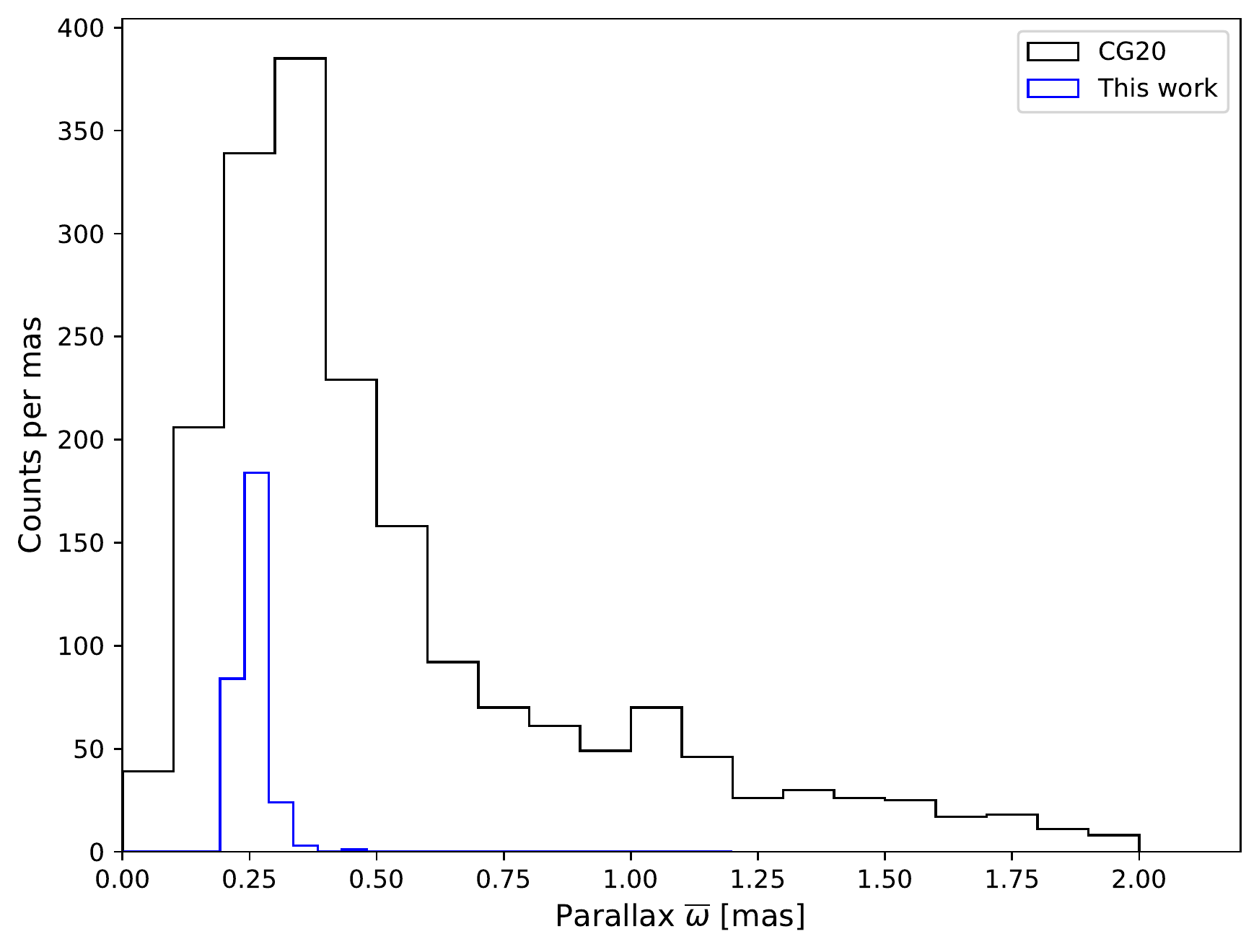}
\caption{Hisogram of parallaxes for the OCs. Black ones  represent the OCs reported in Cantat-Gaudin et al.~\cite{Cantat-Gaudin2020}. Blue ones represent the 46 new OCs found in this work using Gaia EDR3.}
\label{hisogramparallaxes}
\end{figure}

As shown in Figure ~\ref{hisogramparallaxes}, the distribution of our OC candidates indicated that 
the parallaxes of them are mostly in the range of 0.16 and 0.42, which is consistent with the results of previous work.

\begin{figure}[htbp]
\centering
\includegraphics[scale=0.7]{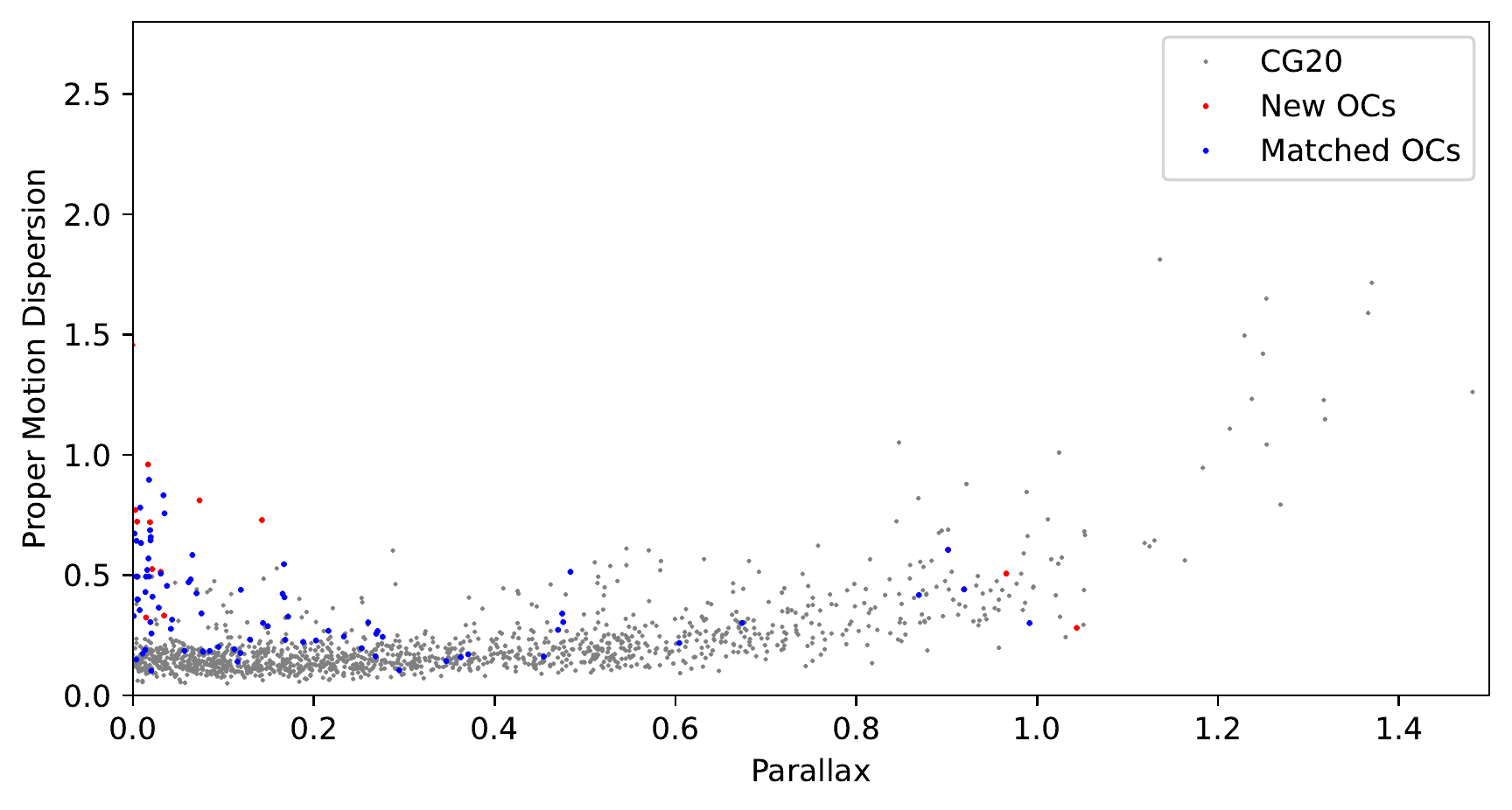}
\caption{Total observed proper motion dispersion using Gaia EDR3 vs. parallax.The x-axis is a log scale for known OCs from the homogeneous OC catalog of CG20 (black dots) compared with the new candidates in this study (red dots).}
\label{pd_parallax}
\end{figure}

In addition, we also compared the proper motion dispersions of our new OCs with those of CG20.
Figure $\ref{pd_parallax}$ shows the distribution of OC candidates identified in the study have similar smaller dispersions to the known OCs, which is the characteristics of a real cluster \citep{Cantat-Gaudin2020, Dias2022}.

\subsection{Classification And Result Analysis}

As a result of our high classification criteria, the number of OCs in Class A and Class B is relatively low in the final classification results. Meanwhile, the fitted models used are more considerate of multi-family situations such as binary stars, stellar rotation, and multiple starbursts. The parameters of the generative theory isochrone used in the fit are sparse, resulting in some potential star clusters not being fitted.

However, analyzed in a different way, the FoF and our identification model yielded 801 candidates. There are 501 cross-matched and 46 newly identified sources, which means that 68.16 percent (547/801) of these sources are successfully identified by our proposed method, which shows its value.

\subsection{Discussion of CMD fitting}
\begin{figure}[htbp]
\centering
\subfigure{
\includegraphics[width=1.6in,height=1.6in]{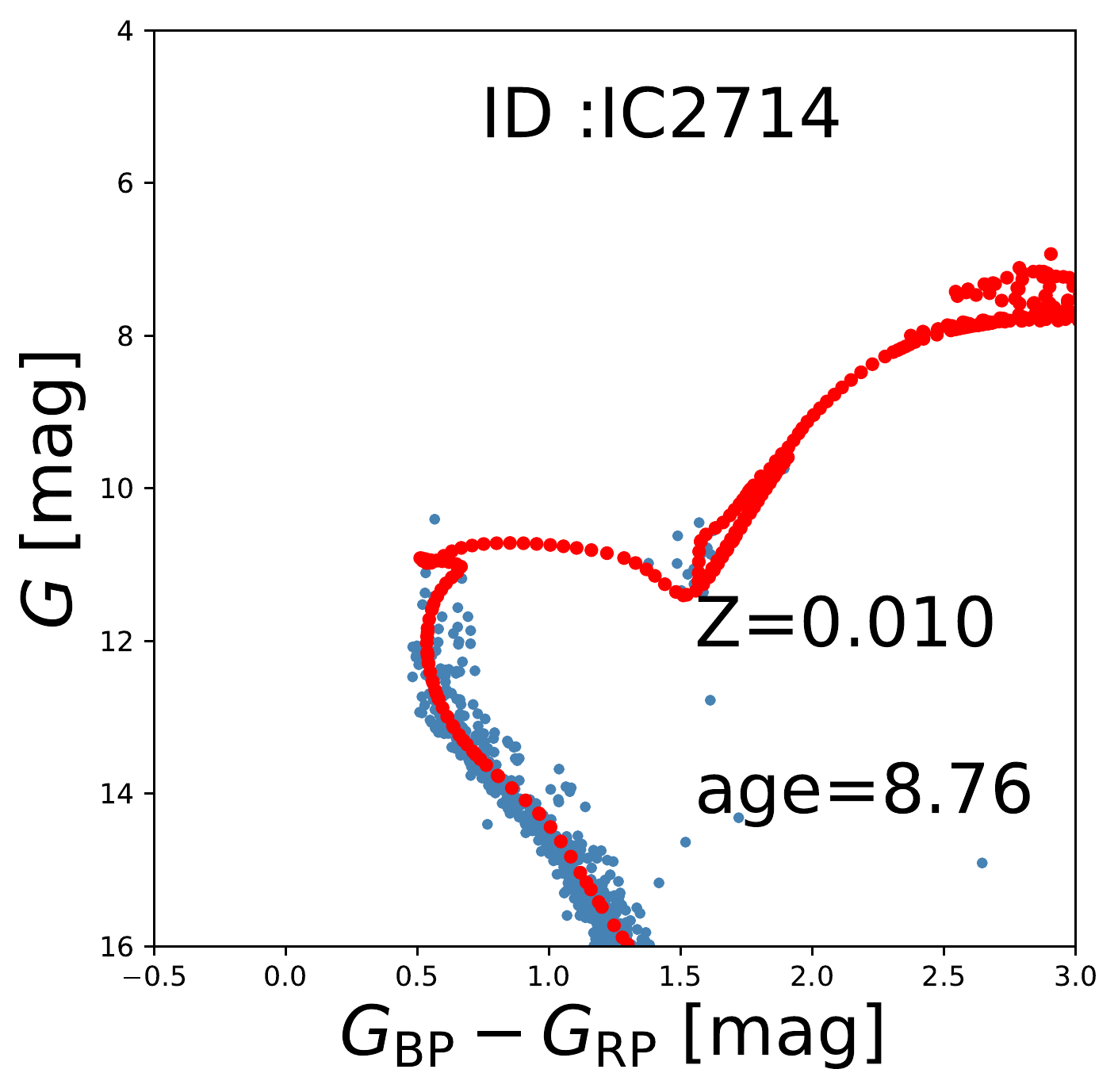}
}  
\subfigure{
\includegraphics[width=1.6in,height=1.6in]{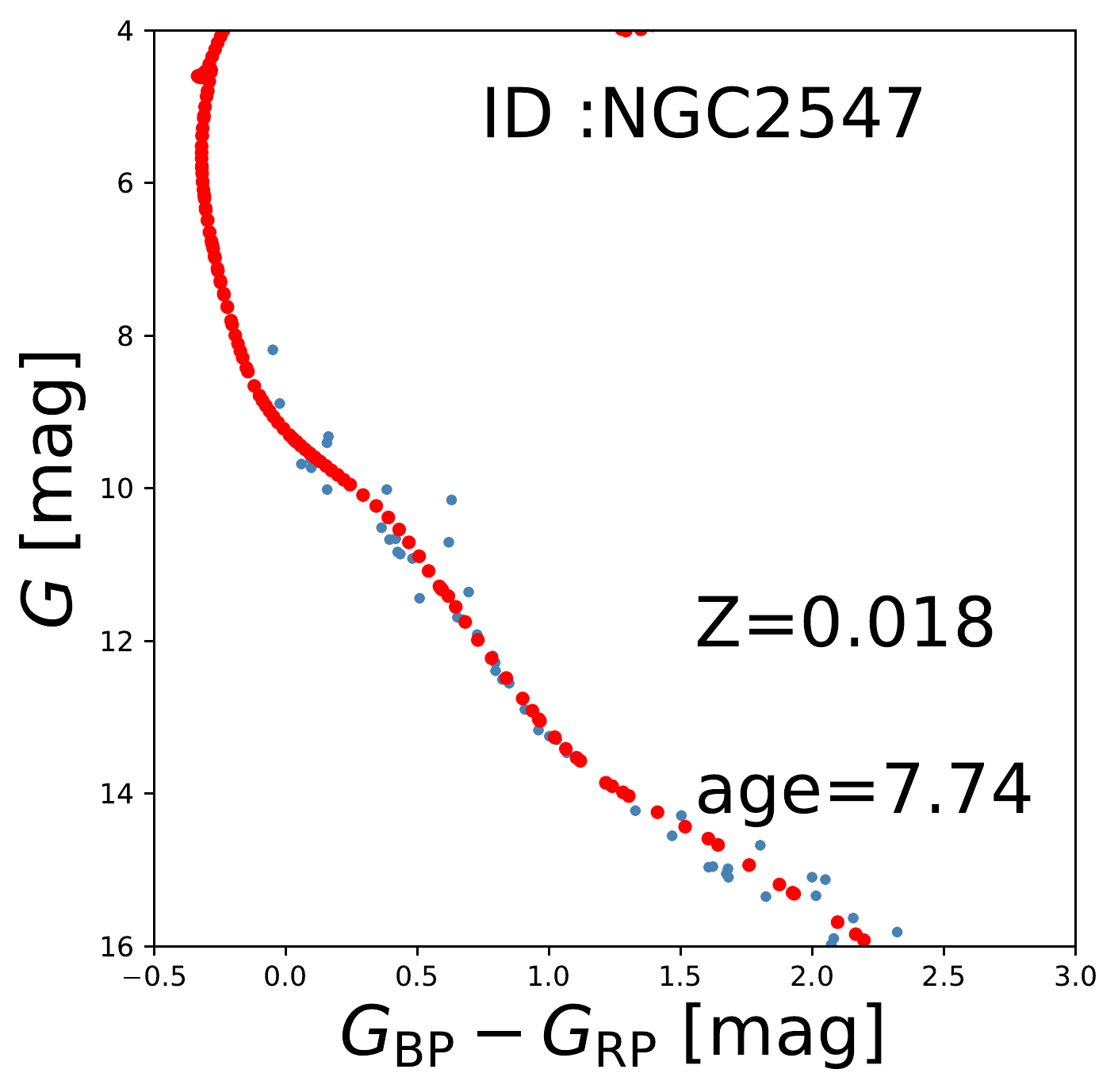}
}  
\subfigure{
\includegraphics[width=1.6in,height=1.6in]{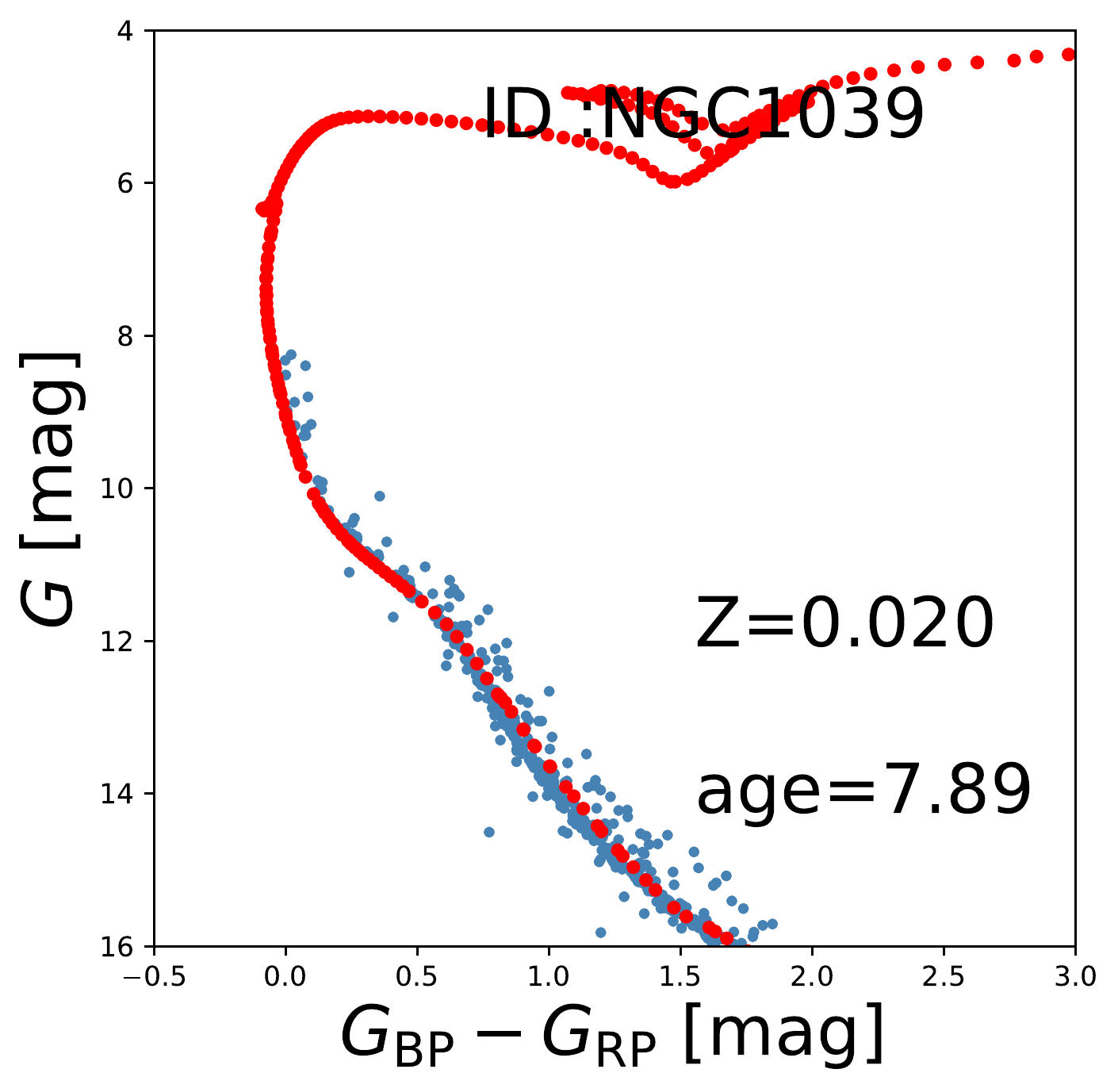}
}  
\subfigure{
\includegraphics[width=1.6in,height=1.6in]{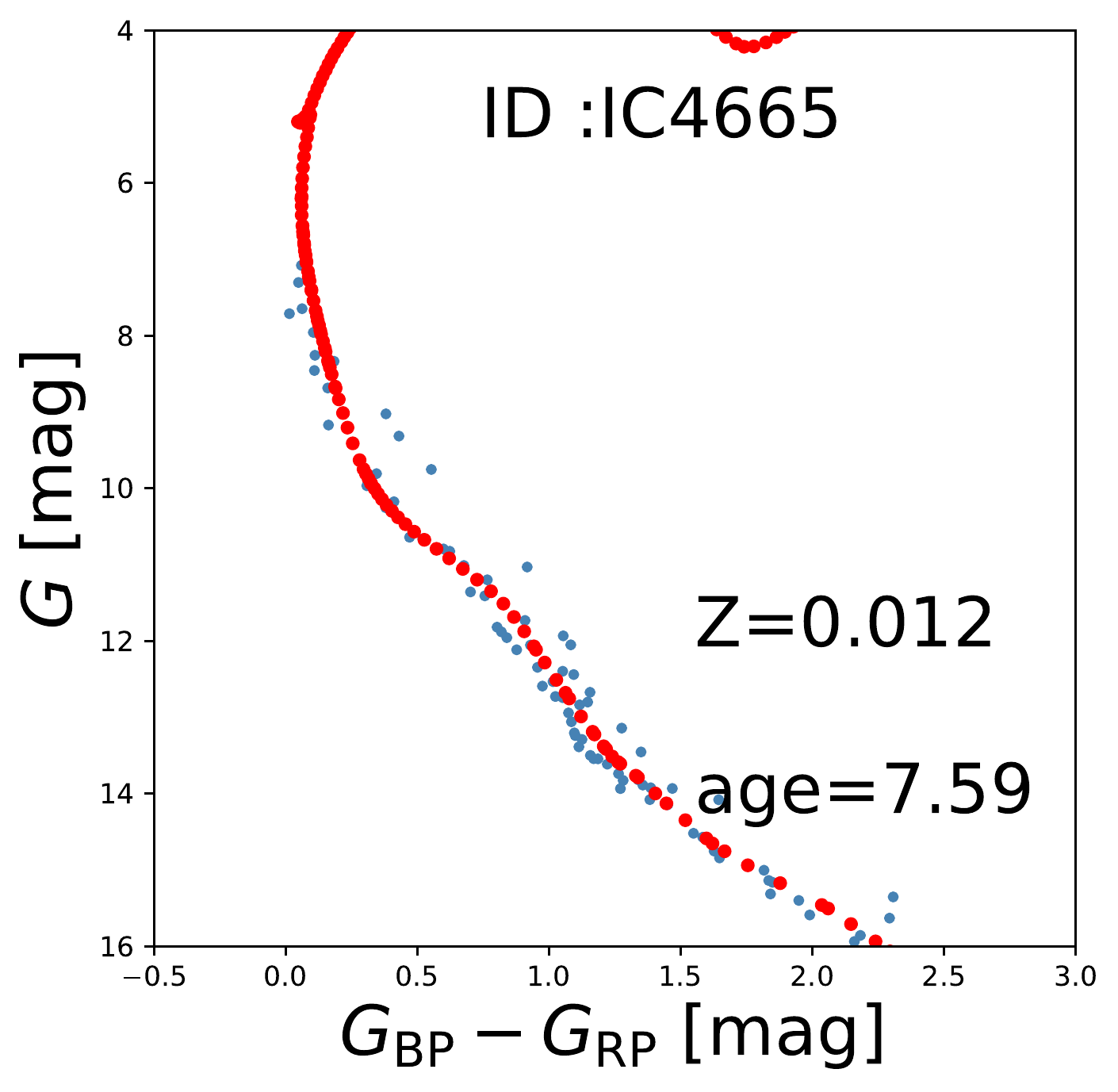}
}  

\caption{ Blue points represent cluster members. The red dotted curve is the best-fitting isochrone. }
\label{fig:validate_cmdfitting}
\end{figure}

To validate our isochrone fitting method, we randomly selected 4 clusters from \citet{Bossini2019}, which have a similar size to our reported 46 OCs. We fitted them with isochrones and methods presented by~\citet{Bossini2019}, respectively. Fig~\ref{fig:validate_cmdfitting} show the fitted results. Overall, the fitting results of two methods are consistent. 
The error between two methods is within acceptable limits. Two methods may use different isochrone libraries, resulting in errors. While~\citet{Bossini2019} used the isochrones library based on Gaia DR2, our method uses the isochrones library updated  by Gaia EDR3 passbands using  the  photometric  calibrations. The fitted parameters of two methods are shown in Table~\ref{tab:validate_fit}.

\begin{table}[htbp]
\centering
	\caption{Comparison Parameters Information of 2 fitting method.}
	\label{tab:validate_fit}
	\begin{tabular}{lccccr}
	\toprule   
	OC\_ID&IC2714&NGC2547&NGC1039 &IC4665&Unit\\ 
	\midrule  
	Bossini2019 &age:8.550, z:0.02&age:7.432, z:0.00&age:8.101, z:0.00&age:7.581, z:-0.03& age(Gyr), z($[\mathrm{Fe} / \mathrm{H}]$) \\
	This Work&age:8.760, z:0.01&age:7.740, z:0.018&age:7.890, z:0.02&age:7.590, z:0.012& age(Gyr), z($[\mathrm{Fe} / \mathrm{H}]$)\\
	Error Range&age:1.26\%, z:0.01&age:3.97\%, z:0.018&age:2.67\%, z:0.02&age:0.11\%, z:0.048&z($[\mathrm{Fe} / \mathrm{H}]$) \\
	\bottomrule  
	\end{tabular}
	\begin{tablenotes}
        \footnotesize
        \item{
        Parameters information of Bossini2019 are derived from~\citet{Bossini2019}.}
      \end{tablenotes}
	
\end{table}

We further inspected the CMD fit results. We noticed that some parameters in the fitting results were not reasonable (e.g., the age of ID1746 was only the age of 4 dex). According to the method of ~\citet{He886}, we also selected the member stars with small errors and re-fitted them using the isochrone-fitting method, and obtained reasonable results.

\subsection{Future works}
We identified 46 reliable clusters among 300 OC candidates. However, we cannot regard the rest of the 254 candidates are not open clusters. It can only be said that the method we proposed in the study cannot accurately identify these 254 candidates. We still suspect that there are Open Cluster samples among these 254 candidates. We need to find other methods in the future.

In addition, multi-view learning should be further introduced in the future. We supposed that the member star consists of three basic sub-views: 2D proper motion, the three-dimensional position, and the magnitude (photometric) sub-view. We completed the probability census in one sub-view, which can be expanded to multiple views and then integrated with complementary information to improve the accuracy of star member identification in the future.

\section{Conclusions}
\label{sec:conclusions}

In this study, we proposed a robust approach to identifying OCs.
For the given OC sample data, a  pyUPMASK and RF hybrid method is first used to remove field stars. Then an identification model based on the RF algorithm and Gaia EDR3 data is used to identify OC candidates. Finally, open cluster candidates are obtained after isochrone fitting and manual visual inspection. Based on the proposed approach, we obtained 46 new reliable open cluster candidates that have not been reported before, which proved that the method proposed in the study is reasonable.

\begin{acknowledgments}

This work is supported by the National SKA Program of China No 2020SKA0110300, Joint Research Fund in Astronomy (U1831204) under cooperative agreement between the National Natural Science Foundation of China (NSFC) and the Chinese Academy of Sciences (CAS). Funds for International Cooperation and Exchange of the National Natural Science Foundation of China (11961141001). National Natural Science Foundation of China (No. 11863002), Yunnan Academician Workstation of Wang Jingxiu (202005AF150025),China Manned Space Project with NO.CMS-CSST-2021-A08 and Sino-German Cooperation Project (No. GZ 1284). This work is also supported by Astronomical Big Data Joint Research Center, co-founded by National Astronomical Observatories, Chinese Academy of Sciences and Alibaba Cloud, and Yunnan Ten Thousand Talents Plan Young and Elite Talents Project.
\end{acknowledgments}

\bibliography{references}{}
\bibliographystyle{aasjournal}

\end{sloppypar}
\end{CJK*}
\end{document}


\appendixpage
\addappheadtotoc
\renewcommand\thefigure{A.\arabic{figure}}    
\renewcommand\thefigure{A.\arabic{figure}}   

\begin{figure*}
\begin{center}
\subfigure{
\includegraphics[width=2.0in,height=2in]{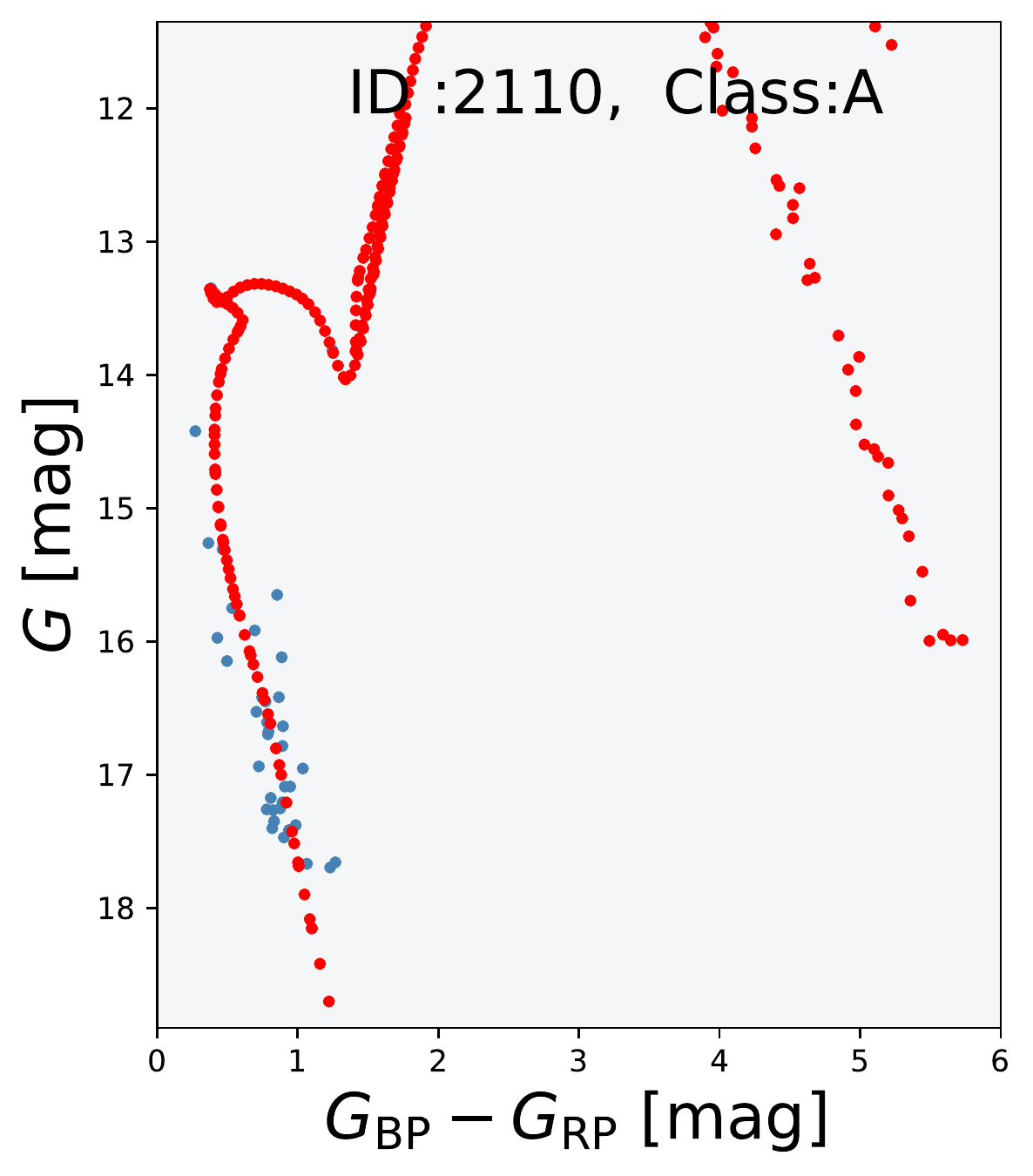}
}
\subfigure{
\includegraphics[width=2in,height=2in]{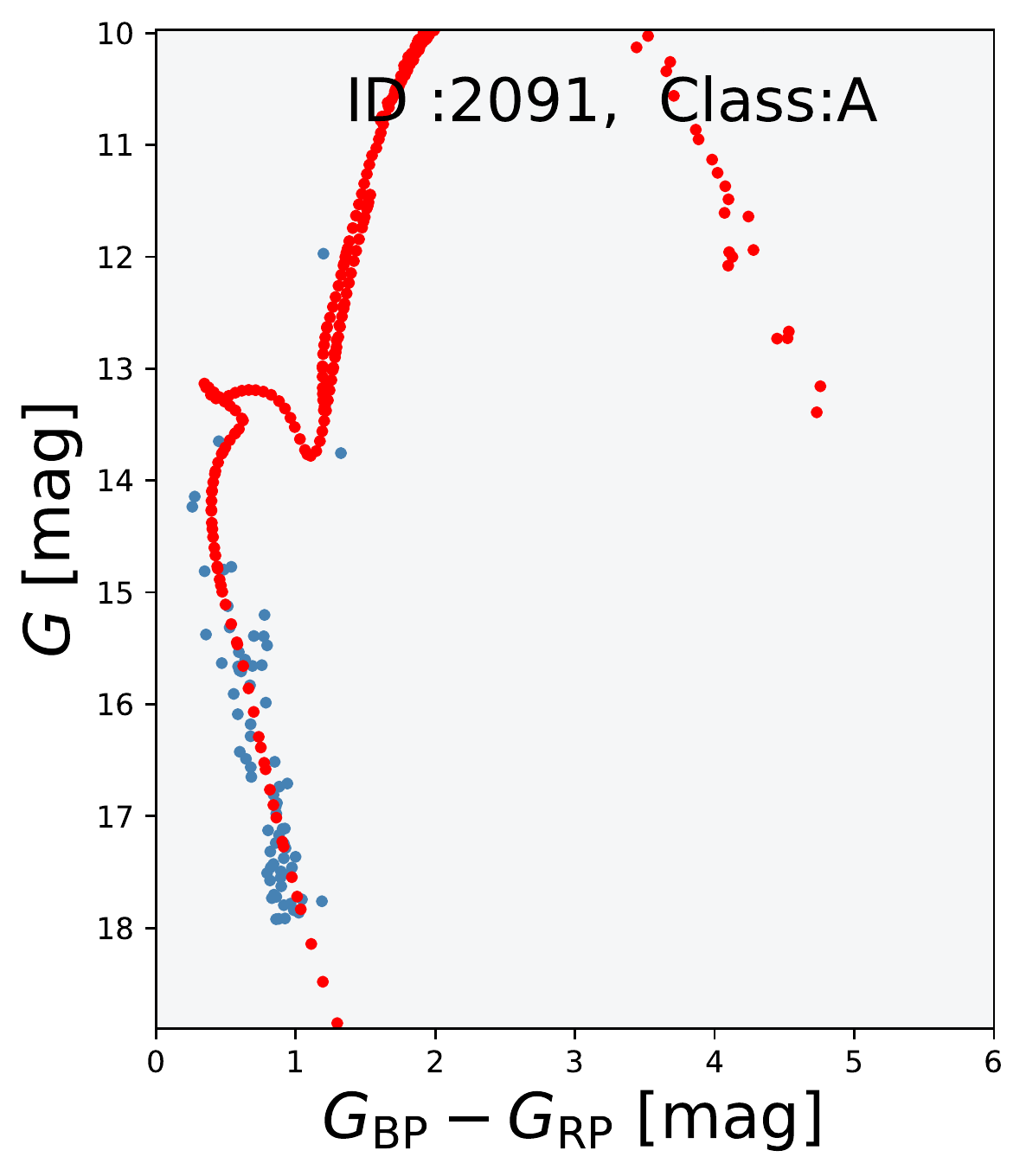}
}
\subfigure{
\includegraphics[width=2in,height=2in]{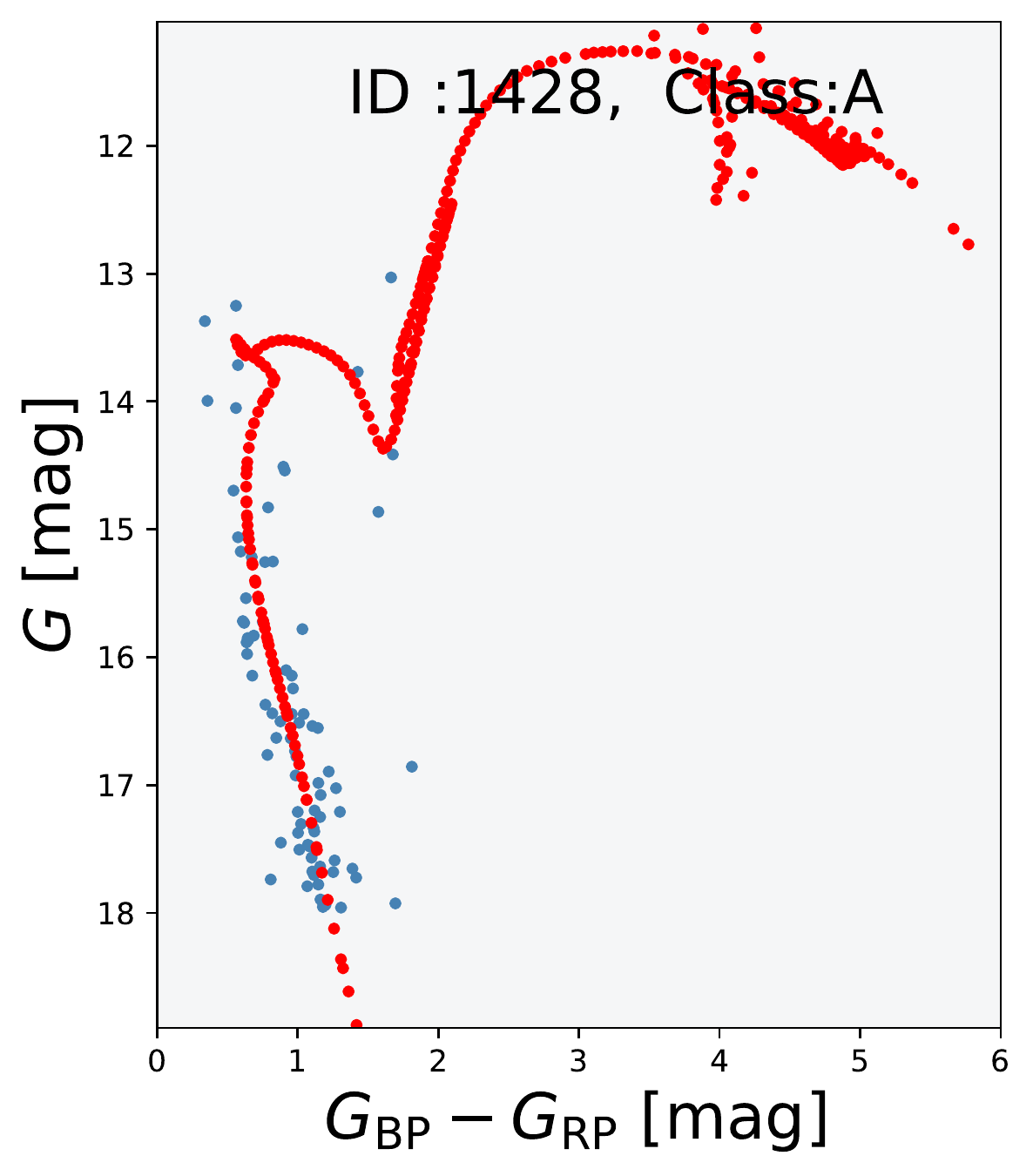}
}
\end{center}
\begin{center}
\subfigure {
\includegraphics[width=2in,height=2in]{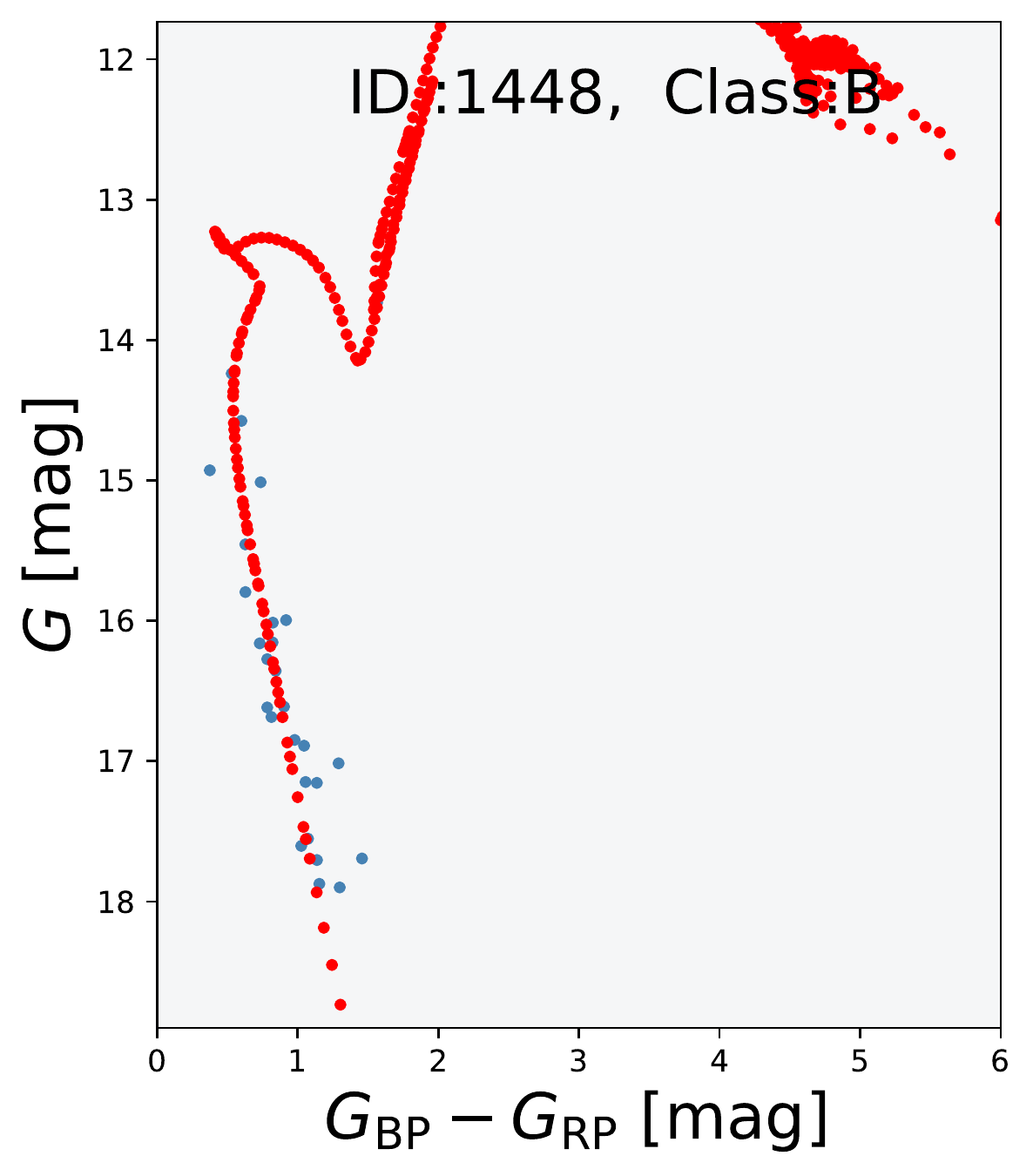}
}
\subfigure  {
\includegraphics[width=2in,height=2in]{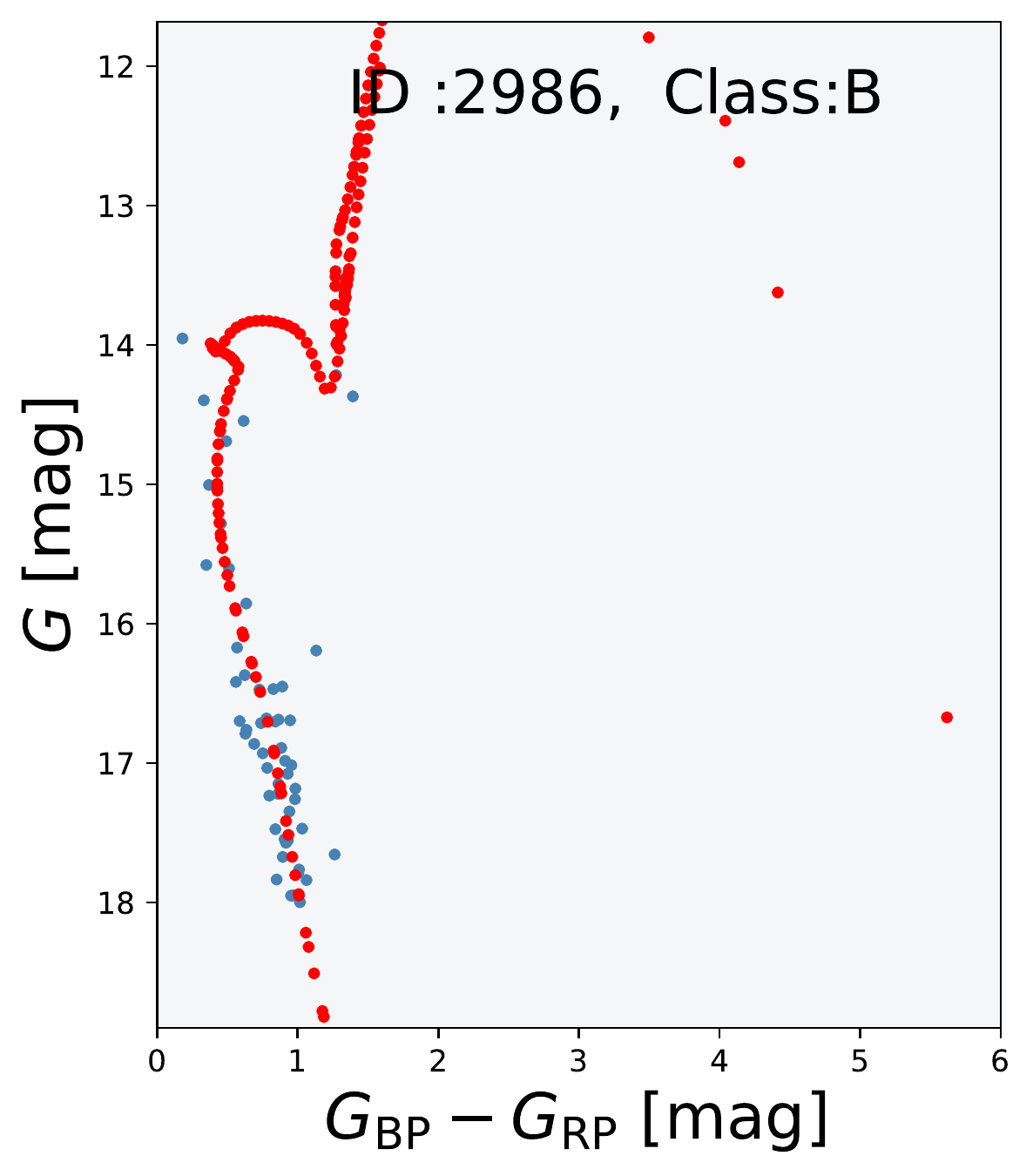}
}
\subfigure {
\includegraphics[width=2in,height=2in]{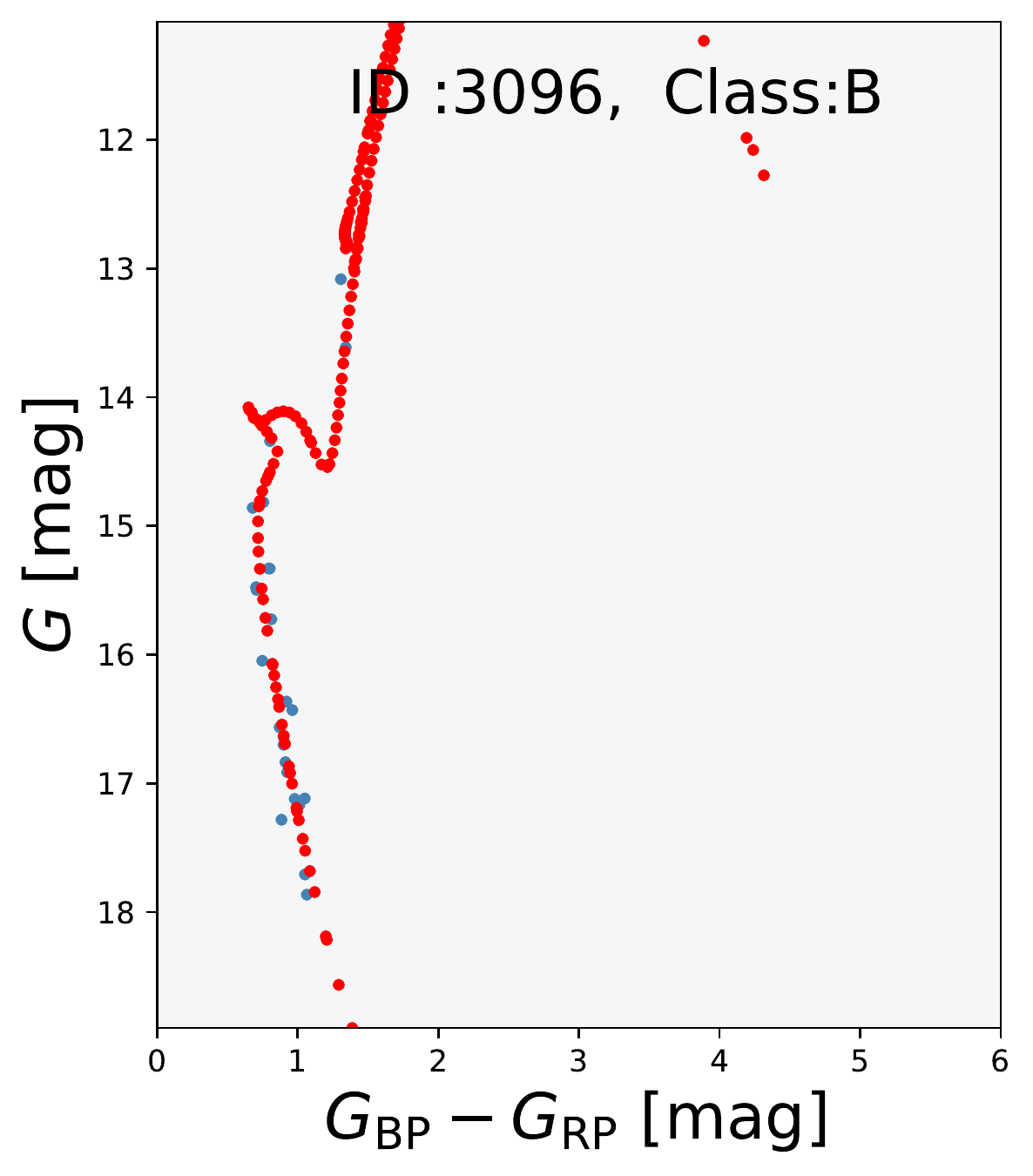}
}
\end{center}
\begin{center}
\subfigure {
\includegraphics[width=2in,height=2in]{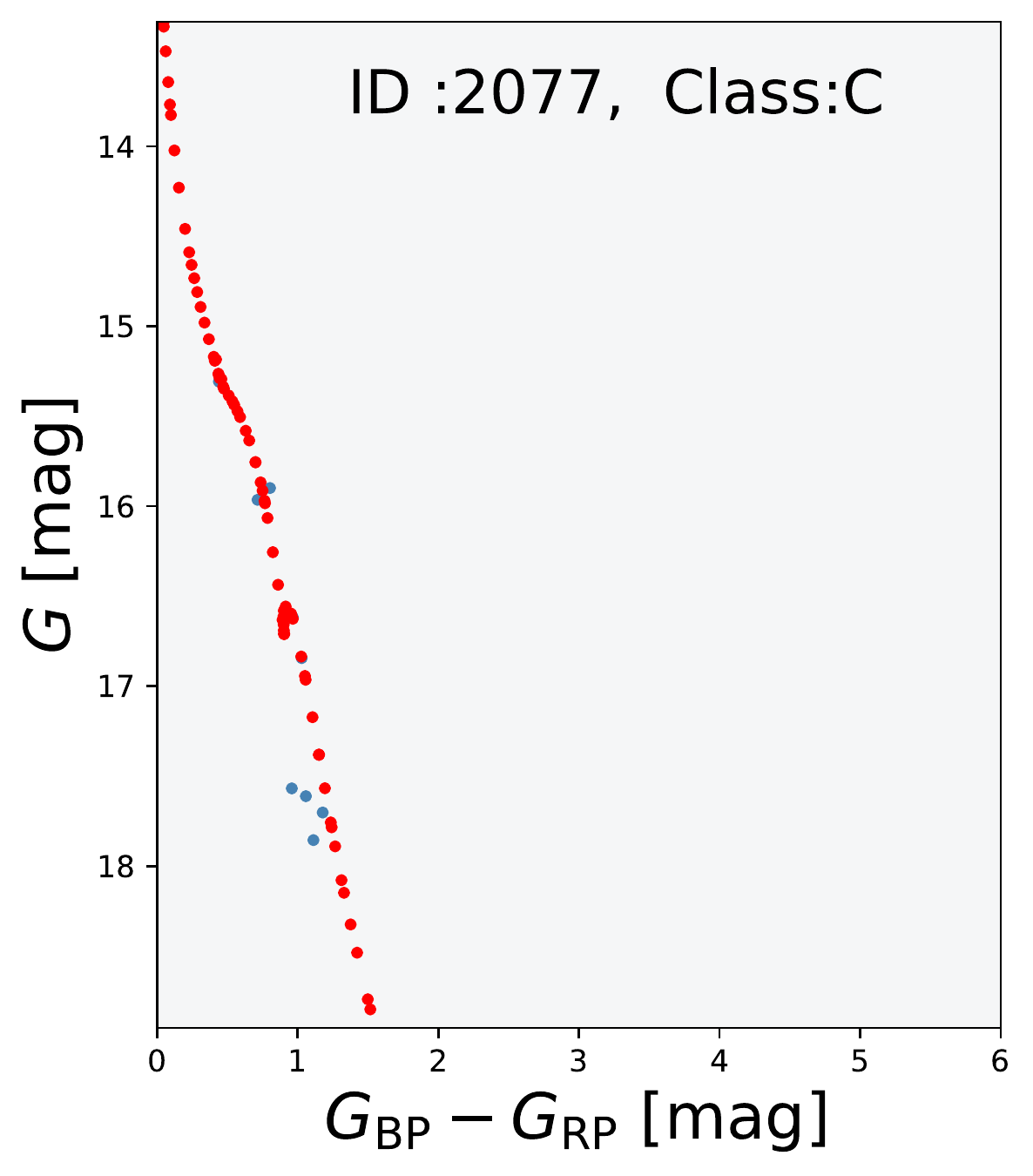}
}
\subfigure  {
\includegraphics[width=2in,height=2in]{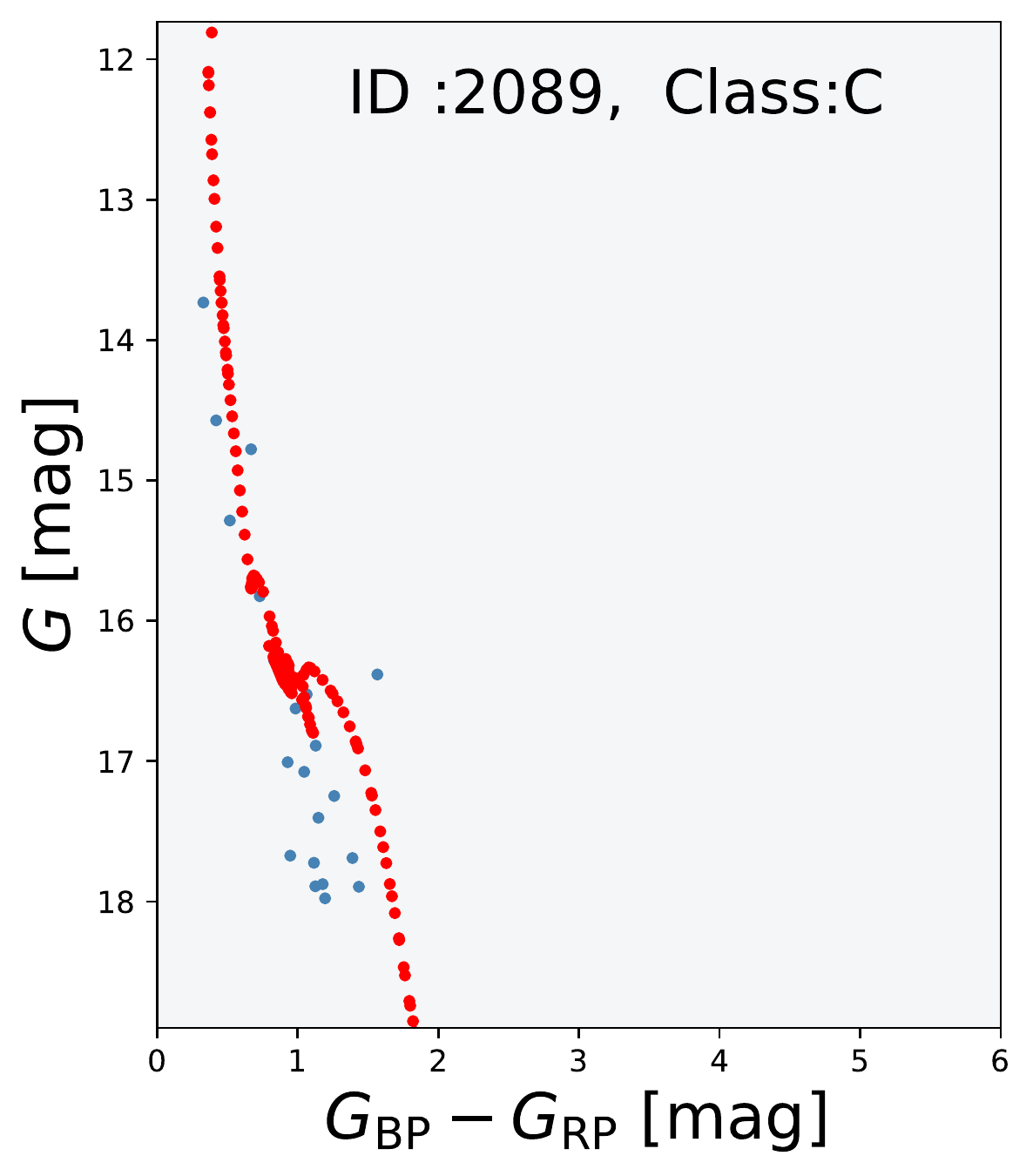}
}
\subfigure  {
\includegraphics[width=2in,height=2in]{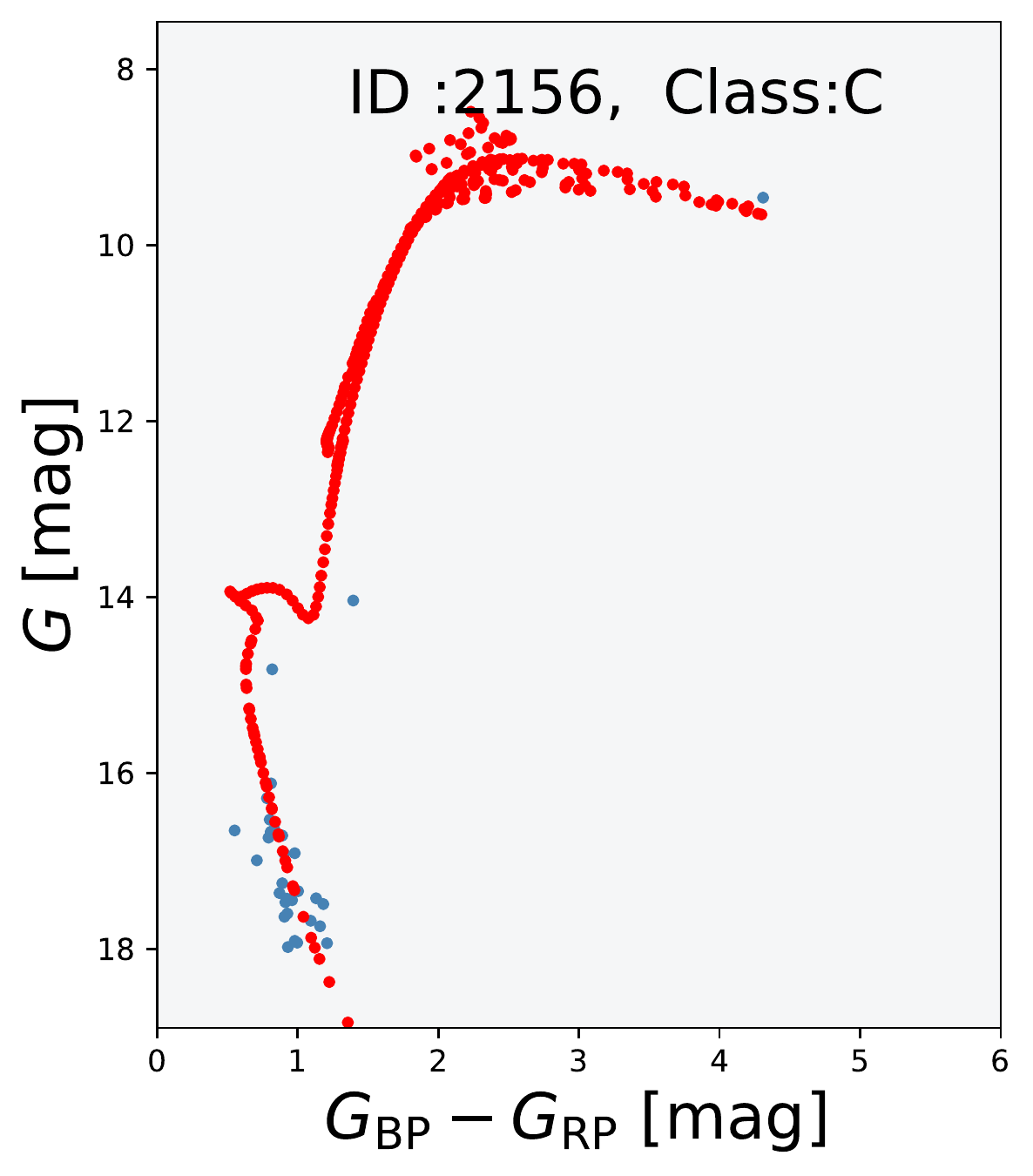}
}
\end{center}

\caption{Class A, B, and C for the new OCs. Black point respect observed stars, and the red one are for best-fit isochrone. The name of each cluster in the catolog are given in the figure.}
\label{fig_class}
\end{figure*} 

\begin{figure*}

\begin{center}
\subfigure {
\includegraphics[width=7.0in,height=1.7in]{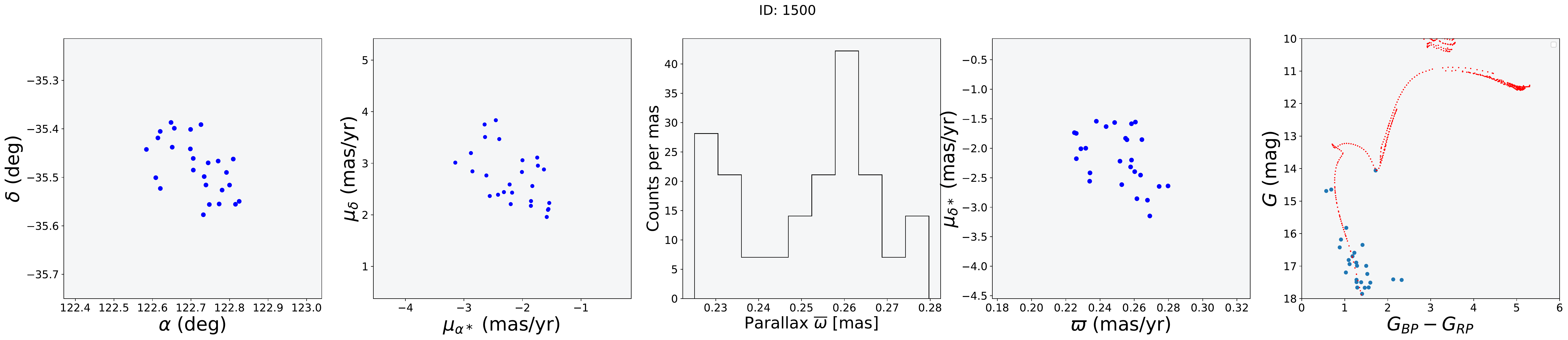}
}

\end{center}
\begin{center}
\subfigure {
\includegraphics[width=7.0in,height=1.7in]{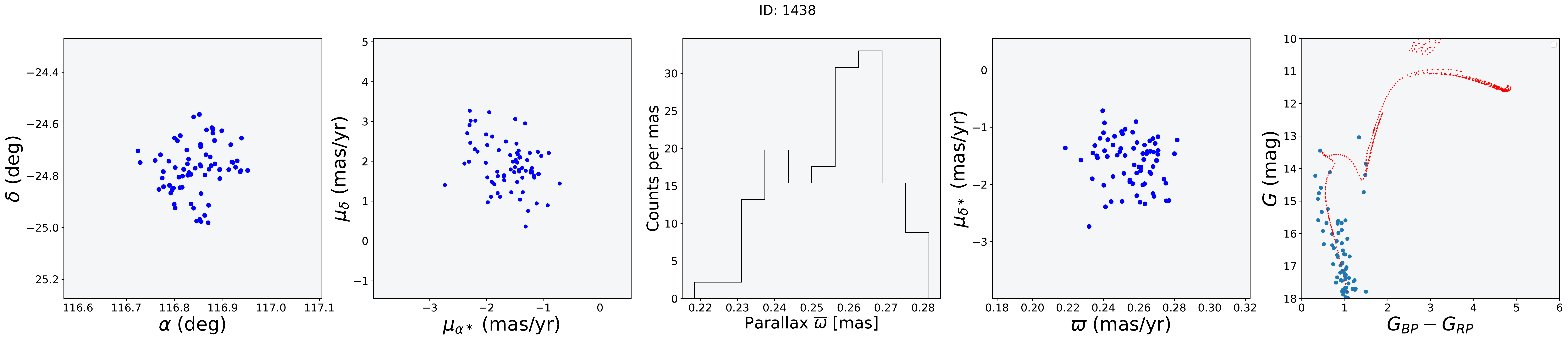}
}

\end{center}  
\begin{center}
\subfigure {
\includegraphics[width=7.0in,height=1.7in]{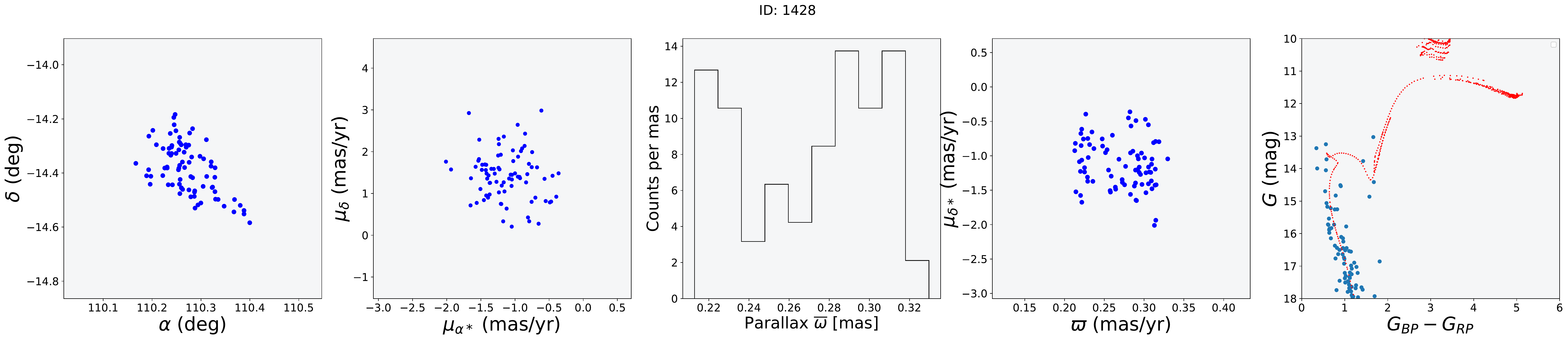}
}
\begin{center}
  \subfigure{
\includegraphics[width=7.0in,height=1.7in]{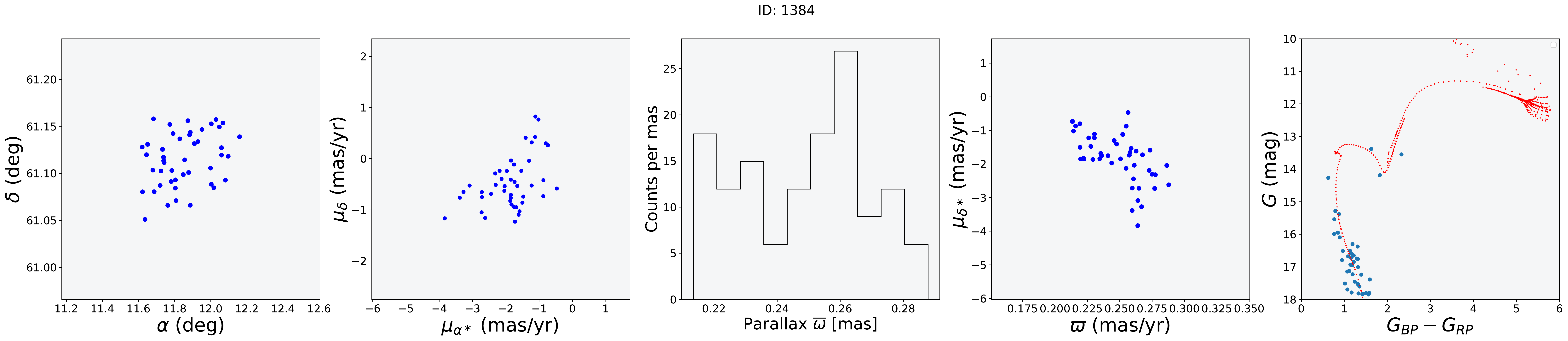}
}  
\end{center}
\end{center}
\caption{Schematic diagram of the 4 OCs identified in the study. From left to right, five sub-diagrams illustrate the position of the OC in ($\alpha$, $\delta$), ($\varpi$, $\mu_{a^{*}}$ ) distribution, parallax distribution, ($\mu_{a^{*}}$, $\mu_{\delta}$ ) distribution, and CMD of OC, respectively. Blue points represent observed stars, and the red line is for best-fit isochrone. ID means cluster-ID.}
\label{fig:5panels}
\end{figure*} 

\begin{figure*}

\begin{center}
  \subfigure{
\includegraphics[width=2in,height=2in]{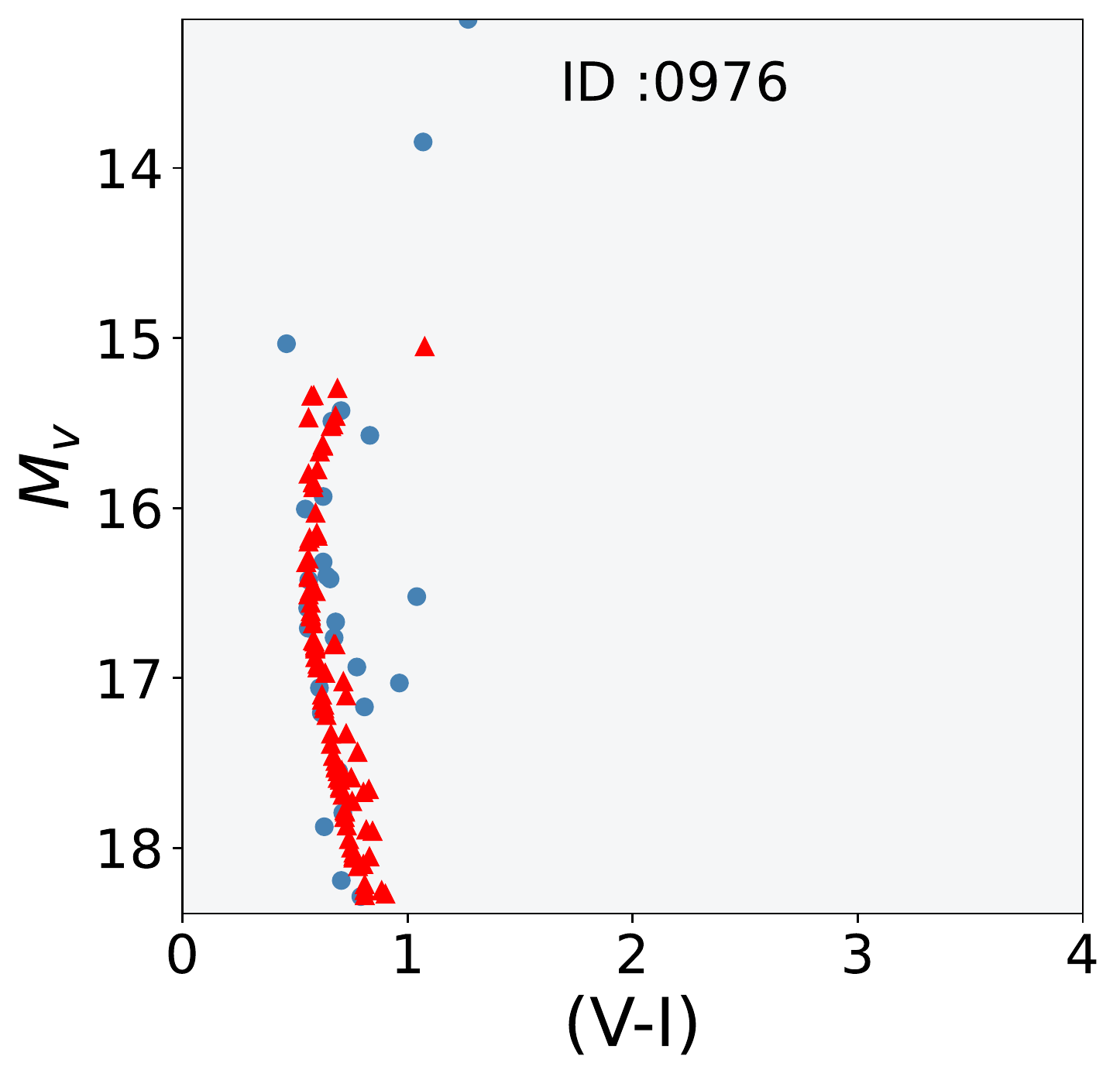}
}  
\subfigure{
\includegraphics[width=2in,height=2in]{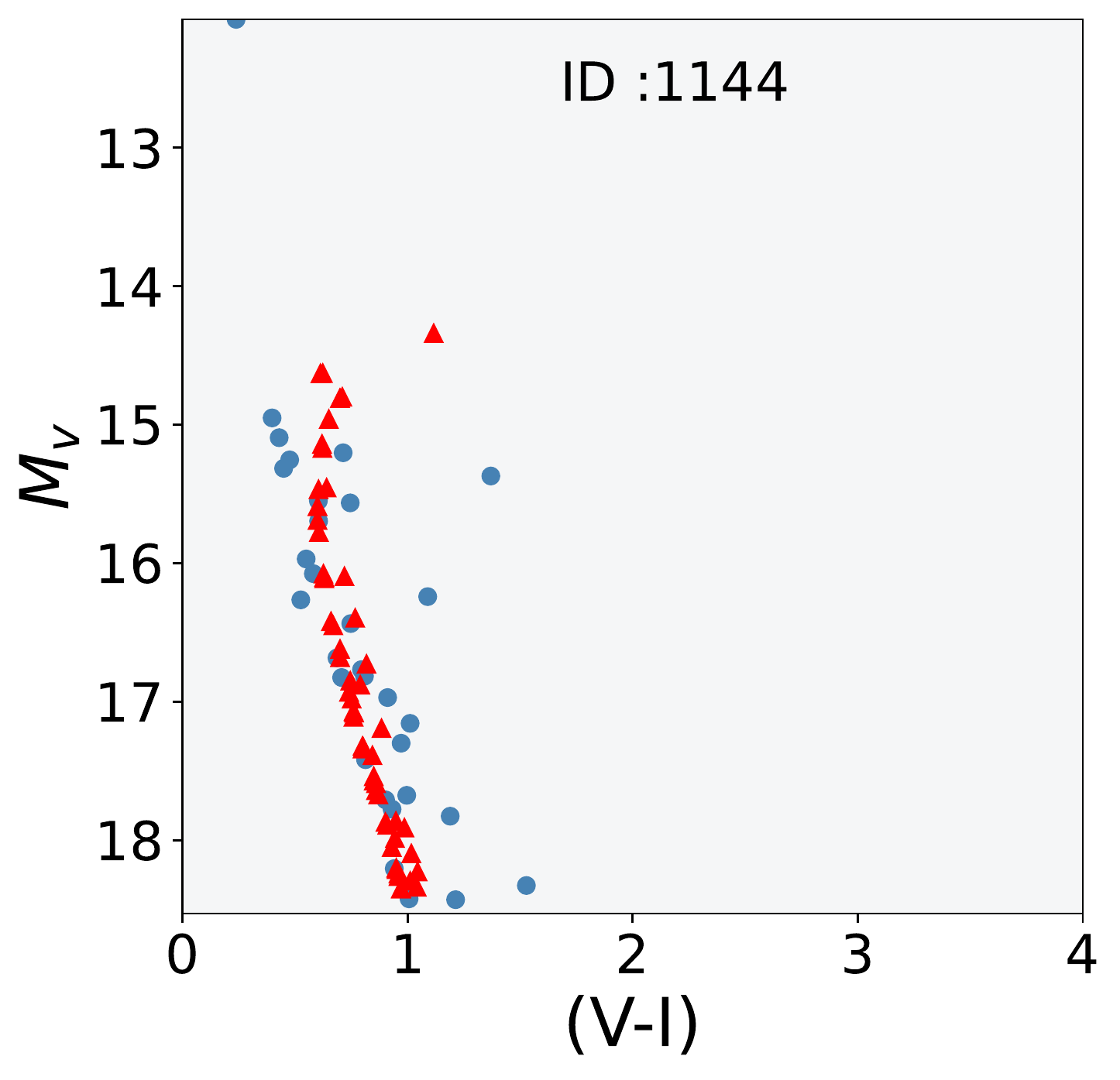}
}  
\subfigure{
\includegraphics[width=2in,height=2in]{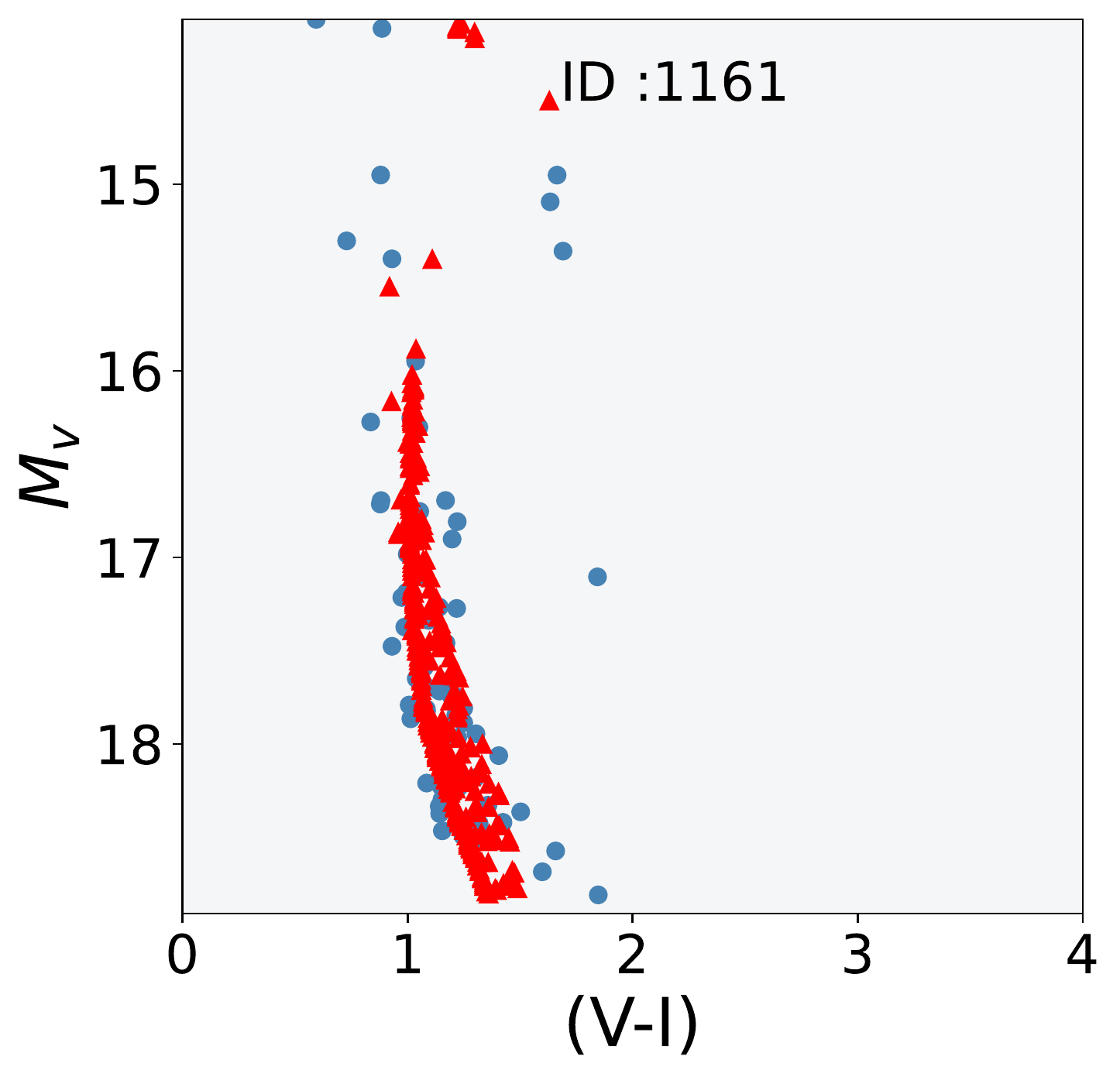}
}  
\end{center}

\begin{center}
 \subfigure{
\includegraphics[width=2in,height=2in]{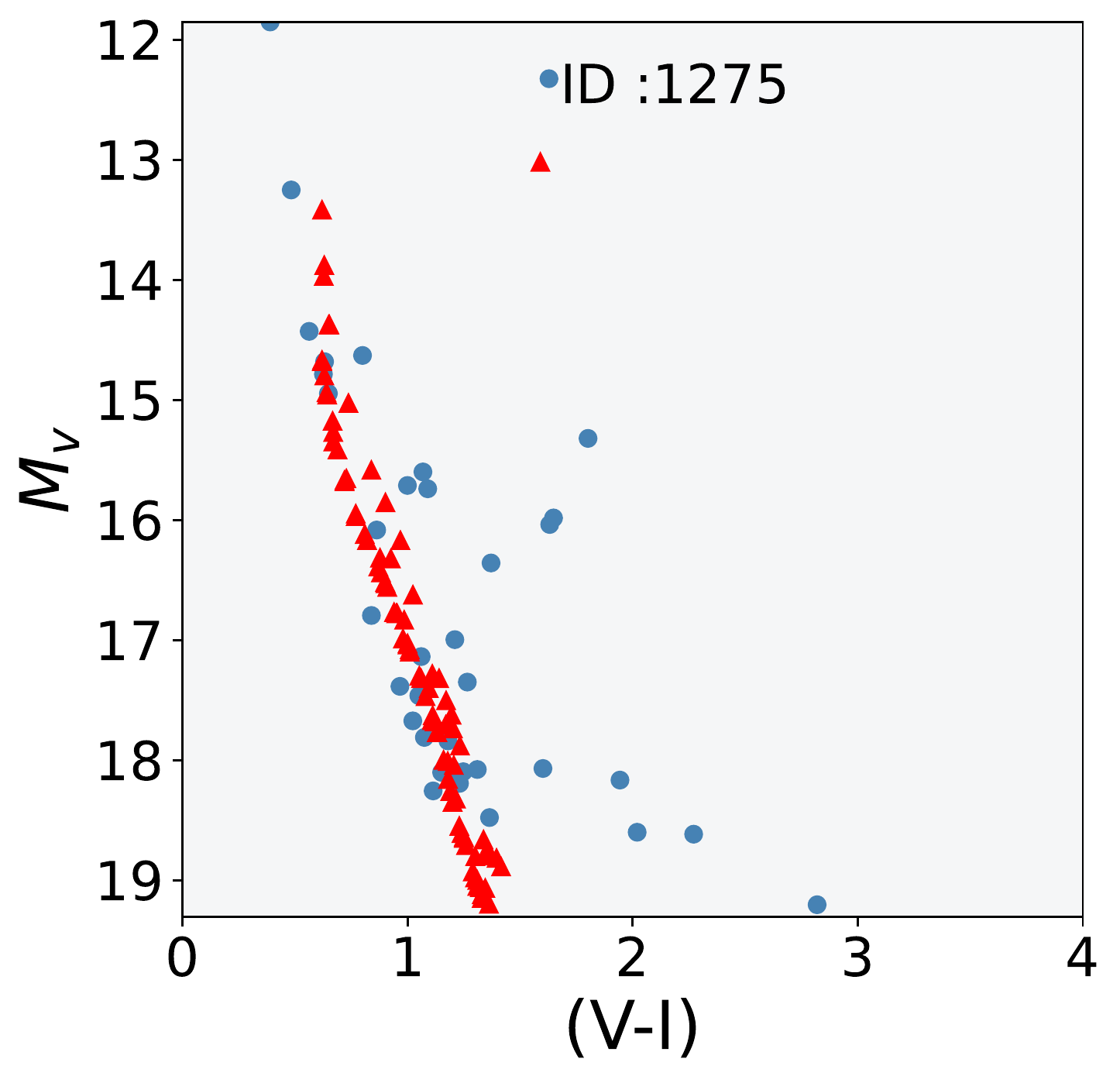}
}  
\subfigure{
\includegraphics[width=2in,height=2in]{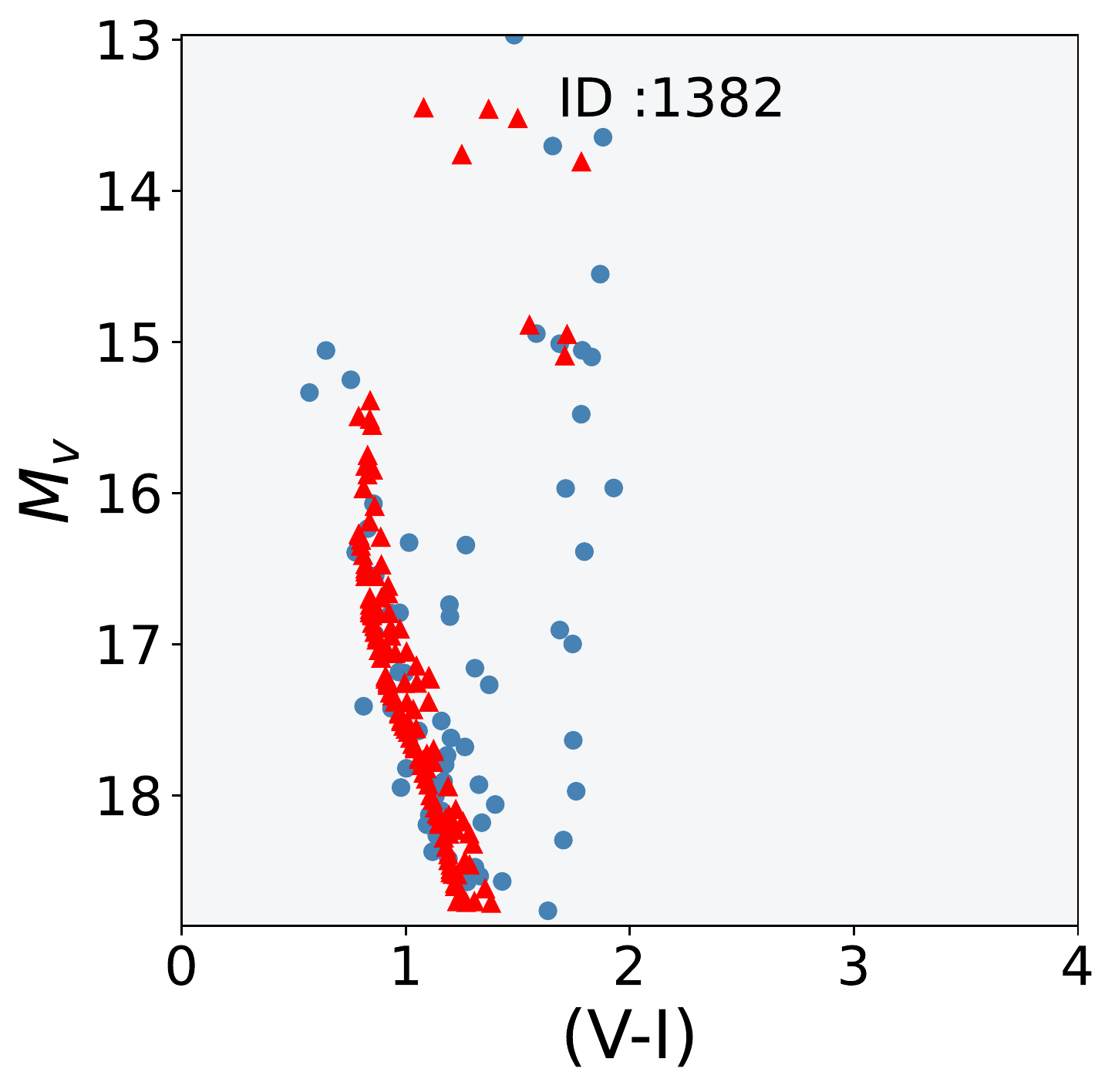}
}  
\subfigure{
\includegraphics[width=2in,height=2in]{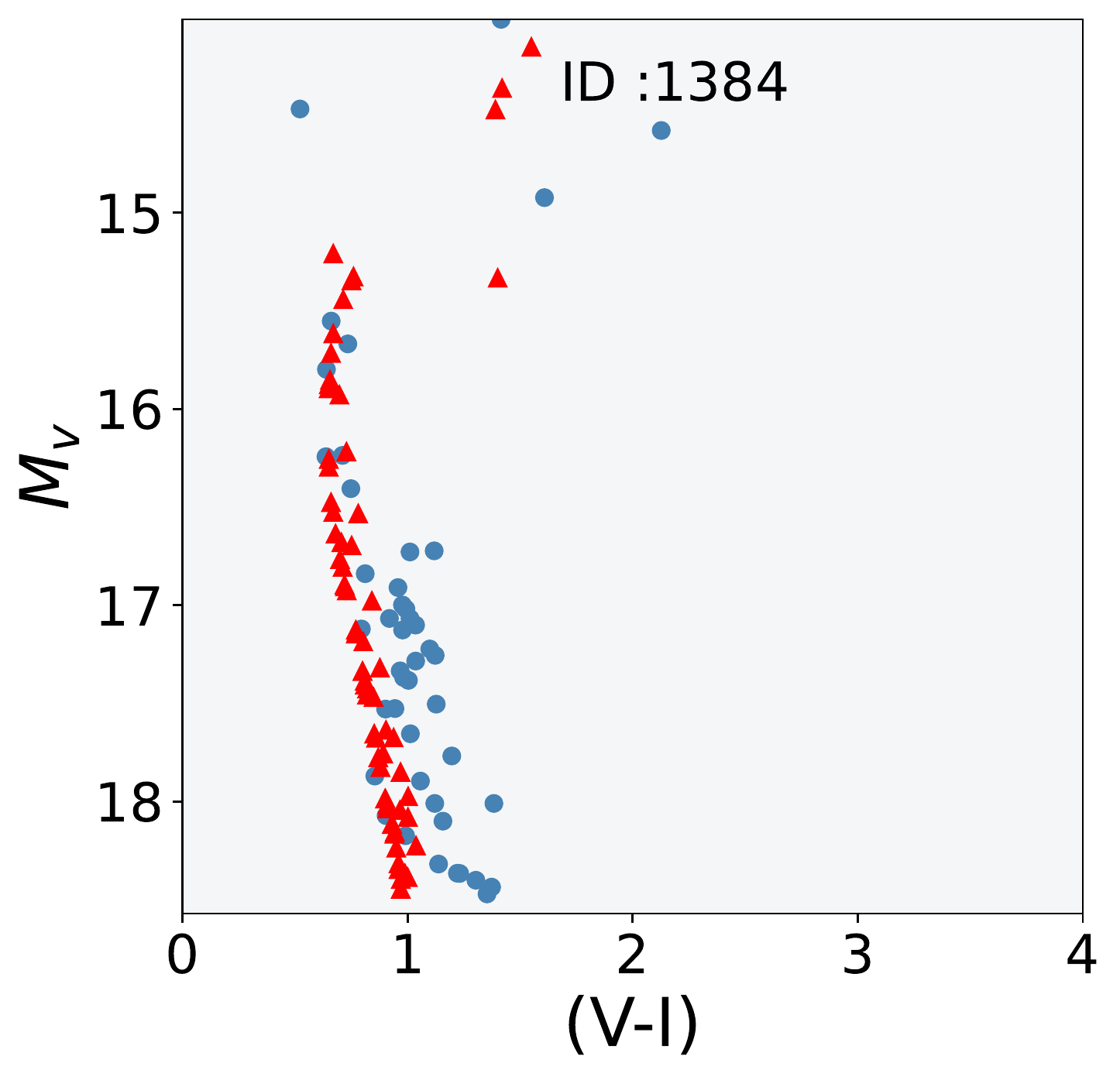}
}

\end{center}
\begin{center}
\subfigure{
\includegraphics[width=2in,height=2in]{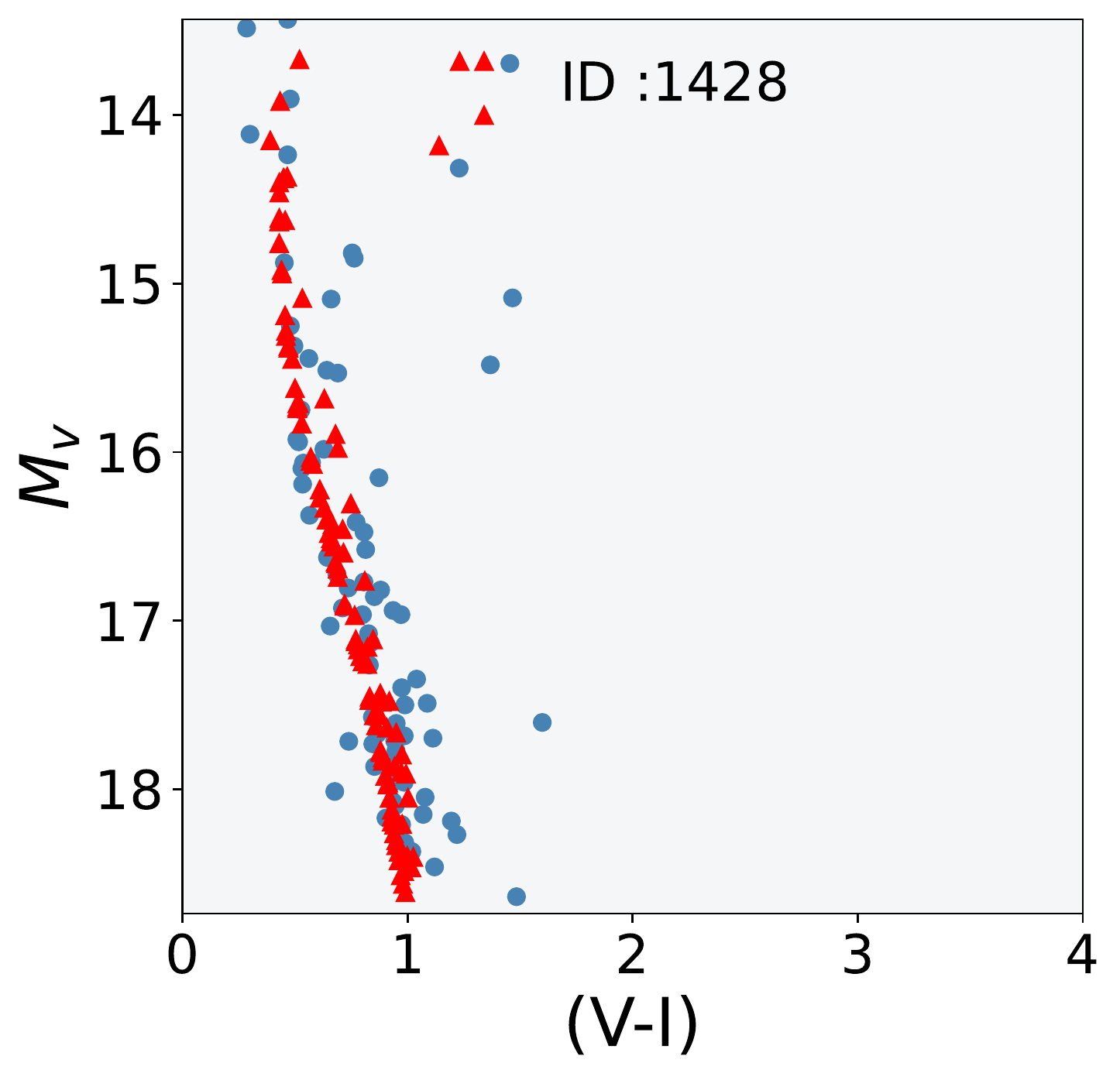}
}  
\subfigure{
\includegraphics[width=2in,height=2in]{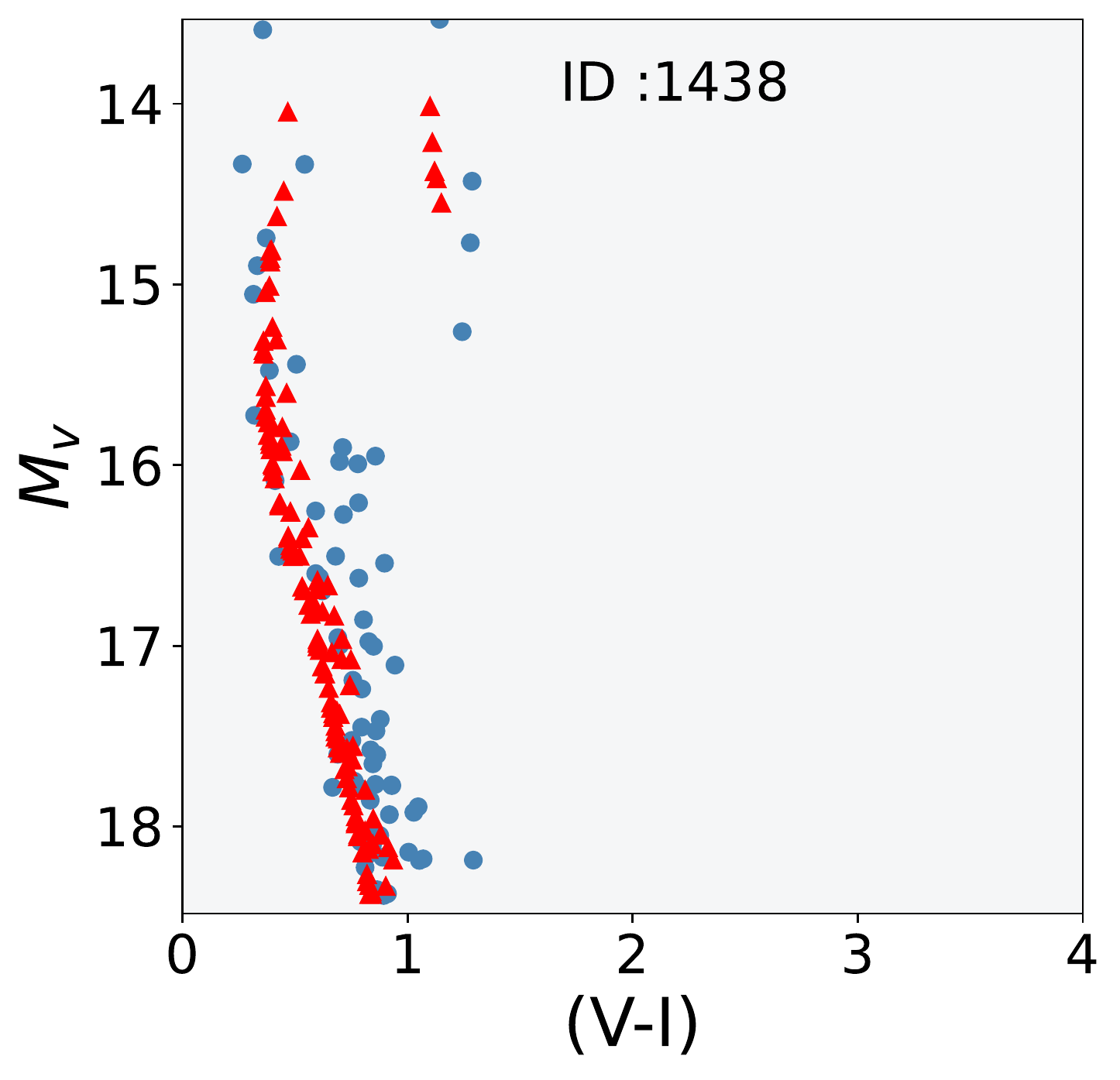}
}  
\subfigure{
\includegraphics[width=2in,height=2in]{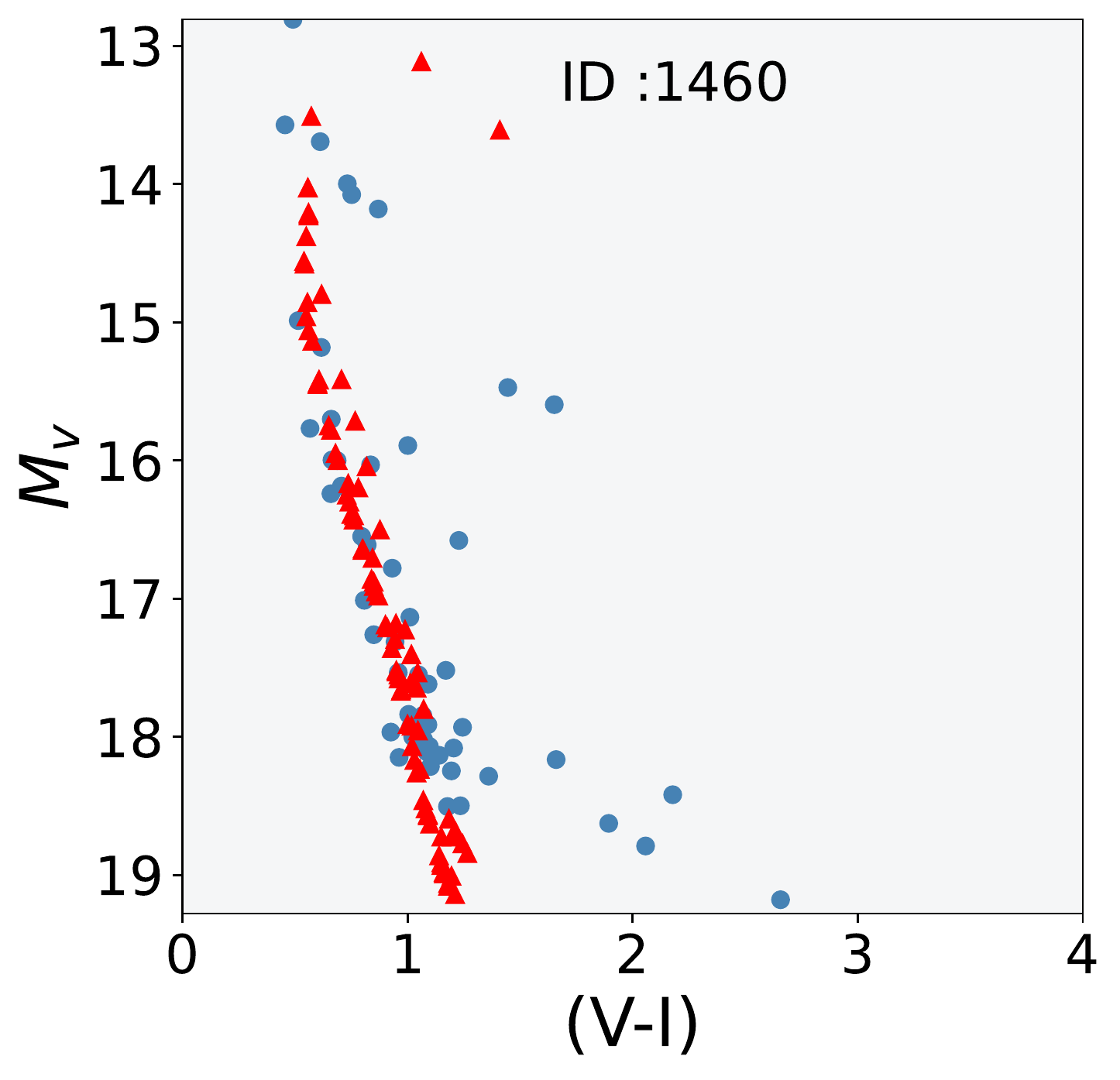}
}  

\end{center}
\begin{center}
\subfigure{
\includegraphics[width=2in,height=2in]{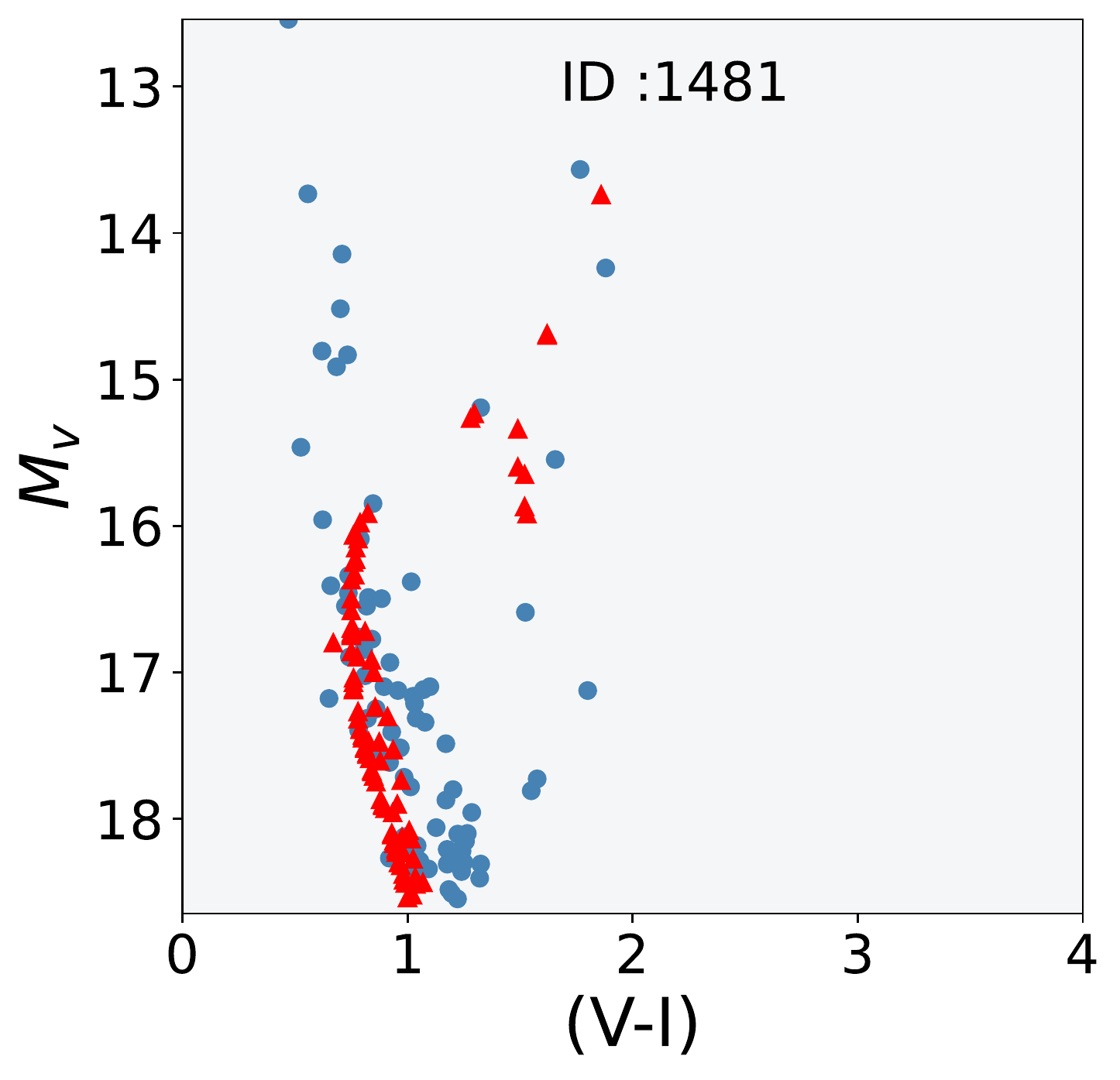}
}  
\subfigure{
\includegraphics[width=2in,height=2in]{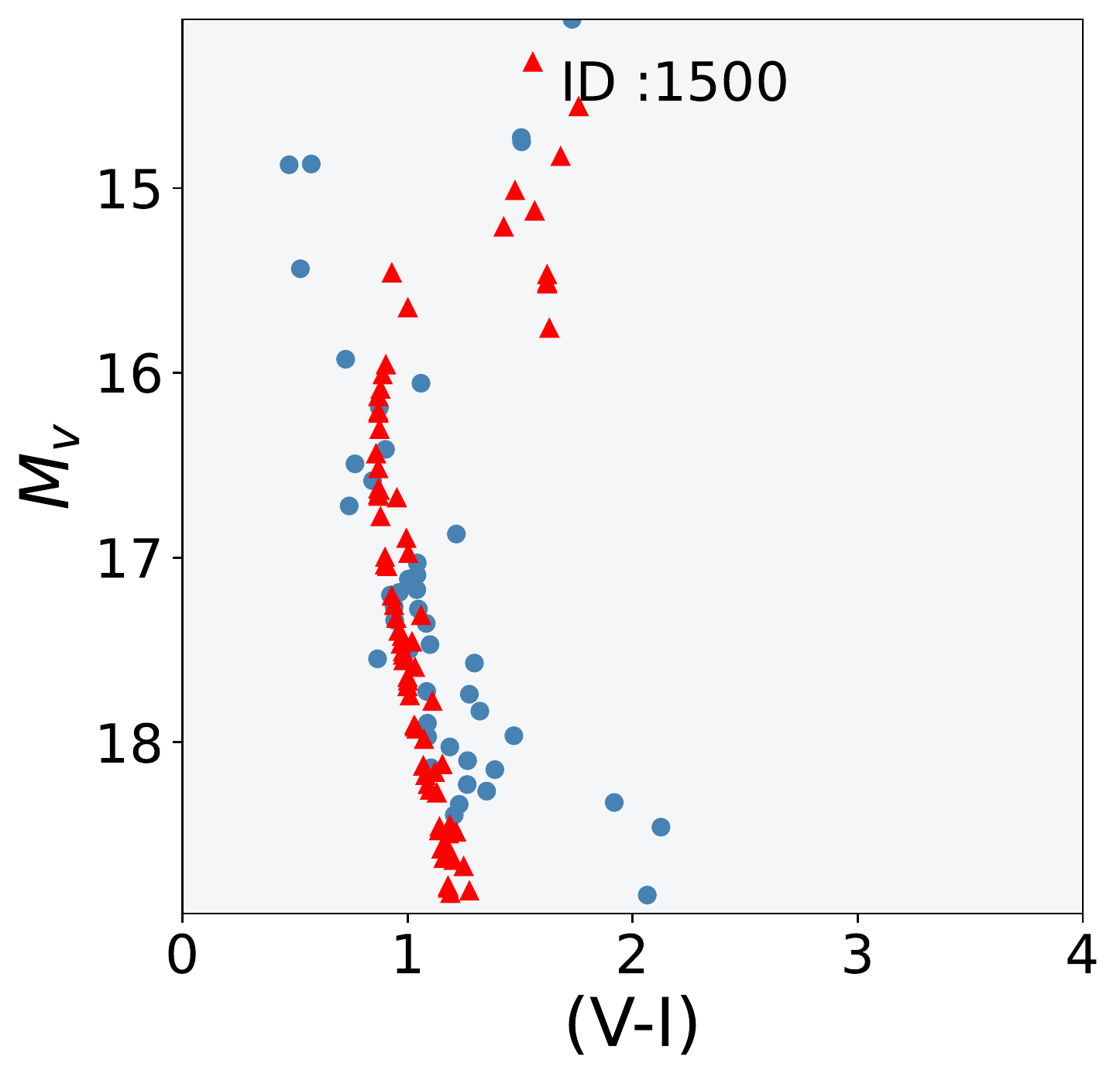}
}  
\subfigure{
\includegraphics[width=2in,height=2in]{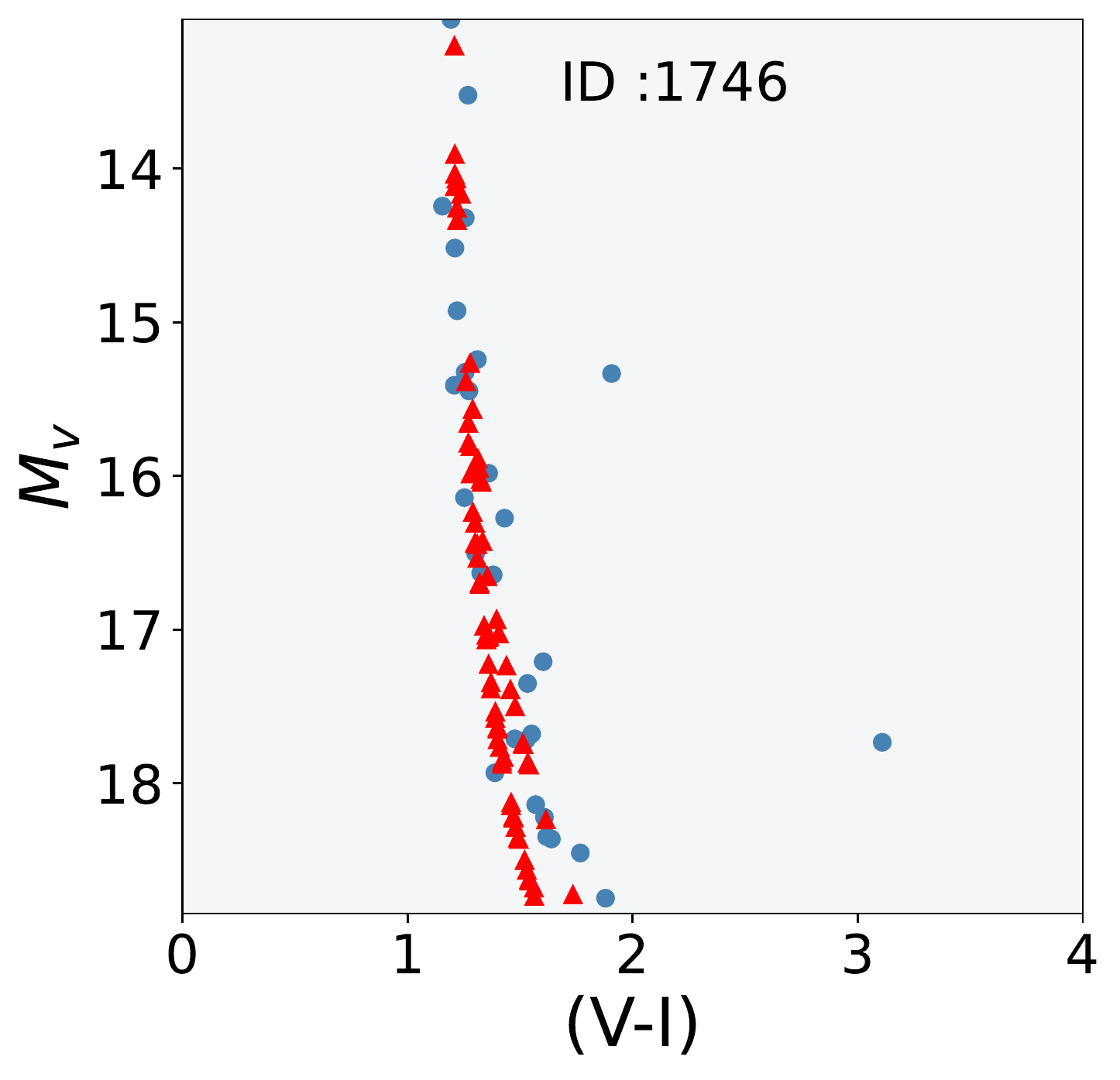}
}  

\end{center}
\caption{Comparison of best-fit and observed CMDs of 12 newly identified OCs with clear main sequences. The blue points represent observed stars, and the  red triangle is for best-fit stellar populations. ID means cluster-ID. The V and I magnitudes are taken here because their uncertainties are relatively small. The magnitudes can be transformed from the G$_{BP}$ and G$_{RP}$ magnitudes using some transformation functions (Riello et al. 2021).}
\label{fig:hr01}
\end{figure*} 

 \begin{figure*}

\begin{center}
  \subfigure{
\includegraphics[width=2in,height=2in]{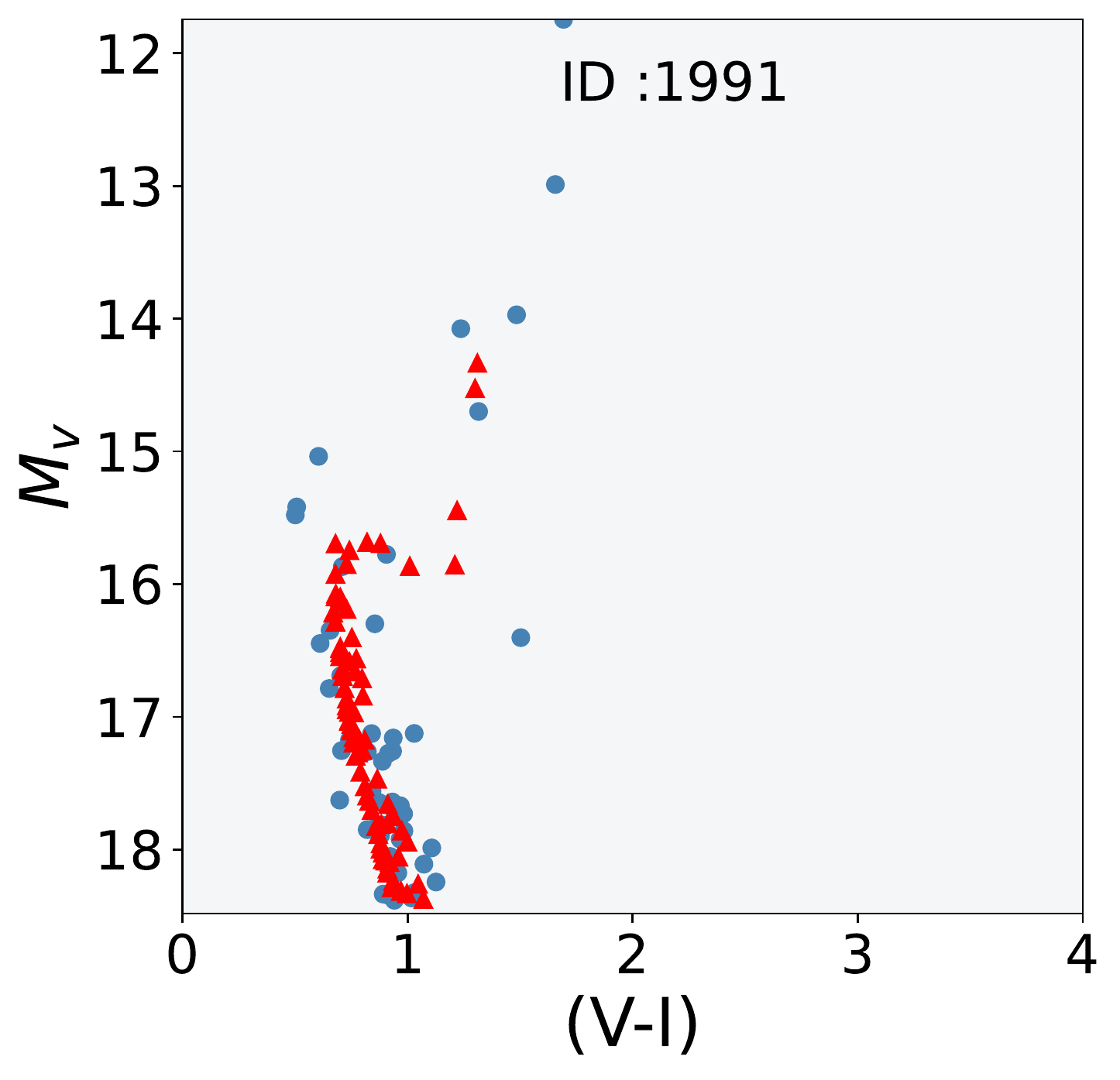}
}  
\subfigure{
\includegraphics[width=2in,height=2in]{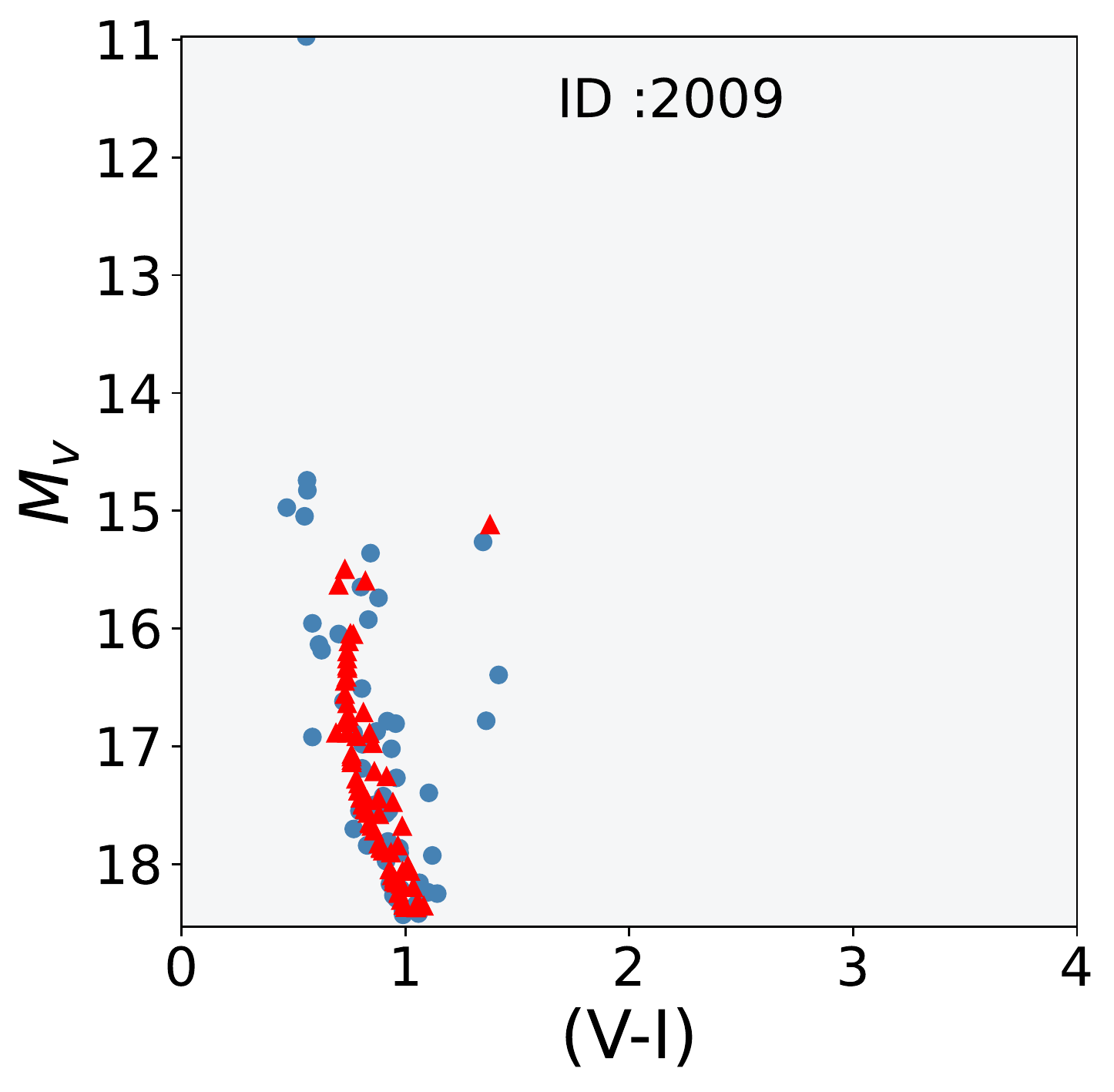}
}  
\subfigure{
\includegraphics[width=2in,height=2in]{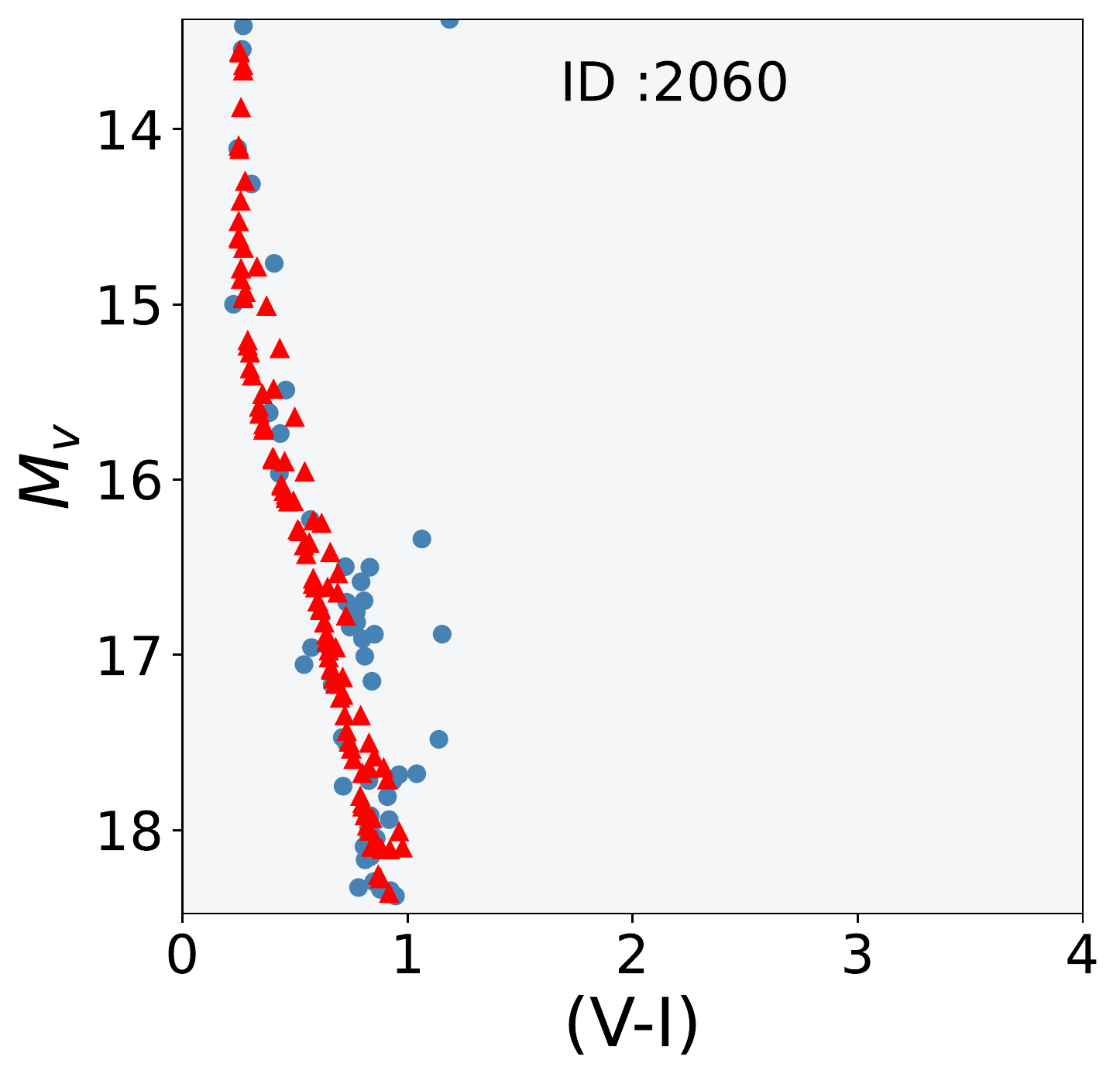}
}  
\end{center}

\begin{center}
 \subfigure{
\includegraphics[width=2in,height=2in]{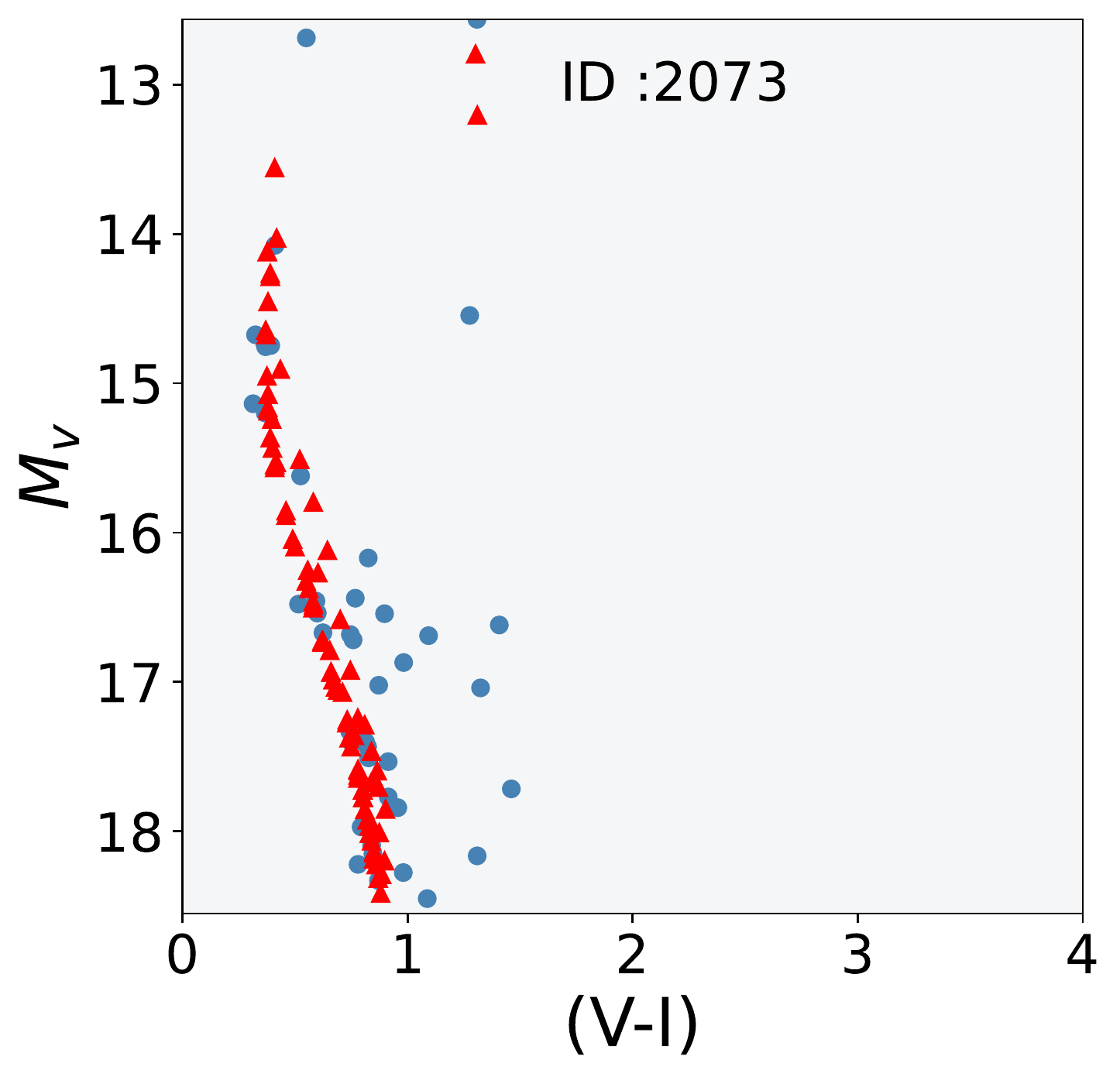}
}  
\subfigure{
\includegraphics[width=2in,height=2in]{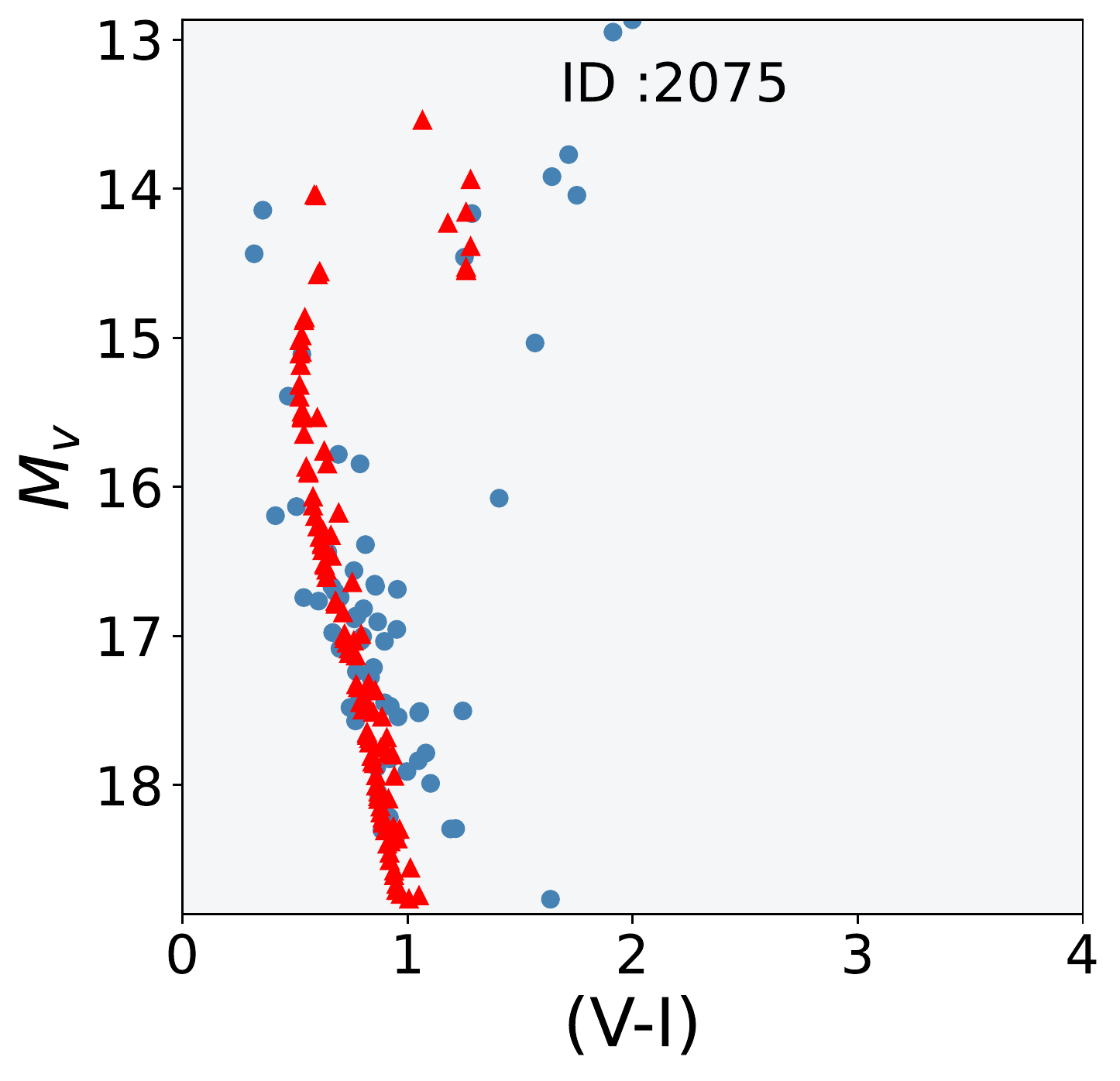}
}  
\subfigure{
\includegraphics[width=2in,height=2in]{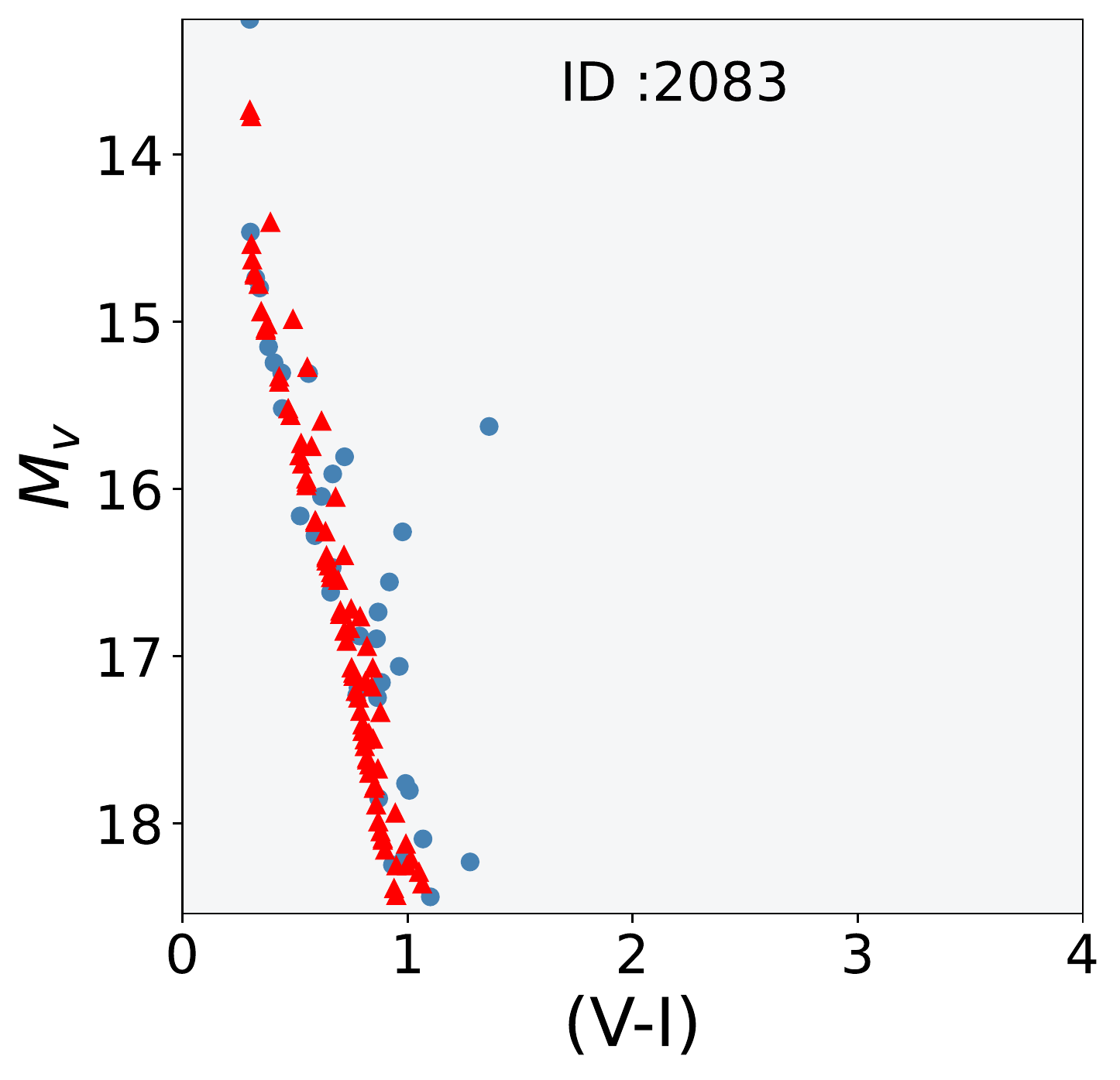}
}

\end{center}
\begin{center}
\subfigure{
\includegraphics[width=2in,height=2in]{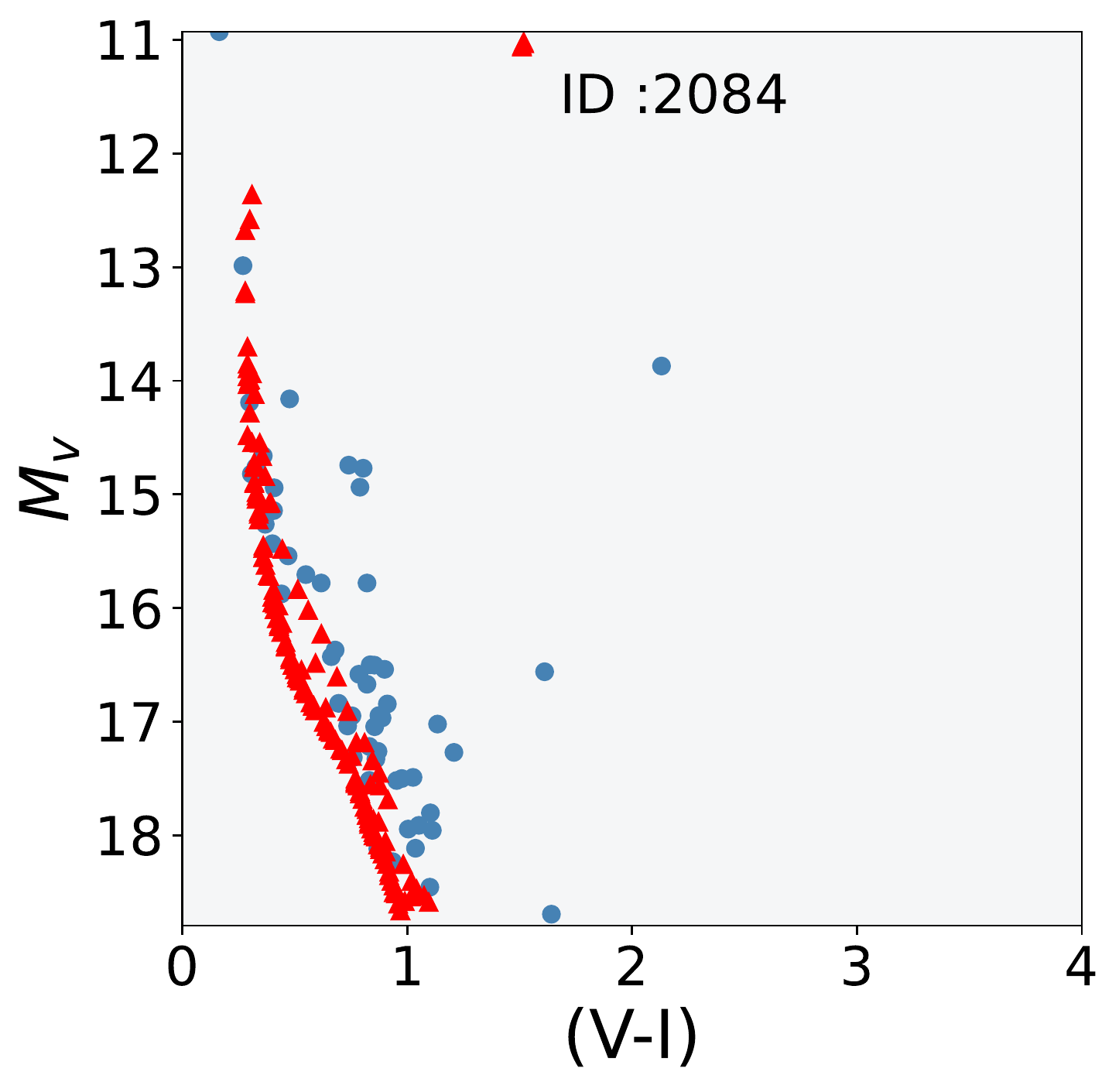}
}  
\subfigure{
\includegraphics[width=2in,height=2in]{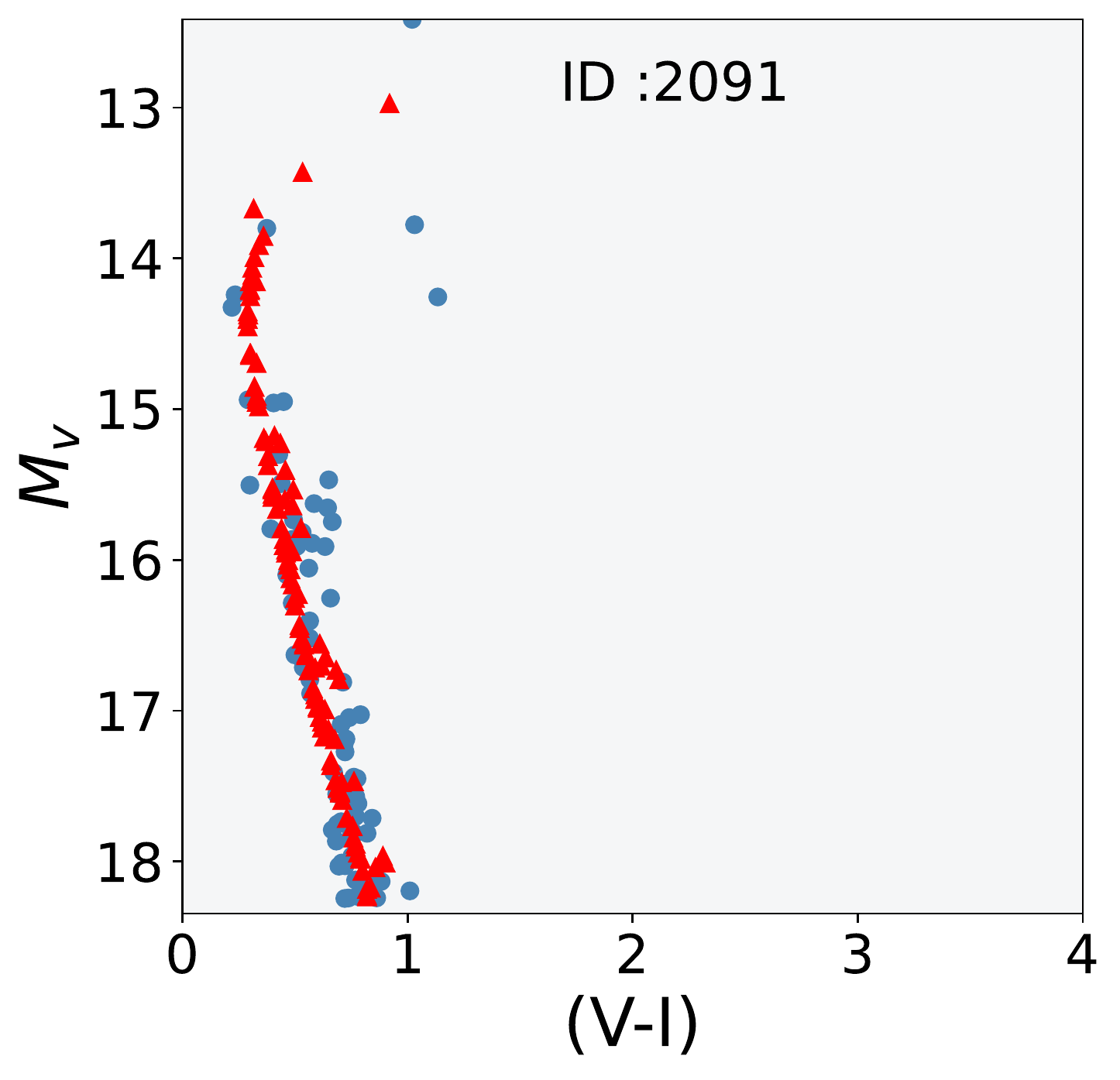}
}  
\subfigure{
\includegraphics[width=2in,height=2in]{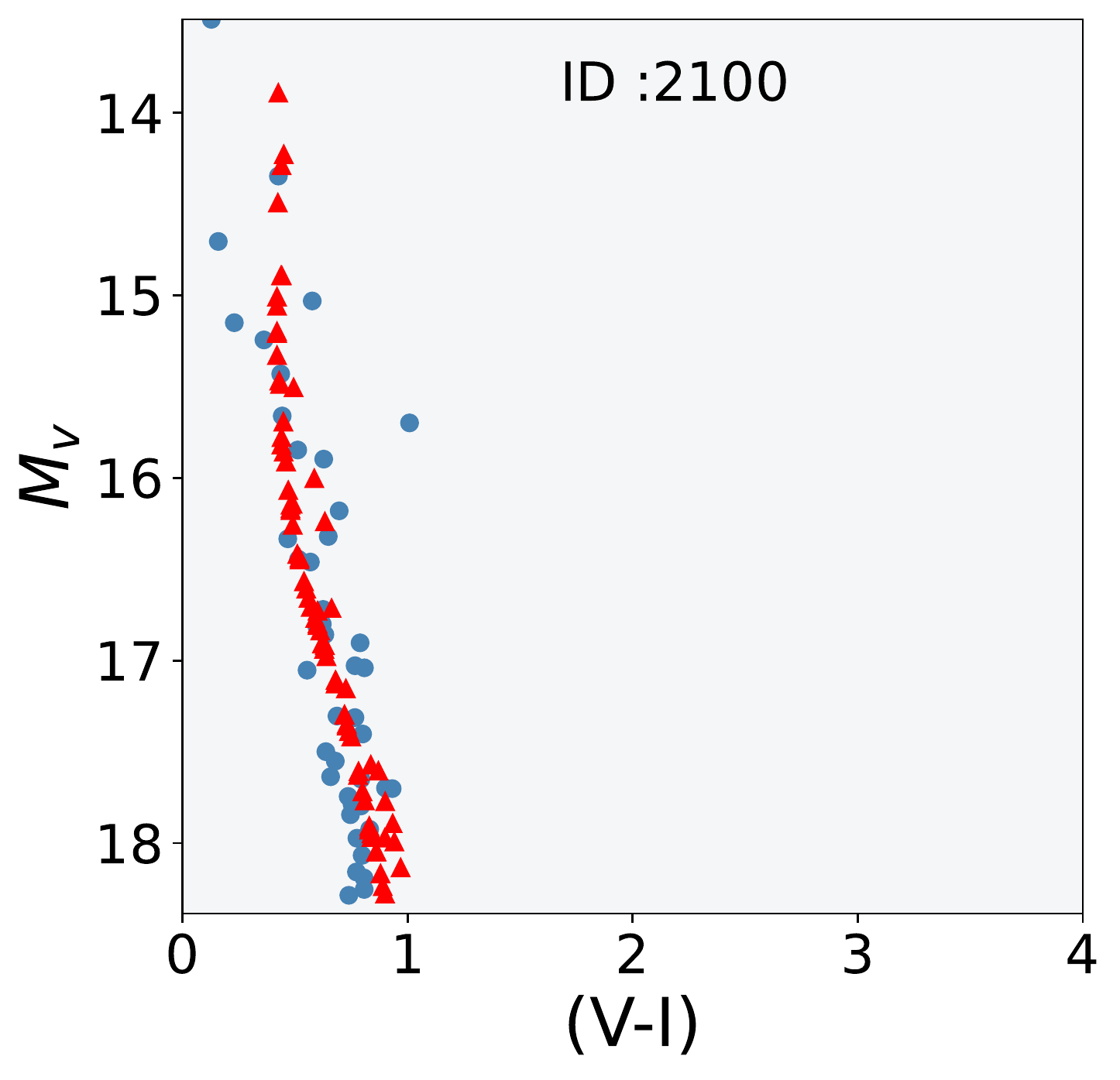}
}  

\end{center}
\begin{center}
\subfigure{
\includegraphics[width=2in,height=2in]{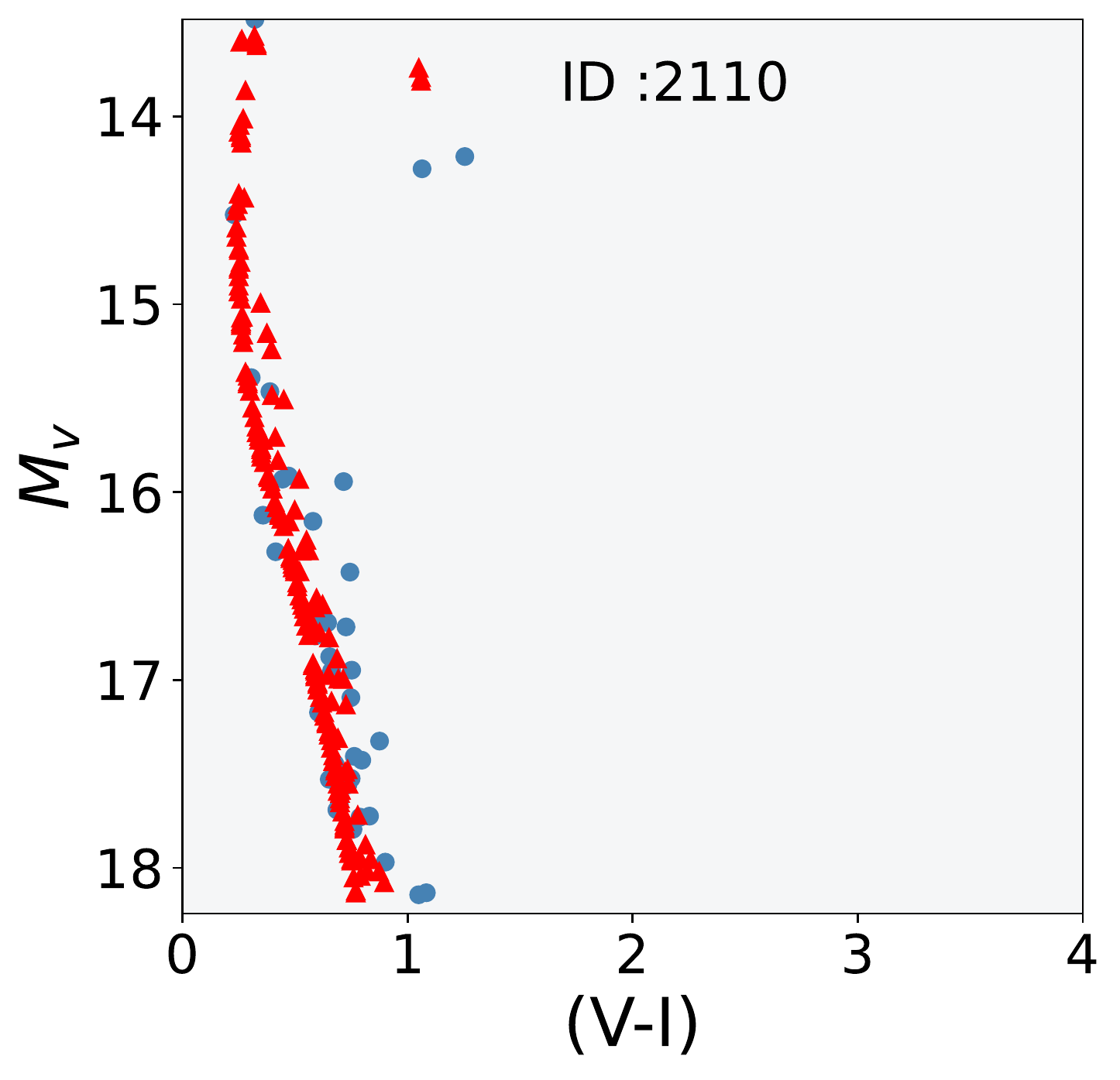}
} 
\subfigure{
\includegraphics[width=2in,height=2in]{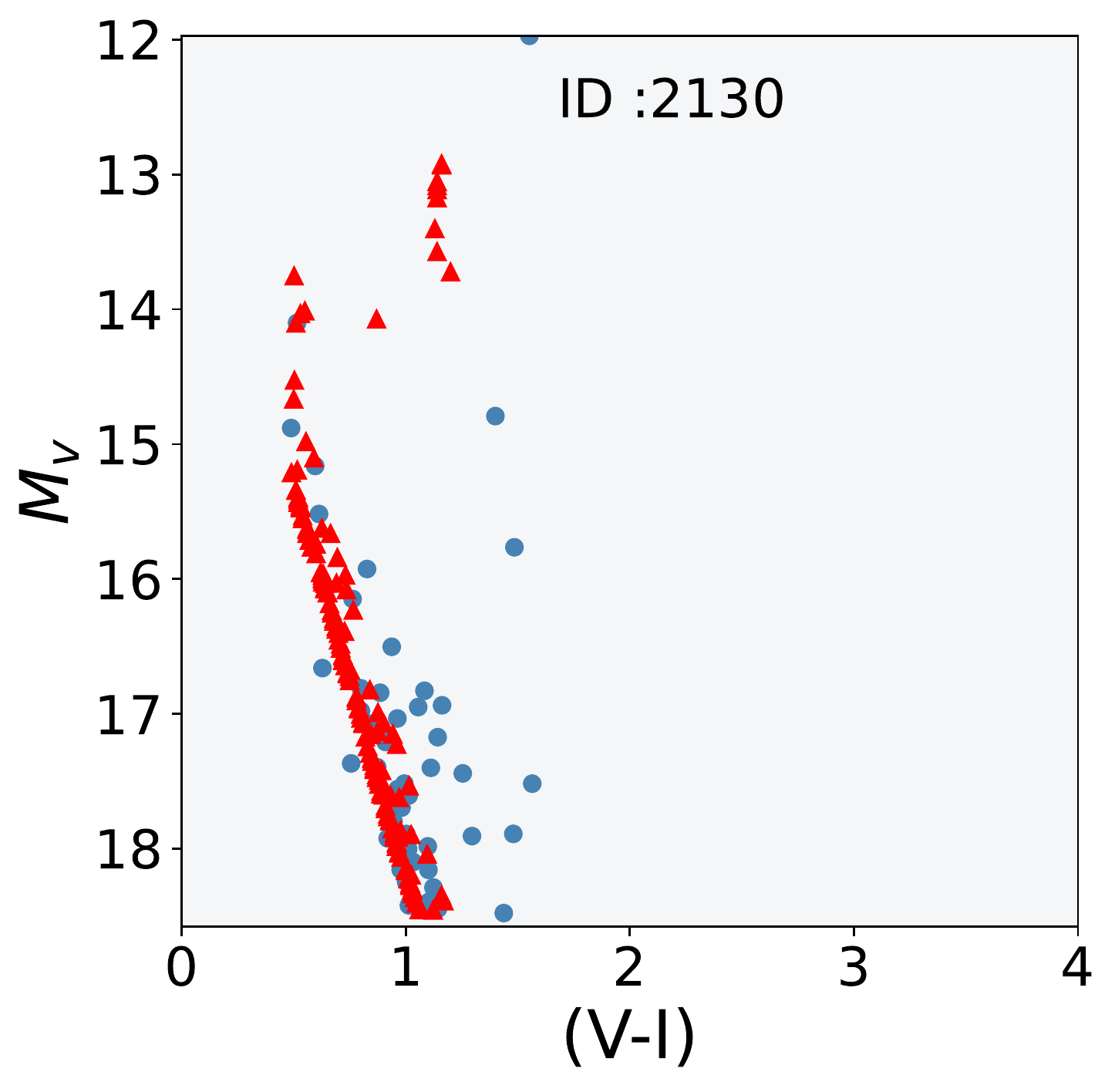}
} 
\subfigure{
\includegraphics[width=2in,height=2in]{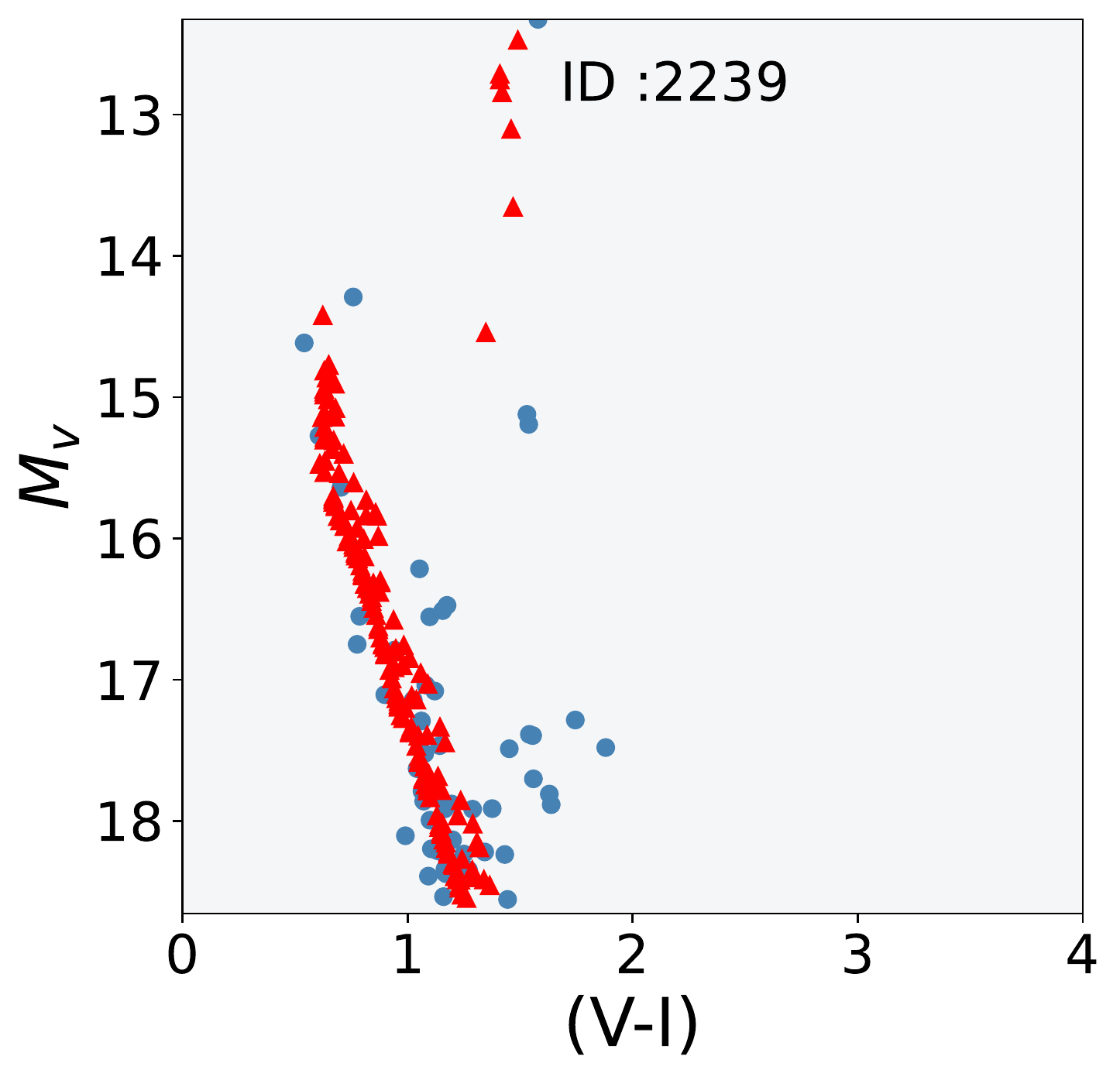}
}

\end{center}
\caption{Same as Fig~\ref{fig:hr01} but for other 12 OCs.
}
\label{fig:hr02}
\end{figure*}

\begin{figure*}

\begin{center}
    \subfigure{
\includegraphics[width=2in,height=2in]{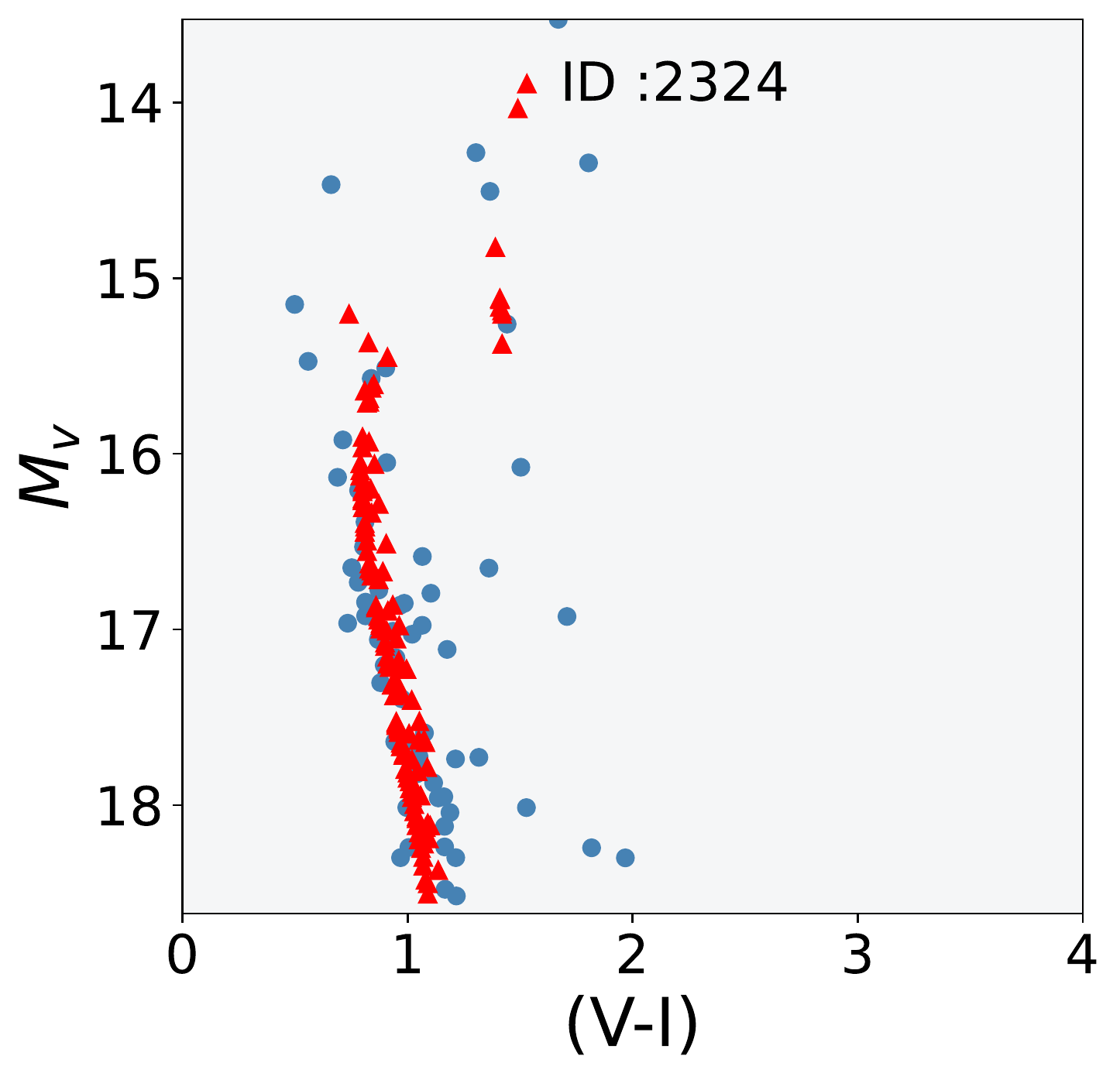}
}  
  \subfigure{
\includegraphics[width=2in,height=2in]{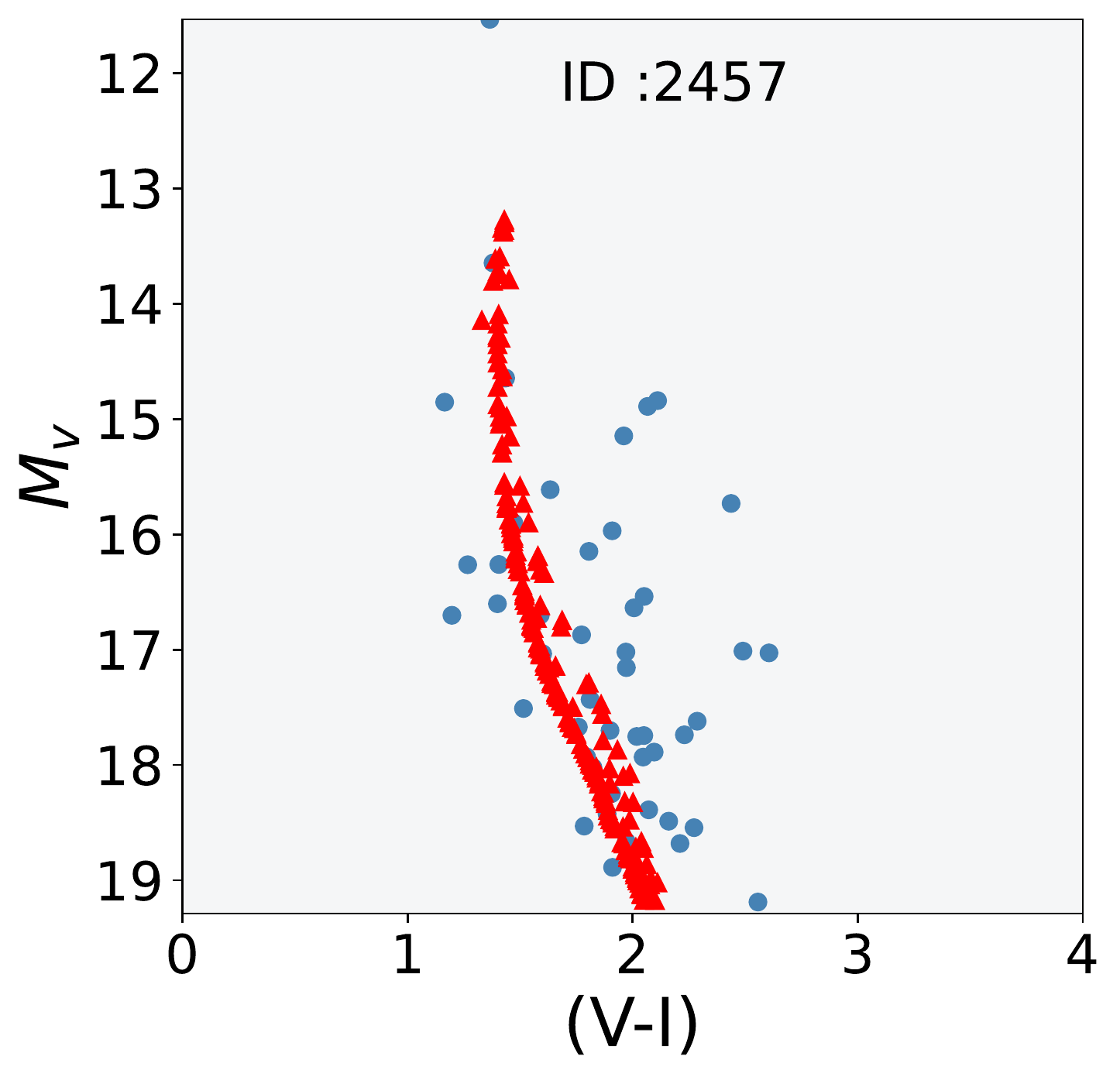}
}  
\subfigure{
\includegraphics[width=2in,height=2in]{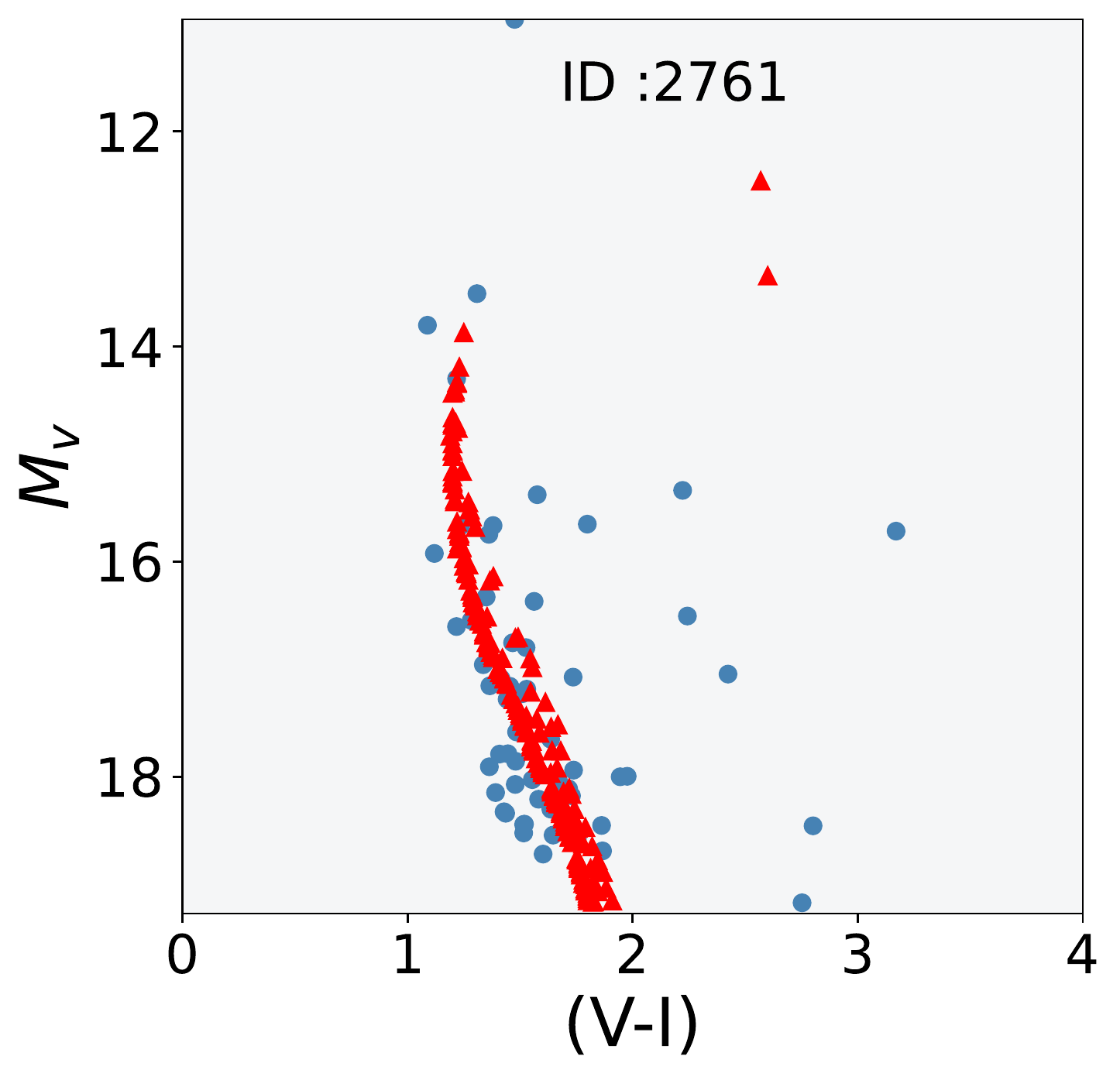}
}  

\end{center}

\begin{center}
\subfigure{
\includegraphics[width=2in,height=2in]{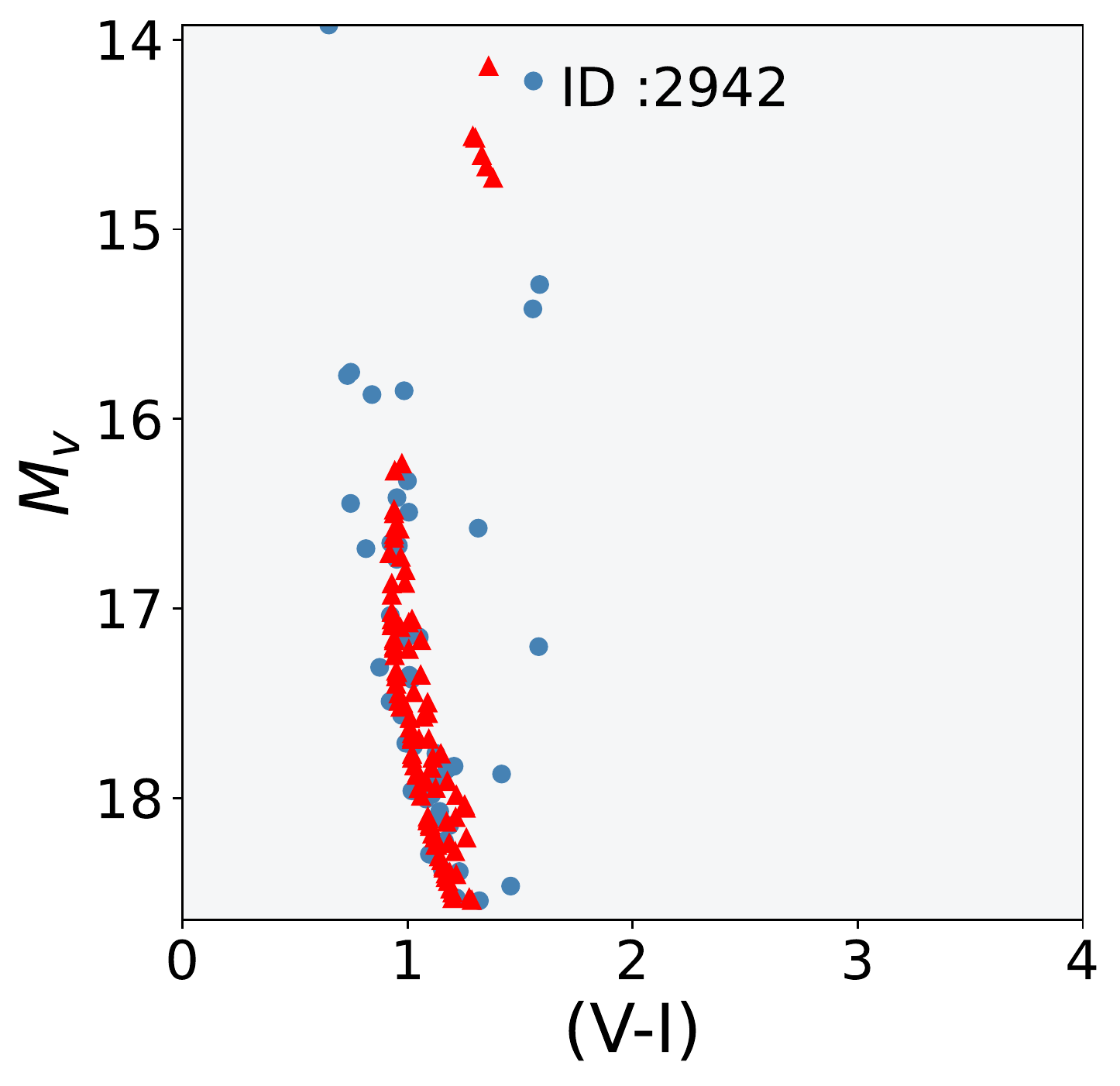}
}  
 \subfigure{
\includegraphics[width=2in,height=2in]{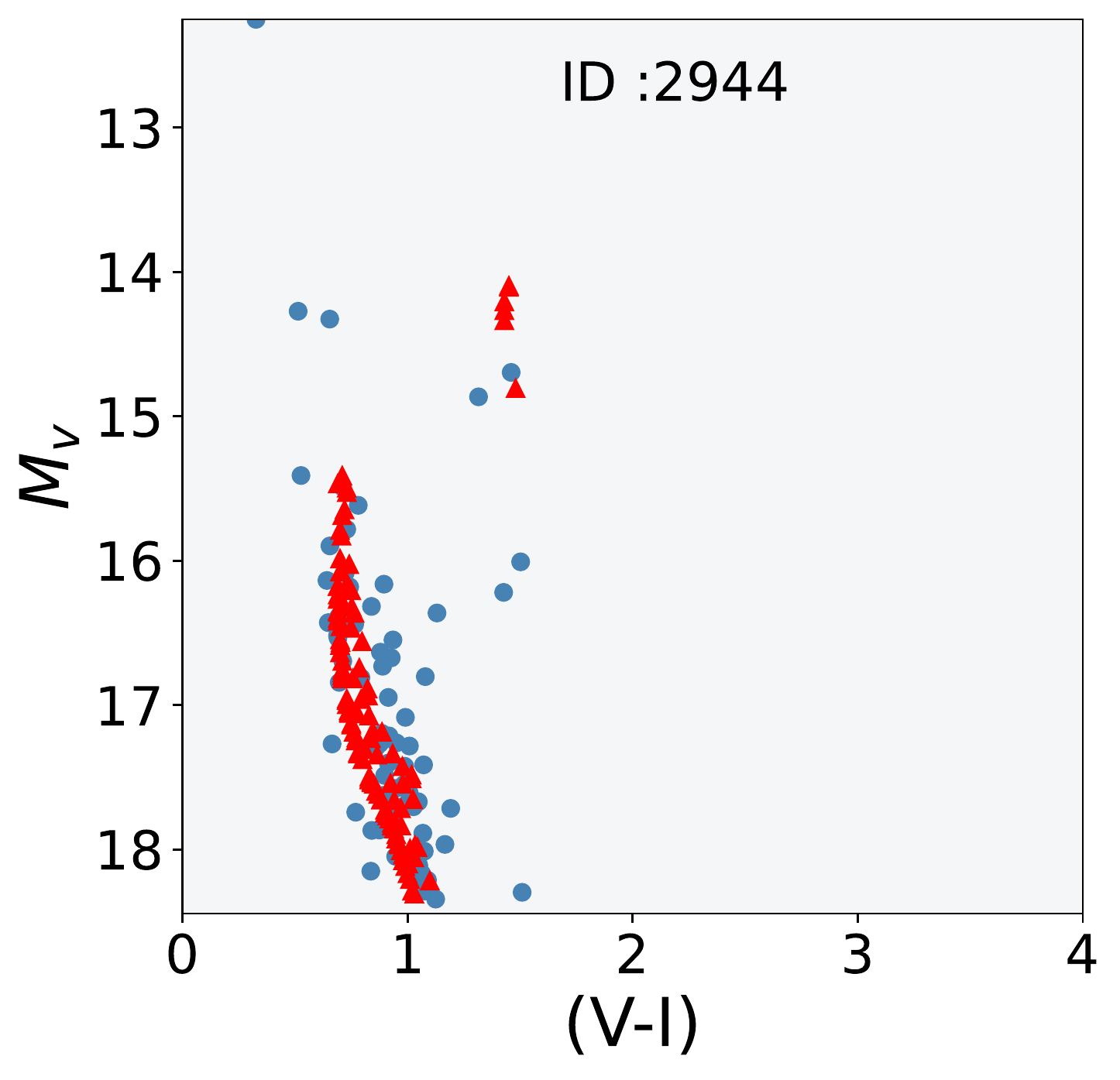}
}  
\subfigure{
\includegraphics[width=2in,height=2in]{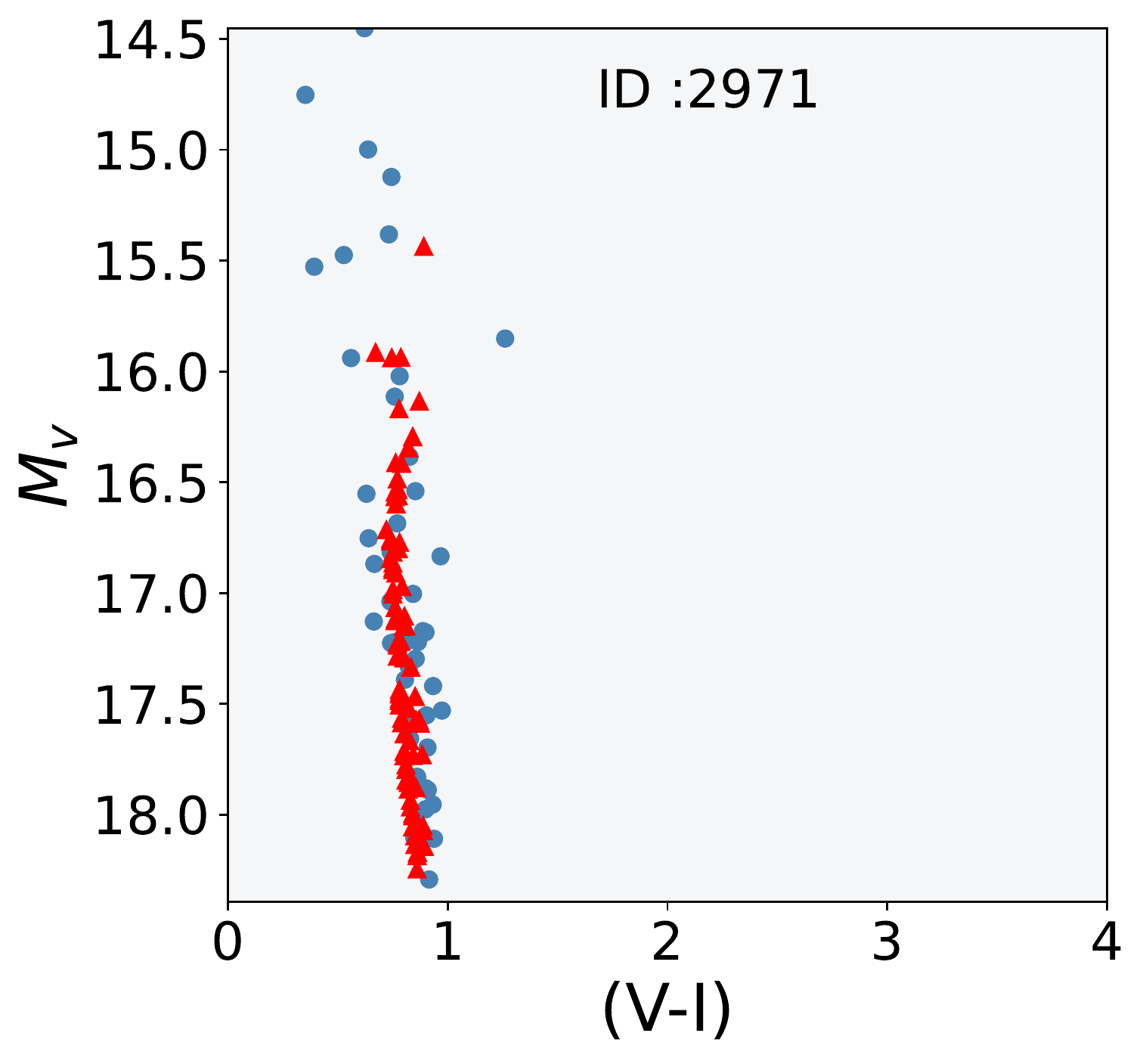}
}

\end{center}
\begin{center}
\subfigure{
\includegraphics[width=2in,height=2in]{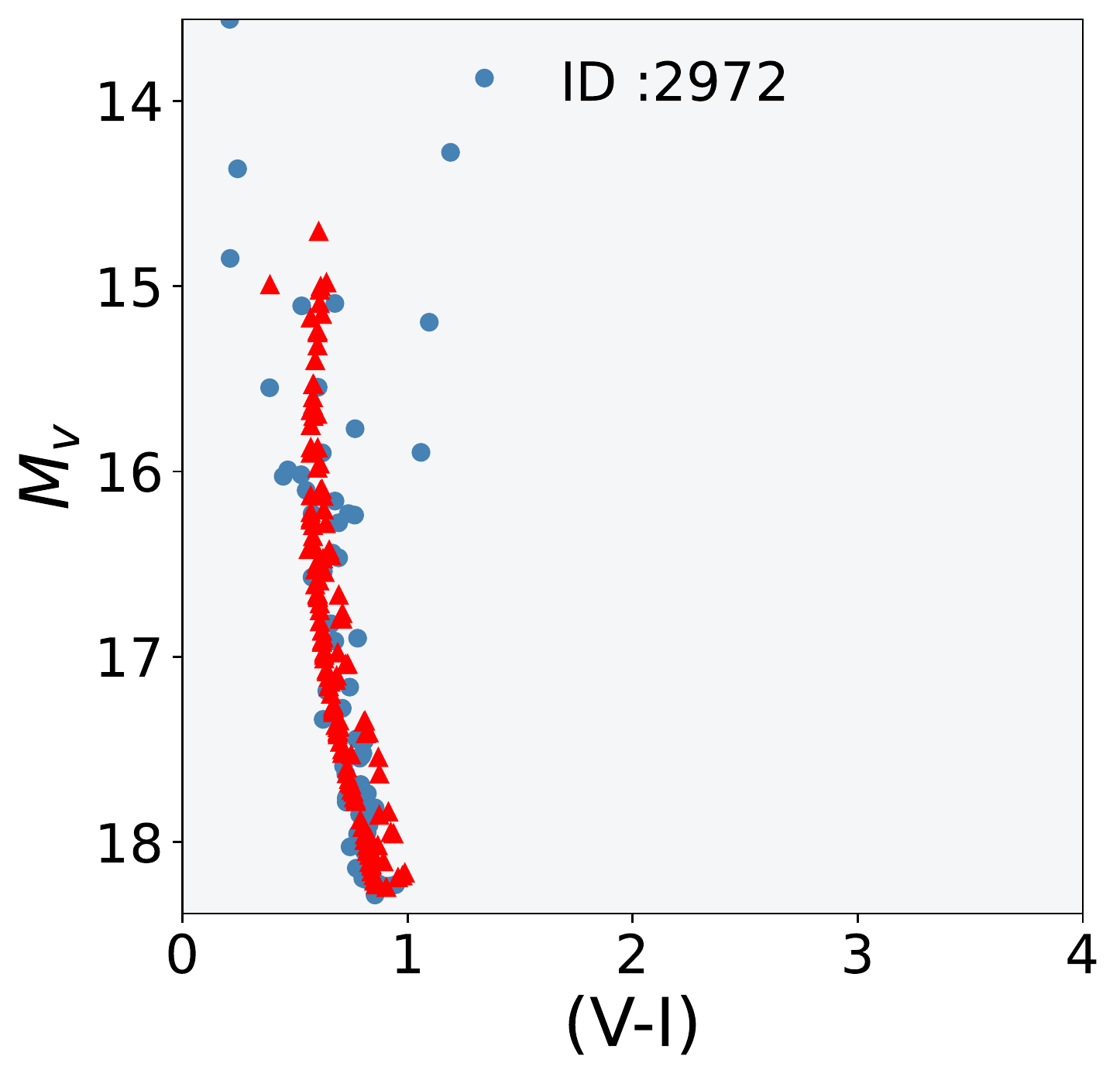}
} 
\subfigure{
\includegraphics[width=2in,height=2in]{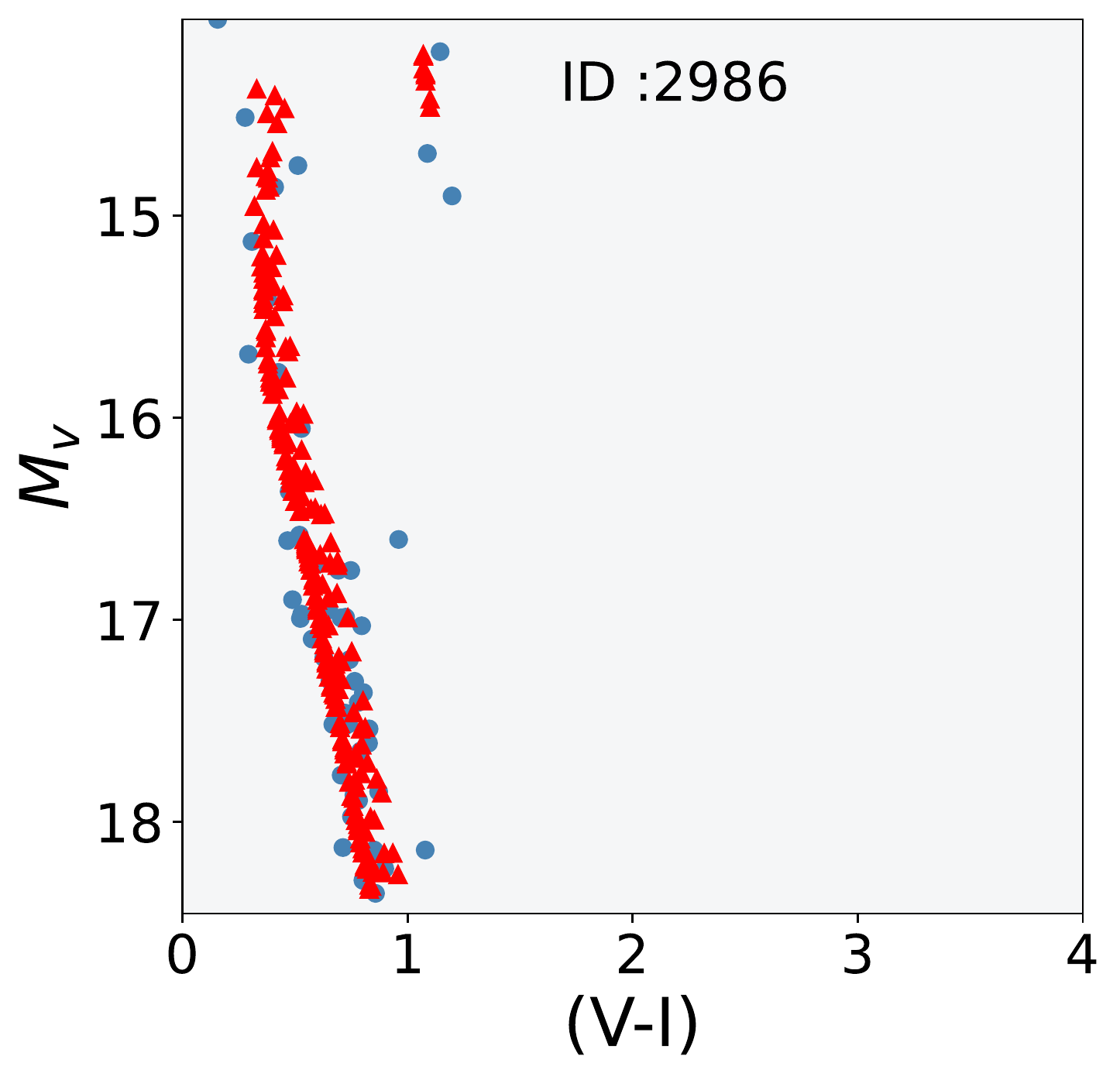}
}  
\subfigure{
\includegraphics[width=2in,height=2in]{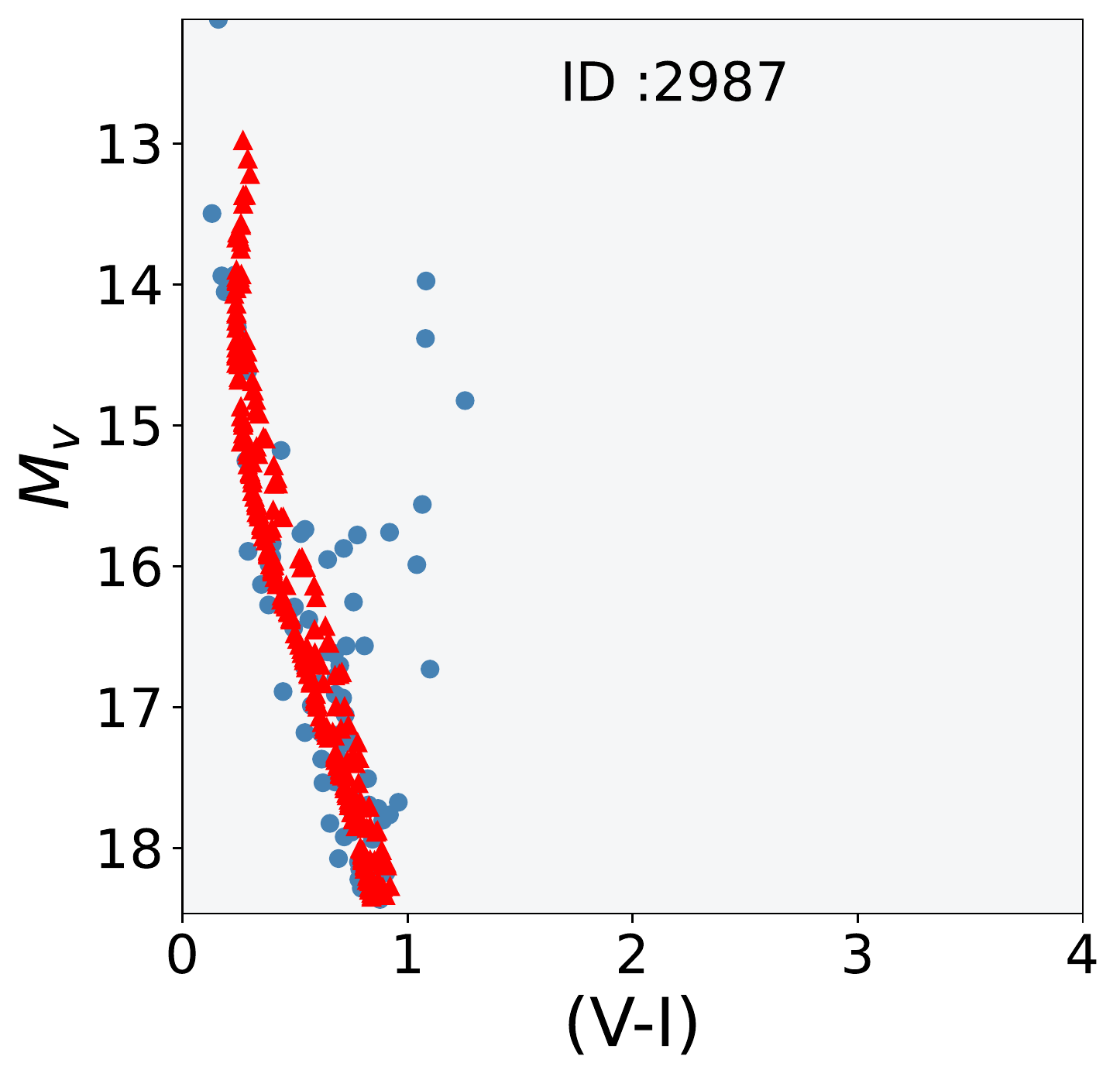}
}

\end{center}
\begin{center}
\subfigure{
\includegraphics[width=2in,height=2in]{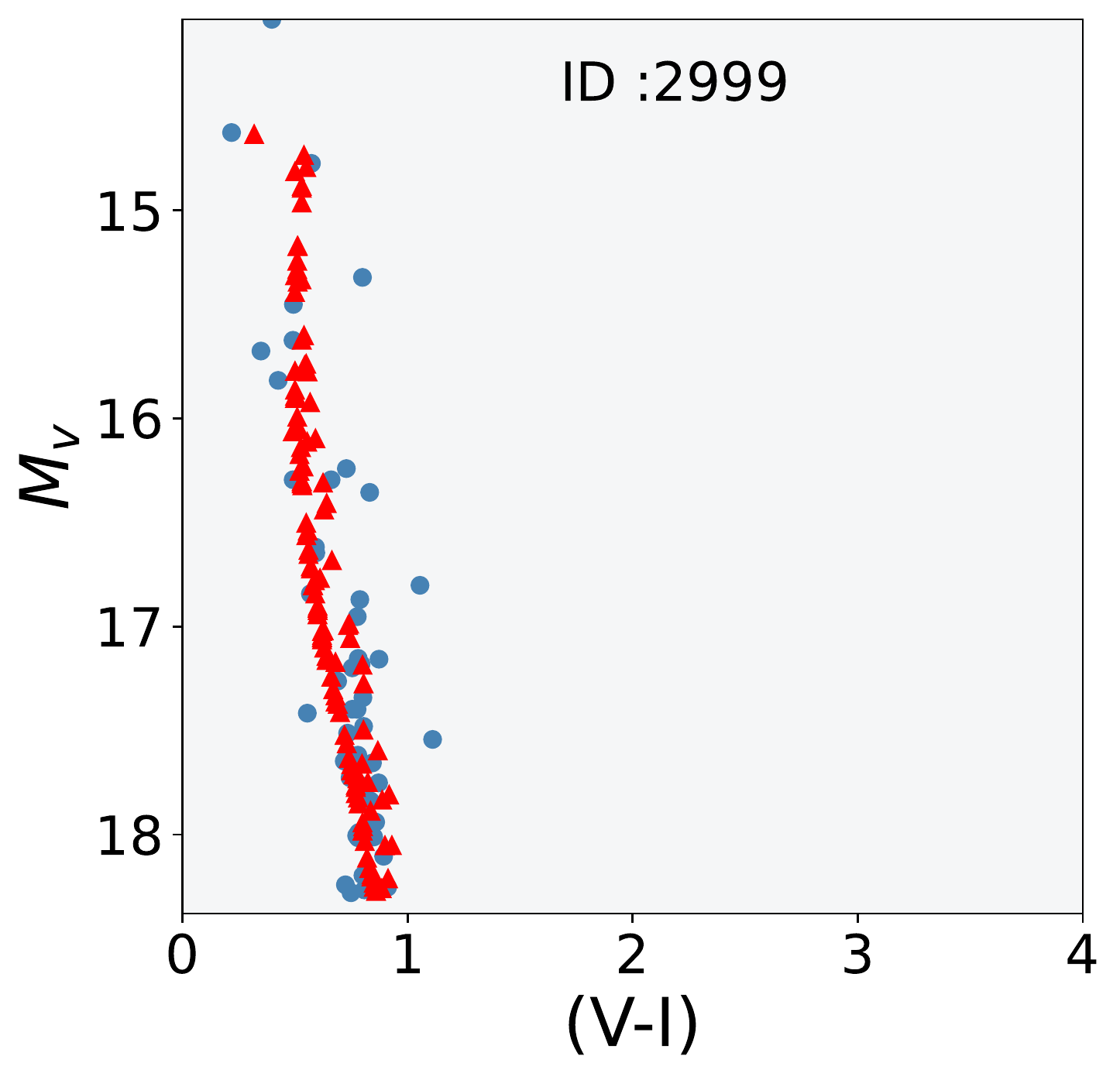}
} 
\subfigure{
\includegraphics[width=2in,height=2in]{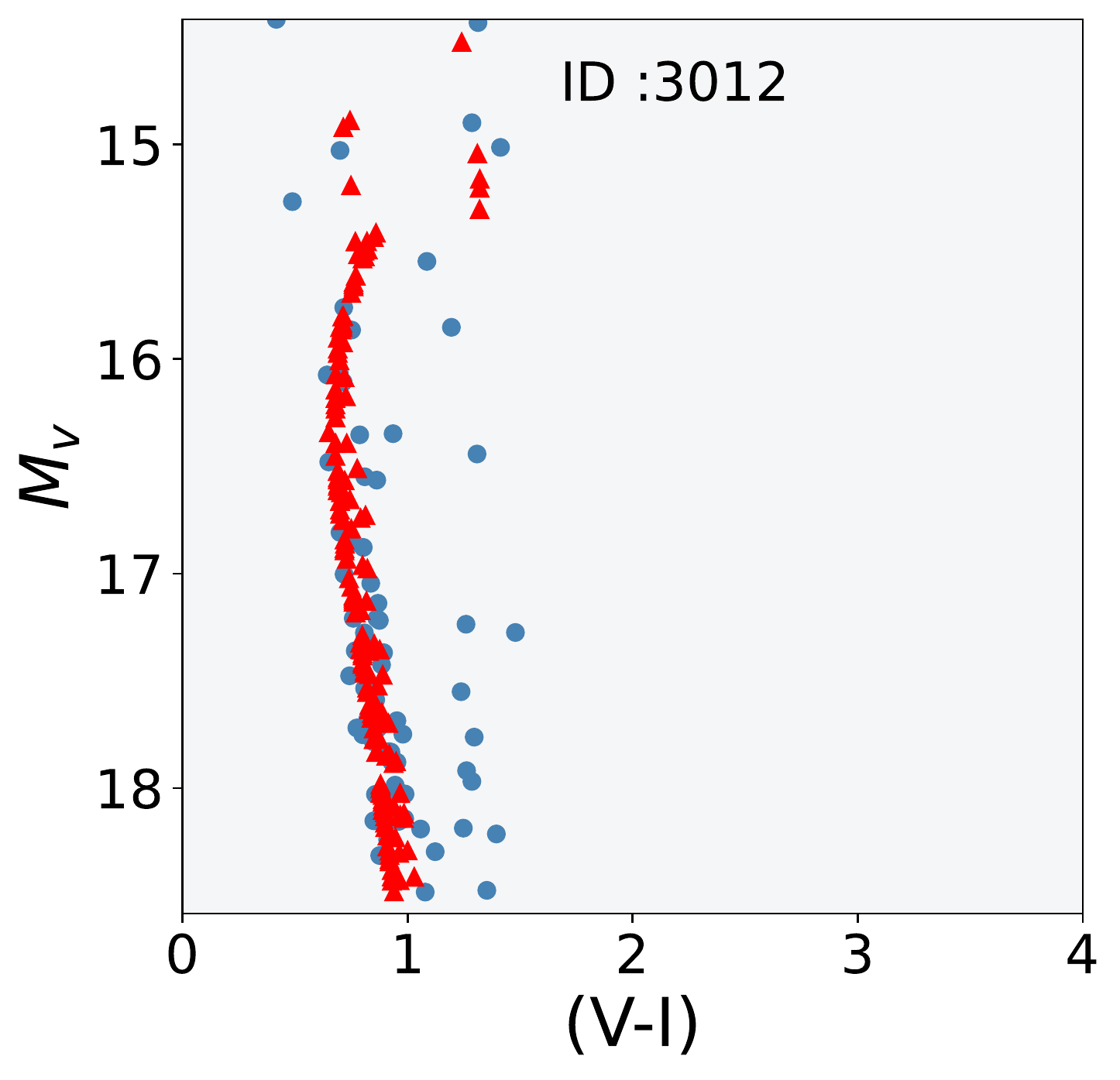}
}  
\subfigure{
\includegraphics[width=2in,height=2in]{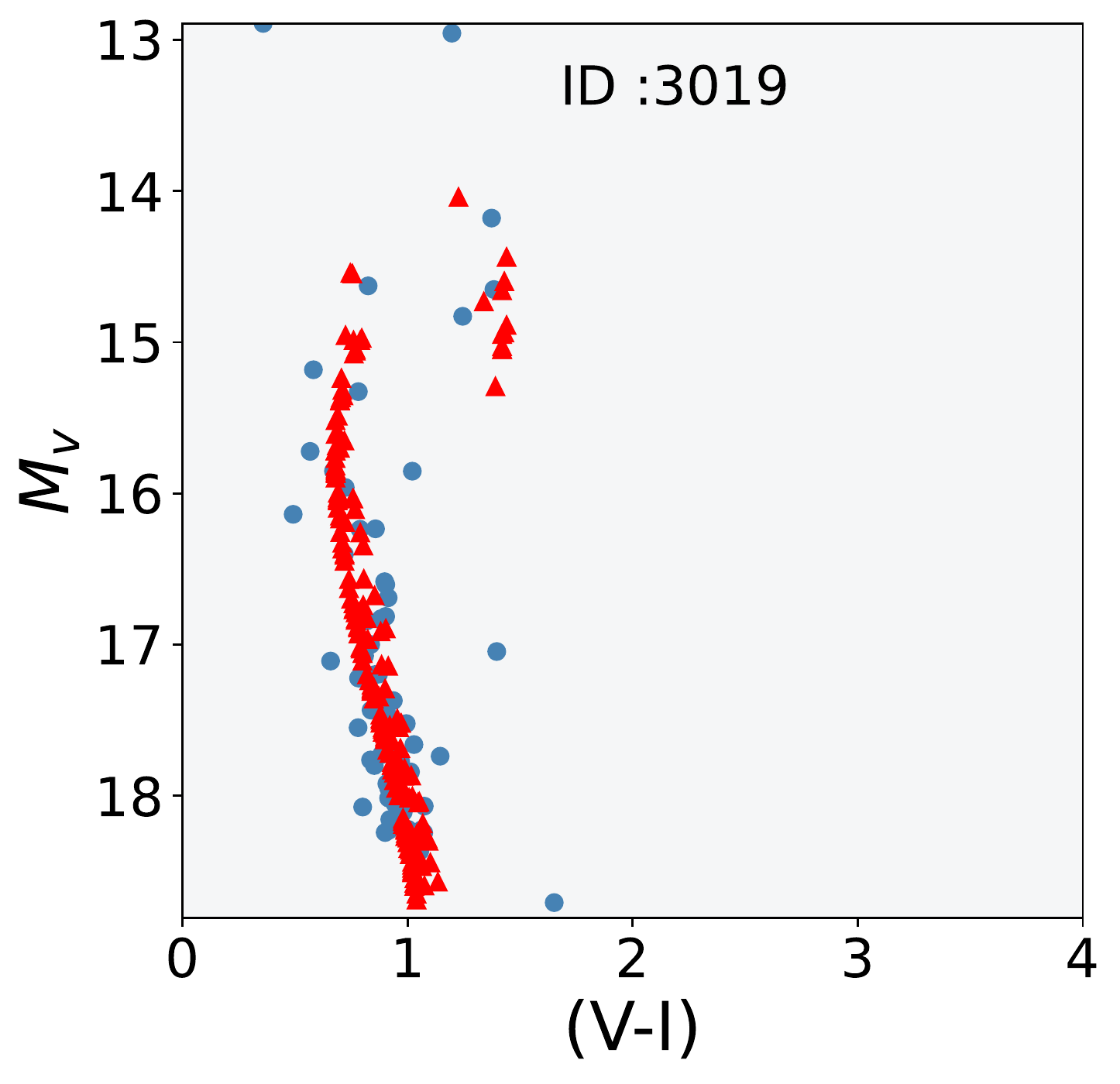}
}

\end{center}
\caption{Same as Fig~\ref{fig:hr01} but for other 12 OCs.
}
\label{fig:hr03}
\end{figure*}

\begin{figure*}

\begin{center}
\subfigure{
\includegraphics[width=2in,height=2in]{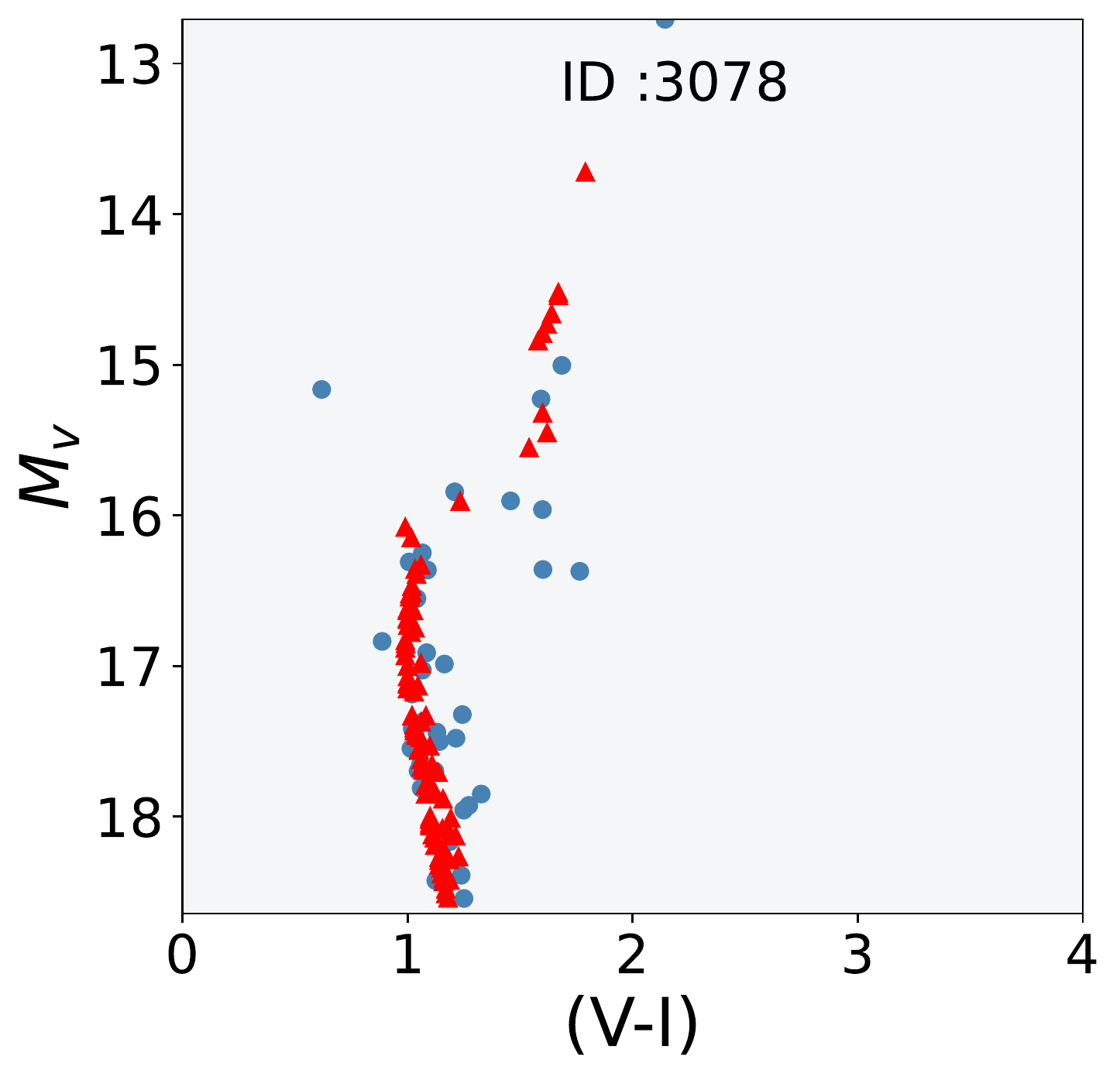}
}  
  \subfigure{
\includegraphics[width=2in,height=2in]{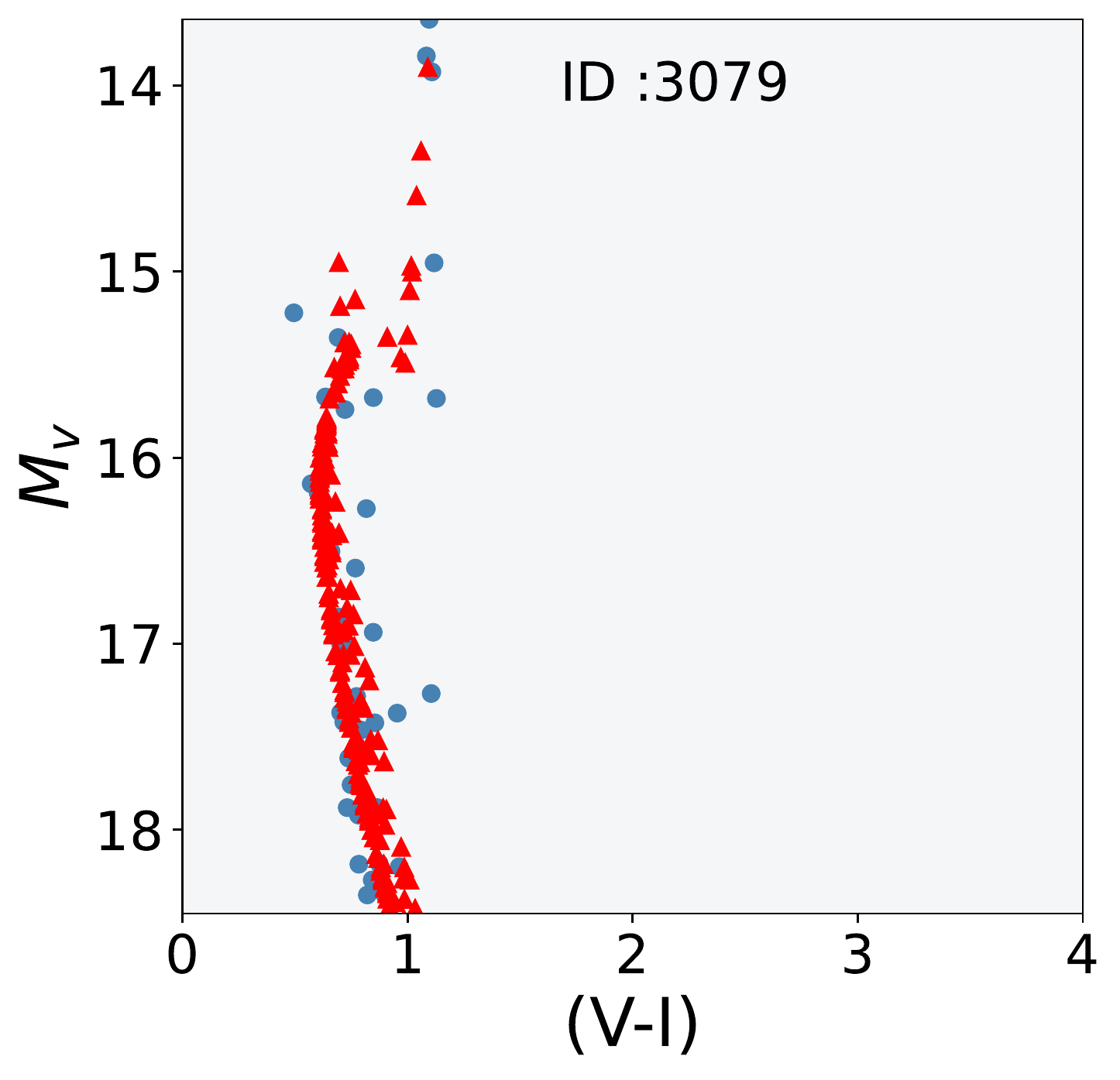}
}  
\subfigure{
\includegraphics[width=2in,height=2in]{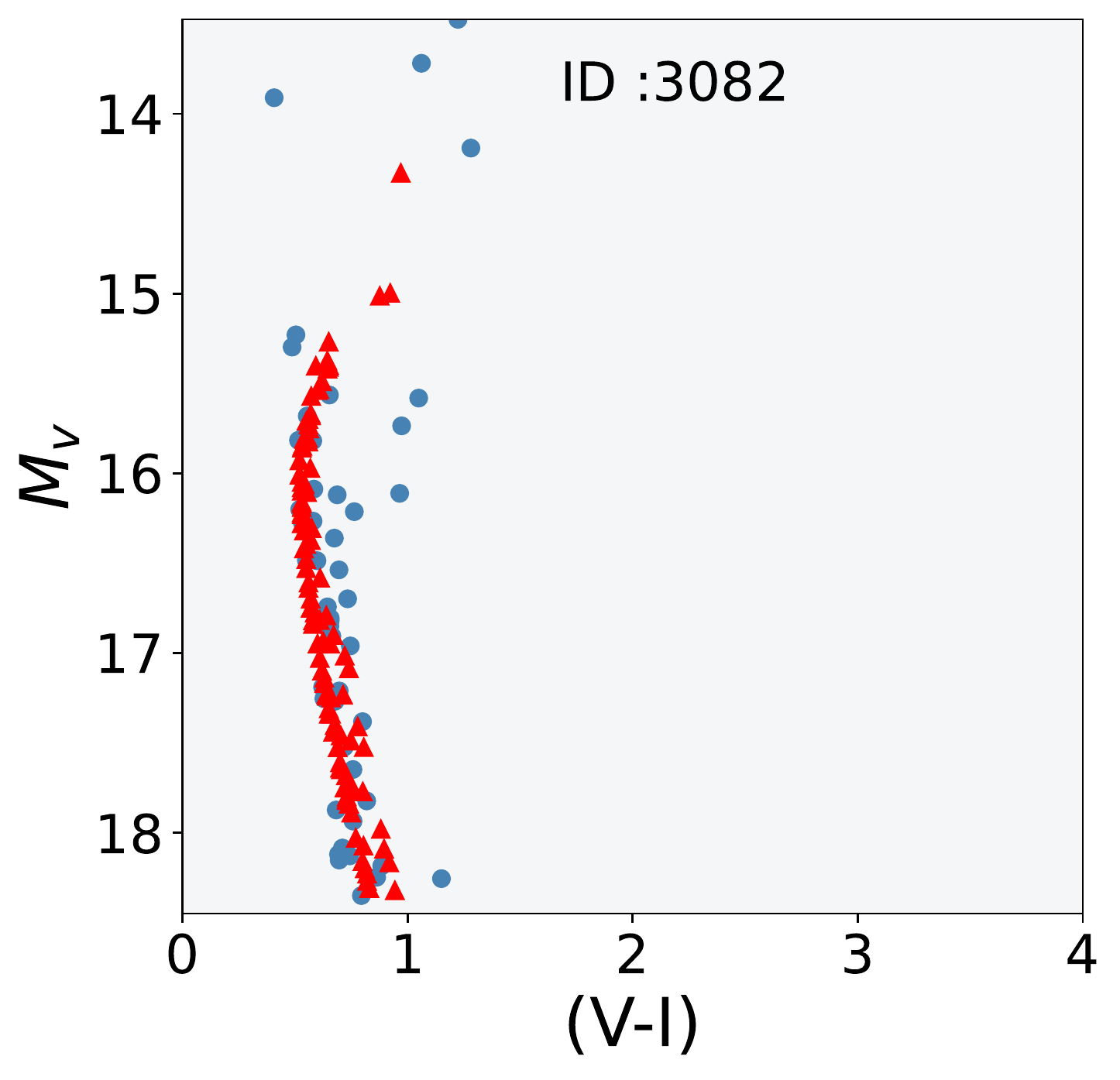}
}  
 
\end{center}

\begin{center}
\subfigure{
\includegraphics[width=2in,height=2in]{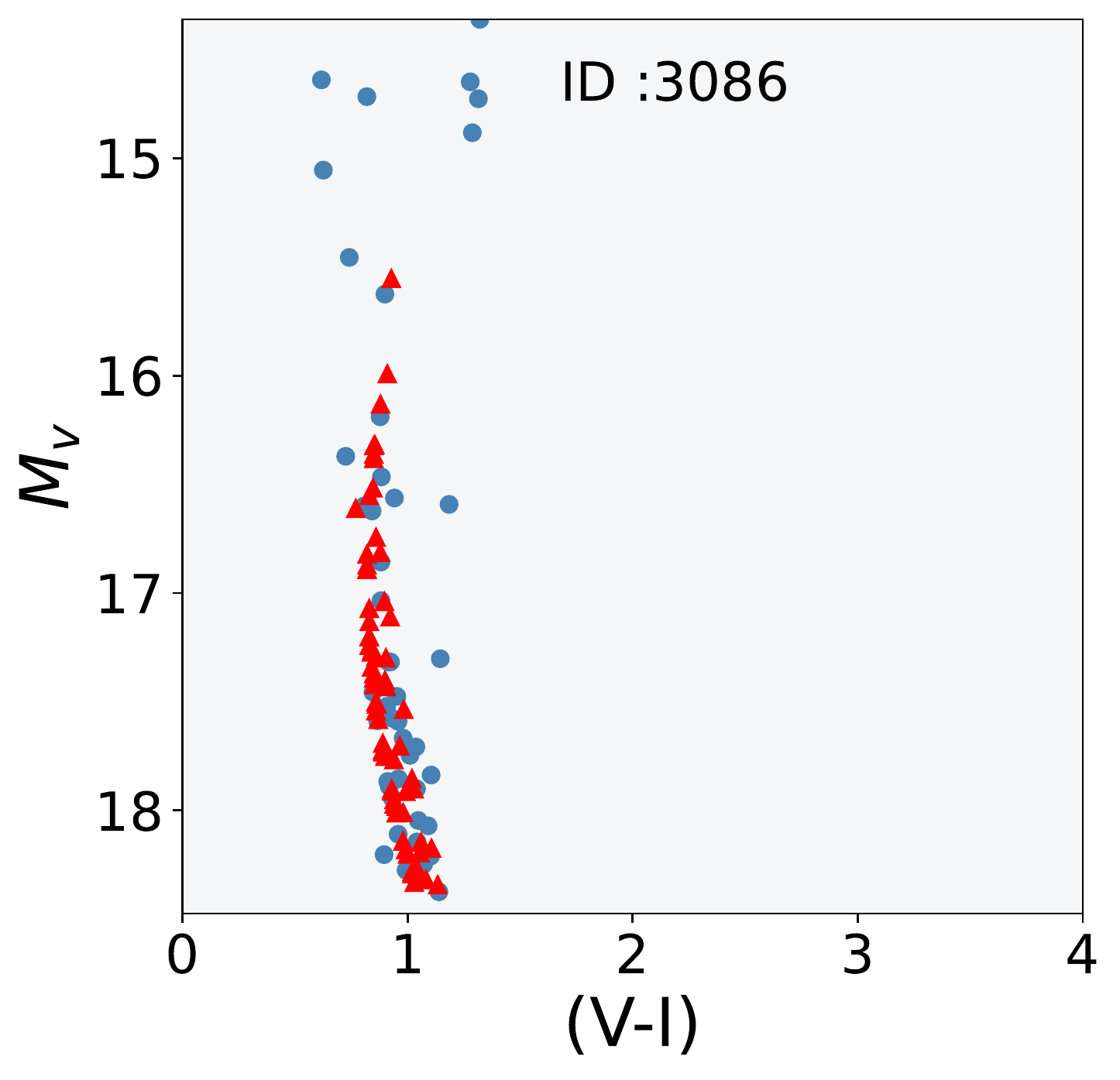}
} 
 \subfigure{
\includegraphics[width=2in,height=2in]{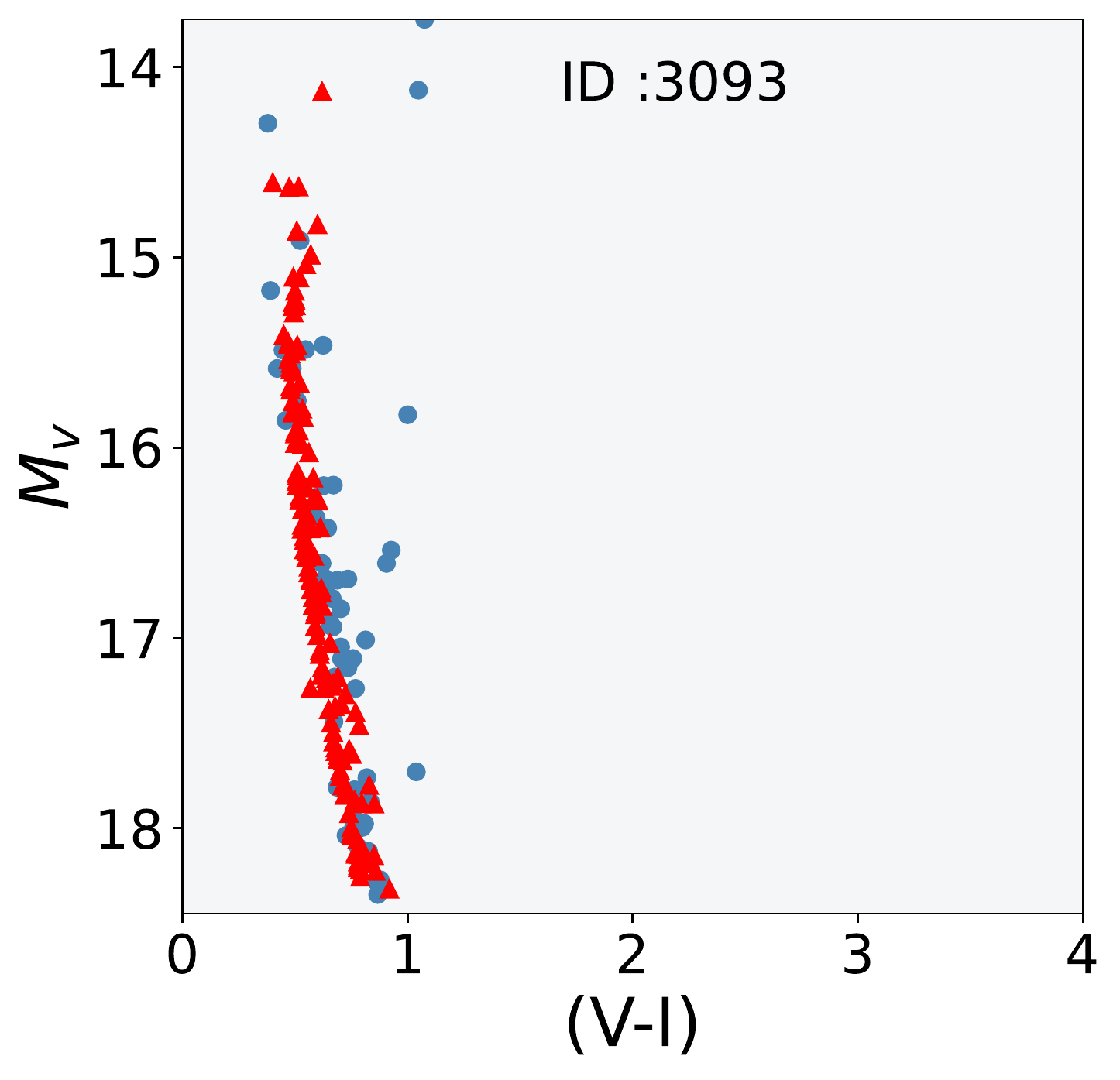}
}  
\subfigure{
\includegraphics[width=2in,height=2in]{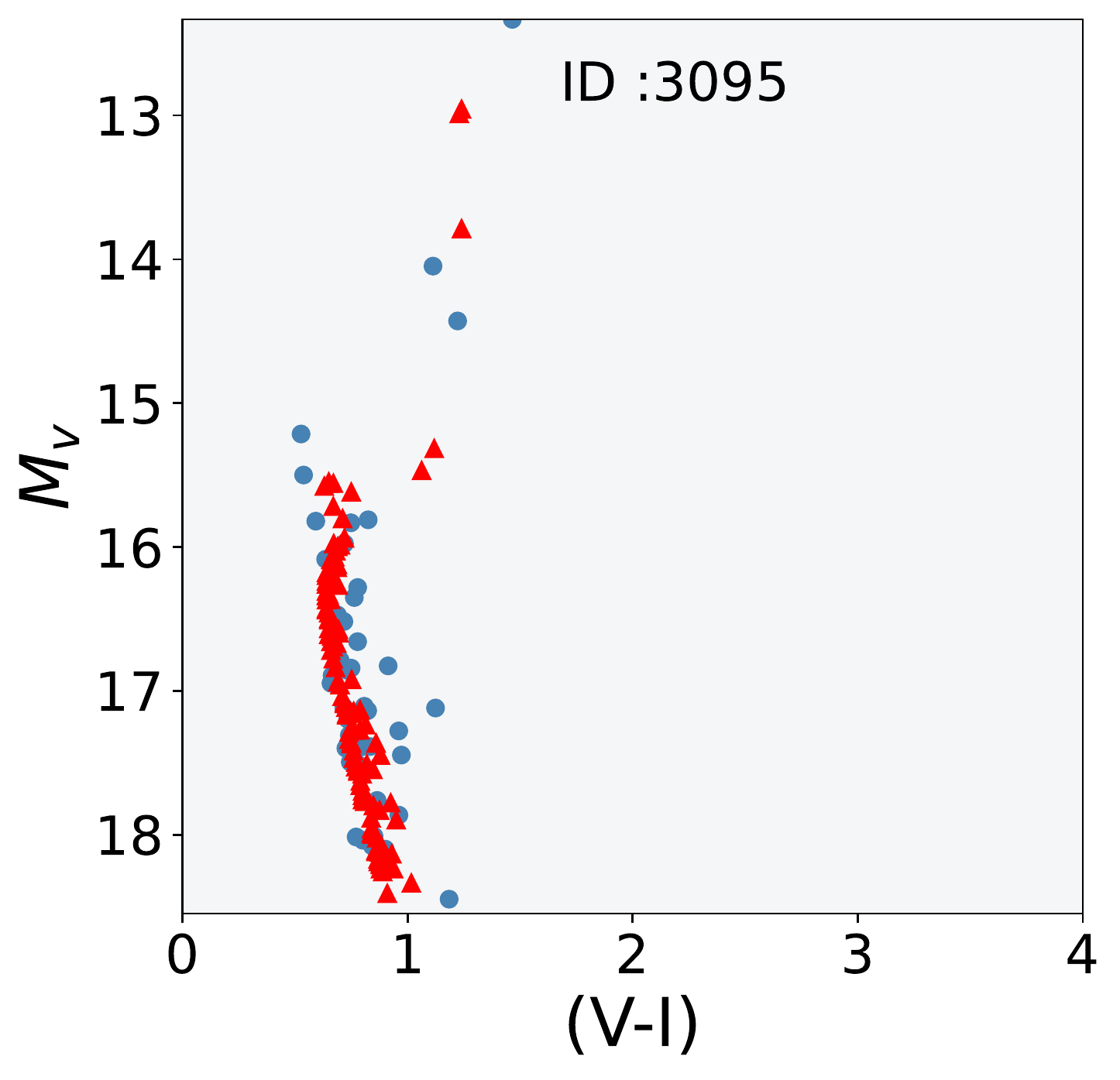}
}

\end{center}

\caption{Same as Fig~\ref{fig:hr01} but for other 6 OCs.
}
\label{fig:hr04}
\end{figure*}

\begin{figure*}
\begin{center}
    \subfigure{
\includegraphics[width=1.6in,height=1.6in]{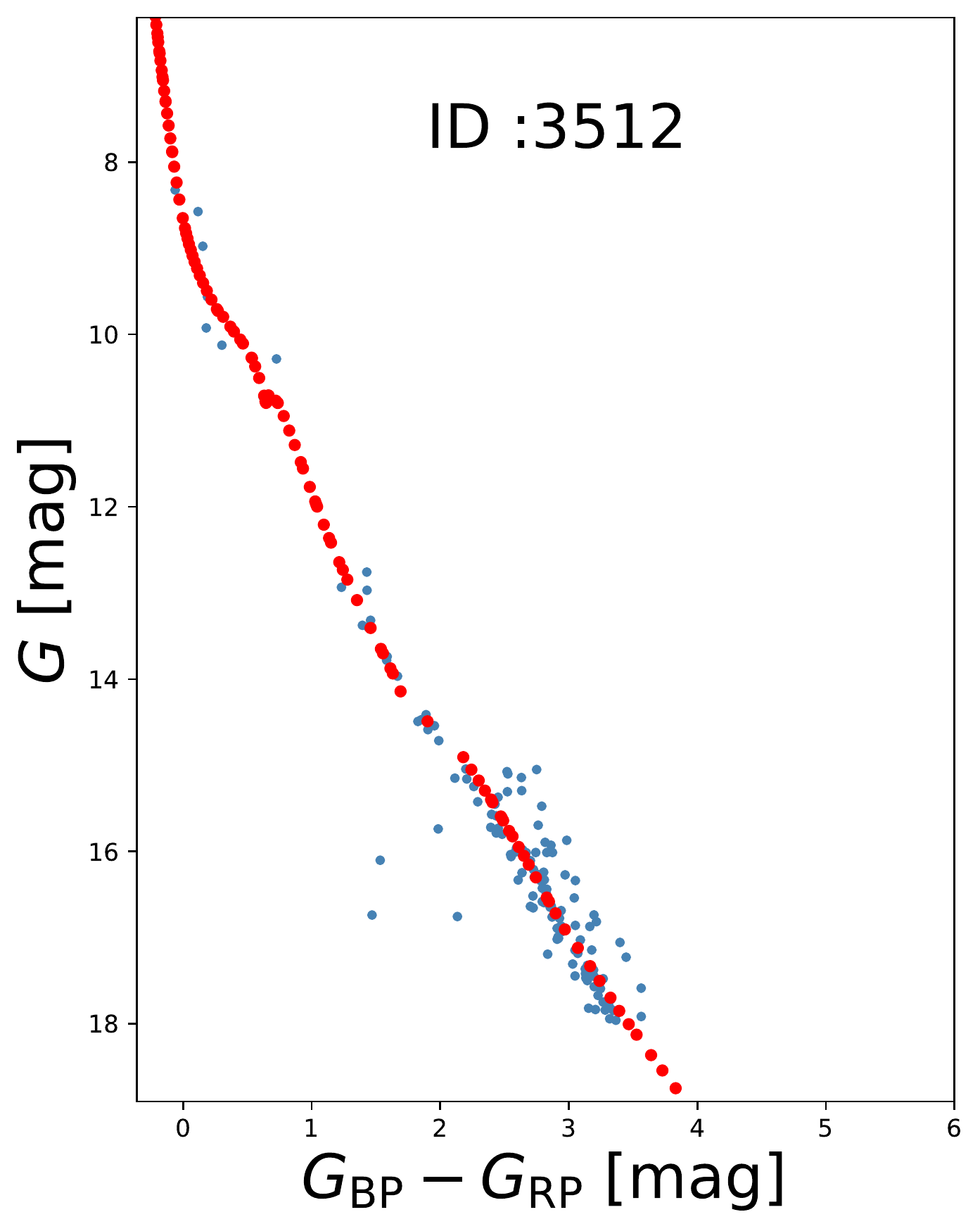}
}  
\subfigure{
\includegraphics[width=1.6in,height=1.6in]{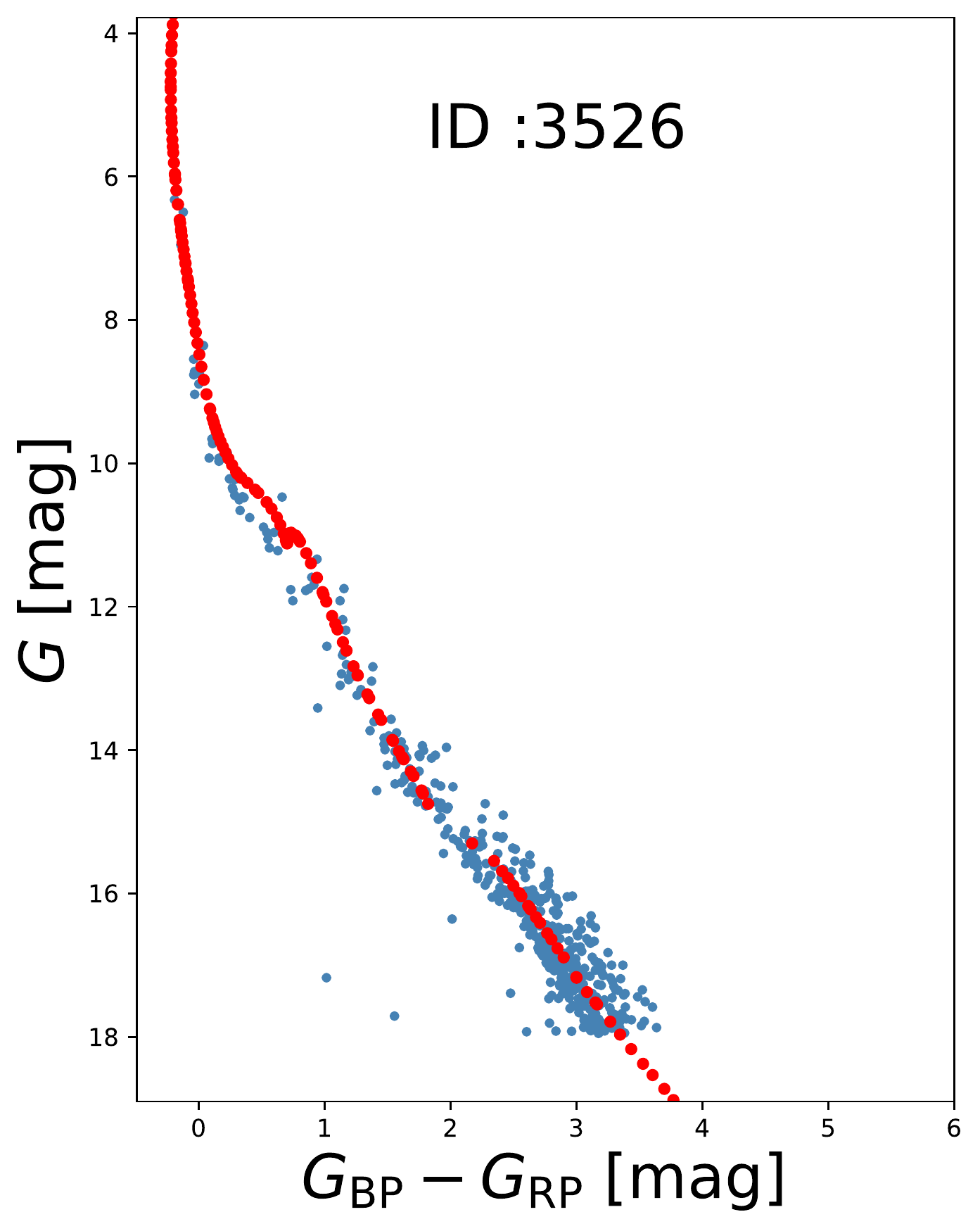}
}  
\subfigure{
\includegraphics[width=1.6in,height=1.6in]{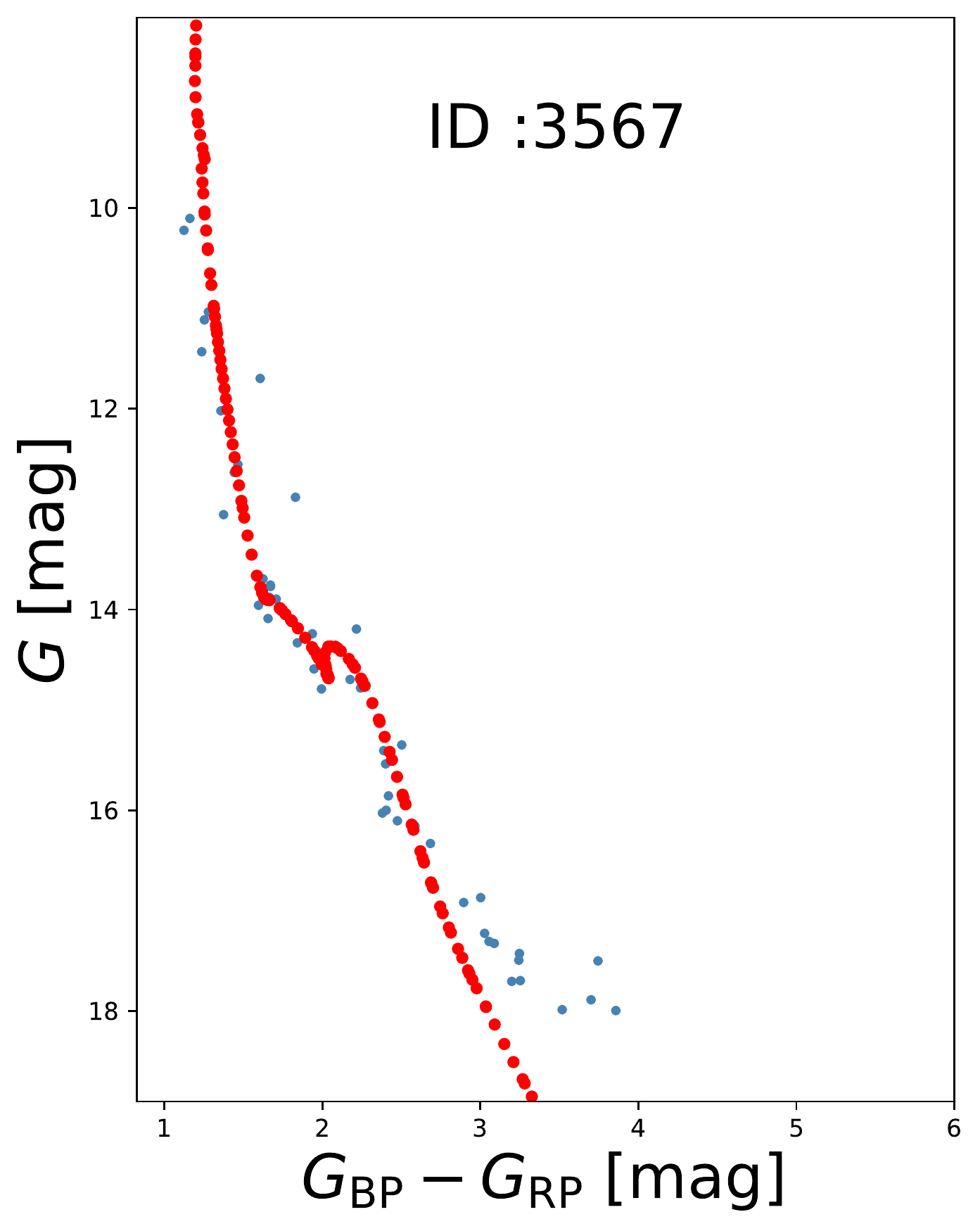}
}  
\subfigure{
\includegraphics[width=1.6in,height=1.6in]{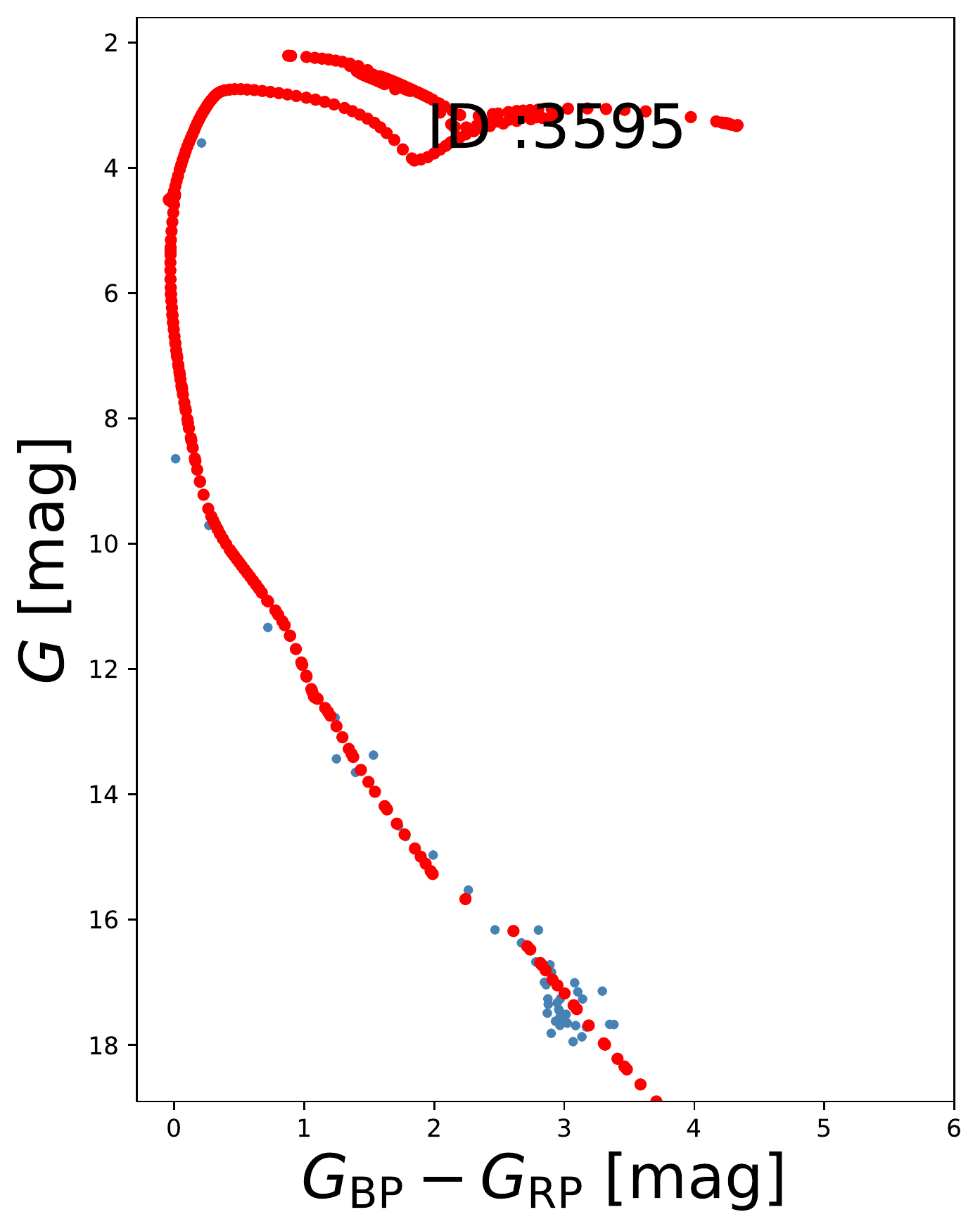}
}  

\end{center}

\caption{ Blue points represent cluster members. The red dotted curve is the best-fitting isochrone. }
\label{fig:4isochrone-fitting-1}
\end{figure*}
